\documentclass[aps,pre,showpacs,superscriptaddress,nofootinbib,twocolumn]{revtex4}

\usepackage{amsmath,amssymb,multirow}
\usepackage[tight]{subfigure}
\usepackage[pdftex]{graphicx}
\usepackage[boxed]{algorithm}
\usepackage{algorithmic}
\usepackage{color}

\renewcommand{\leq}{\leqslant}
\renewcommand{\geq}{\geqslant}

\begin{document}

\title{Network Geometry Inference using Common Neighbors}

\author{Fragkiskos Papadopoulos}
\affiliation{Department of Electrical Engineering, Computer Engineering and Informatics,
Cyprus University of Technology, Saripolou 33, Limassol 3036, Cyprus}
\author{Rodrigo Aldecoa}
\affiliation{Northeastern University, Department of Physics, Boston, MA, USA}
\author{Dmitri Krioukov}
\affiliation{Northeastern University, Department of Physics, Department of Mathematics,
Department of Electrical\&Computer Engineering, Boston, MA, USA}

\begin{abstract}
We introduce and explore a new method for inferring hidden geometric coordinates of nodes in complex networks based on the number of common neighbors between the nodes. We compare this approach to the HyperMap method, which is based only on the connections (and disconnections) between the nodes, i.e., on the links that the nodes have (or do not have).  We find that for high degree nodes the common-neighbors approach yields a more accurate inference than the link-based method, unless heuristic periodic adjustments (or ``correction steps") are used in the latter. The common-neighbors approach is computationally intensive, requiring $O(t^4)$ running time to map a network of $t$ nodes, versus $O(t^3)$ in the link-based method. But we also develop a hybrid method with $O(t^3)$ running time, which combines the common-neighbors and link-based approaches, and explore a heuristic that reduces its running time further to $O(t^2)$, without significant reduction in the mapping accuracy. We apply this method to the Autonomous Systems (AS) Internet, and reveal how soft communities of ASes evolve over time in the similarity space. We further demonstrate the method's predictive power by forecasting future links between ASes. Taken altogether, our results advance our understanding of how to efficiently and accurately map real networks to their latent geometric spaces, which is an important necessary step towards understanding the laws that govern the dynamics of nodes in these spaces, and the fine-grained dynamics of network connections.
\end{abstract}

\pacs{89.75.Fb; 02.40.-k; 02.50.Tt}

\maketitle

\section{Introduction}
\label{sec:intro}

The main premise of preferential attachment~\cite{BarAlb99} is that popularity is attractive~\cite{DoMe00pop}, but similarity is also attractive~\cite{McPh01}. Combined together these two attractive forces, popularity and similarity, have shown to form hidden hyperbolic geometries that drive evolution of networks~\cite{PaBoKr11}. Since these geometries are hidden, effective, or latent, they must be inferred from the network structure. Specifically what must be inferred are node coordinates in these underlying hyperbolic spaces. Existing approaches~\cite{hypermap_ton,BoPa10} to such inference are based on the connections (and disconnections) between the nodes, i.e., on the links that the nodes have (or do not have). Connected nodes are attracted to each other, while disconnected nodes repel, and these approaches are placing nodes into a hyperbolic space based on these attraction and repulsion forces.  Both approaches in~\cite{hypermap_ton,BoPa10} are based on \emph{Maximum Likelihood Estimation}. The approach in~\cite{BoPa10} embeds a given network topology into the hyperbolic plane by maximizing the likelihood that the topology is produced by the equilibrium hyperbolic network model~\cite{KrPa10}, while the approach in~\cite{hypermap_ton} embeds the network by maximizing the likelihood that the topology is produced by the hyperbolic model of growing networks~\cite{PaBoKr11}. Both approaches produce similar results, even though there are fundamental differences between them. In this paper, we build on the latter approach~\cite{hypermap_ton}, which is more recent and simpler to implement.

The work in~\cite{PaBoKr11} shows that trade-offs between popularity and similarity shape the structure and dynamics of growing complex networks, and that these tradeoffs in network dynamics give rise to hyperbolic geometry. The growing network model in~\cite{PaBoKr11} is essentially a model of random geometric graphs growing in hyperbolic spaces. Synthetic graphs grown according to this simple model simultaneously exhibit many common structural and dynamical characteristics of real networks. We call the model in~\cite{PaBoKr11} the Popularity$\times$Similarity Optimization (PSO) model.

Given the ability of the PSO model to construct synthetic growing networks that resemble real networks across a wide range of structural and dynamical characteristics, the work in~\cite{hypermap_ton} showed how to reverse this synthesis, and given a real network, how to map (embed) the network into the hyperbolic plane, in a way congruent with the PSO model. Specifically, the mapping method of~\cite{hypermap_ton}, called \emph{HyperMap}, replays the network's geometric growth, estimating at each time step the hyperbolic coordinates of new nodes by maximizing the likelihood of the network snapshot in the model. In the inferred polar coordinates of nodes, the radial coordinate $r$ can be associated with node popularity, while the angular coordinate $\theta$ is the node coordinate in the similarity space abstracted by a circle. HyperMap has been applied to the Autonomous Systems (AS) topology of the real Internet in~\cite{hypermap_ton}, where it was shown that: (i) the method can identify soft communities of ASes belonging to the same geographic region, even though the method is completely geography-agnostic; (ii) the method can predict \emph{missing} links between ASes with high precision, outperforming popular existing methods; and (iii) that the method can construct a highly navigable Internet map---greedy forwarding in the map can reach destinations with more than 90\% success probability and low stretch.

Here we introduce and explore a new method for inferring the node similarity coordinates, and release its implementation to public~\cite{hypermap-cn}. This method differs from the one in~\cite{hypermap_ton} in that it is not based on the links that the nodes have or do not have. Instead, it is based on the \emph{number of common neighbors} between the nodes. The method is inspired by the observation that the number of common neighbors between two nodes is a measure of similarity between the nodes; in general, the more the common neighbors between two nodes the more similar the two nodes are, i.e., the smaller their similarity distance~\cite{sarkar11, maksim_bipartite}. We call the approach in~\cite{hypermap_ton} the \emph{link-based approach}, and the approach considered here the \emph{common-neighbors approach}. We compare the two approaches and find that for high degree nodes the common-neighbors approach yields a more accurate inference than the link-based method, unless heuristic periodic adjustments (or ``correction steps"~\cite{hypermap_ton}) are used in the latter. On the other hand, the common-neighbors approach is computationally intensive, requiring $O(t^4)$ running time to map a network of $t$ nodes, versus $O(t^3)$ in the link-based method.

Based on these above observations, we then introduce a hybrid method with $O(t^3)$ running time, which combines the common-neighbors and link-based approaches, and explore a heuristic that can reduce its running time further to $O(t^2)$, without significantly sacrificing the embedding accuracy. We apply this method to snapshots of the real Internet to reveal how soft communities of ASes evolve over time in the similarity space.  We also demonstrate the method's predictive power by forecasting \emph{future} links between ASes. Taken altogether, our results advance our understanding of how to efficiently and accurately map real networks to their latent hyperbolic spaces, which is an important necessary step towards understanding the laws that govern the dynamics of nodes in these spaces, and the fine-grained dynamics of network connections.

The rest of the paper is organized as follows. In Section~\ref{sec:preliminaries}, we review the extended PSO (E-PSO) model from~\cite{hypermap_ton} and the details of the HyperMap method that we need in this paper. In Section~\ref{sec:common_neighbors_mapping}, we show how the angular coordinates of nodes can be inferred using the common-neighbors approach, and describe the hybrid method. In Section~\ref{sec:embedding_speedup}, we describe how to speedup the method, and in Section~\ref{sec:validation}, we validate our results in synthetic networks. In Section~\ref{sec:internet}, we apply the hybrid method to the AS Internet. In Section~\ref{sec:other_work}, we discuss other relevant work, and in Section~\ref{sec:conclusion}, we conclude with a discussion of open problems and future work.

\section{Preliminaries}
\label{sec:preliminaries}

The E-PSO model of growing networks has been introduced in~\cite{hypermap_ton} for HyperMap development purposes. As its name suggests, this model is a modification of the PSO model in~\cite{PaBoKr11}. The E-PSO model constructs growing networks using external links only, while being equivalent to the generalized PSO model in~\cite{PaBoKr11} that uses both external and internal links~\cite{hypermap_ton}. External links connect new nodes to existing nodes, while internal links appear between existing nodes only. Given a single snapshot of the topology of a real network, there is no way to distinguish external links from internal links. The E-PSO model sidesteps this obstacle, and helps to map a given real network topology by replaying its geometric growth, treating all links in the topology \emph{as if} they were external~\cite{hypermap_ton}. Below, we first review the E-PSO model, and then proceed to HyperMap, which is based on this model. We limit the exposition only to the basic details that we need in the rest of the paper.

\subsection{The E-PSO model}
\label{sec:e_pso}

The E-PSO model has five input parameters $m  \geq 0$, $L \geq 0$, $\beta\in(0,1]$, $T\in[0,1)$, and $\zeta>0$.
Parameters $m$ and $L$ are the rates at which external and internal links appear. (We will explain shortly how we compute them in a real network.)
These two parameters appear inside Eq.~(\ref{eq:m_i_t}) below, and define the average node degree in the network, $\bar{k} \approx 2(m+L)$. Parameter $\beta$ defines the exponent $\gamma=1+1/\beta\geq 2$ of the power-law degree distribution $P(k) \sim k^{-\gamma}$ in the network. Temperature $T$ controls the average clustering~$\bar{c}$~\cite{Dorogovtsev10-book} in the network, which is maximized at $T=0$ and nearly linearly decreases to zero with $T\in[0,1)$. Parameter $\zeta=\sqrt{-K}$ where $K$ is the curvature of the hyperbolic plane. As it is going to be explained, changing $\zeta$ rescales the node radial coordinates. This rescaling parameter does not affect any topological properties of networks generated by the model. Therefore, it can be set to any value in the model, e.g., $\zeta=1$, without loss of generality.  Having these parameters and the final size of the network $t > 0$ specified, the E-PSO model constructs a growing scale-free network up to $t$ nodes according to the following {\bf E-PSO model definition}:
\begin{enumerate}
\item[(1)] initially the network is empty;
\item[(2)] coordinate assignment and update:
\begin{enumerate}
\item at time $i=1,2,\ldots,t$, new node $i$ is added to the hyperbolic plane at polar coordinates $(r_i, \theta_i)$, where radial coordinate $r_i=\frac{2}{\zeta}\ln{i}$, while the angular coordinate $\theta_i$ is sampled uniformly at random from $[0, 2\pi]$;
\item each existing node $j=1,2,\ldots,i-1$, moves increasing its radial coordinate according to
$r_j(i)=\beta r_j +(1-\beta)r_i$;
\end{enumerate}
\item[(3)] creation of edges: node $i$ connects to each existing node $j=1,2,\ldots,i-1$ with probability $p_{ij} \equiv p(x_{ij})$ given by
\begin{equation}
\label{eq:p_x_ji}
p(x_{ij})=\frac{1}{1+e^{\frac{\zeta}{2T}(x_{ij}-R_i)}}.
\end{equation}
\end{enumerate}
In the last expression, $x_{ij}$ is the hyperbolic distance between nodes $i$ and $j$~\cite{Bonahon09-book},
\begin{eqnarray}
\label{eq:x_ji}
\nonumber \cosh{\zeta x_{ij}}&=&\cosh{\zeta r_i}\cosh{\zeta r_j(i)}\\
&&-\sinh{\zeta r_i} \sinh {\zeta r_j(i)} \cos{\theta_{ij}}\\
\nonumber\quad\textnormal{where}~\theta_{ij}&=&\pi-|\pi-|\theta_i-\theta_j||,
\end{eqnarray}
while $R_i$ is given by
\begin{equation}
\label{eq:R_i}
R_i=r_i-\frac{2}{\zeta}\ln\left[\frac{2T}{\sin{T\pi}}\frac{I_i}{\bar{m}_i(t)}\right],
\end{equation}
with $I_i=\frac{1}{1-\beta}(1-i^{-(1-\beta)})$. Equation~(\ref{eq:R_i}) is derived from the condition that the expected number of old nodes $j < i$ that $i$ connects to, denoted by $\bar{m}_i(t)$, is
\begin{align}
\label{eq:m_i_t}
\bar{m}_i(t)&=m+\frac{2L (1-\beta)}{(1-t^{-(1-\beta)})^2(2\beta-1)}\nonumber\\
&\times\left[\left(\frac{t}{i}\right)^{2\beta-1}-1\right]\left[1-i^{-(1-\beta)}\right].
\end{align}

The radial coordinate of a node abstracts its popularity. The smaller the radial coordinate of a node, the more popular the node is, and the more likely it attracts new connections. The angular distance between two nodes abstracts their similarity. The smaller this distance, the more similar the two nodes are, and the more likely they are connected. The hyperbolic distance $x_{ij}$ is then a single-metric representation of a combination of the two attractiveness attributes, radial popularity and angular similarity. The connection probability $p(x_{ij})$ is a decreasing function of $x_{ij}$, meaning that new connections take place by optimizing trade-offs between popularity and similarity~\cite{PaBoKr11}. It has been shown~\cite{hypermap_ton} that the E-PSO model can reproduce not only the degree distribution and clustering of real networks like the AS Internet, but also several other important properties. Given the ability of the model to construct growing synthetic networks that resemble real networks,~\cite{hypermap_ton} then showed how to reverse this synthesis, and given a real network, how to map (embed) the network into the hyperbolic plane, in a way congruent with the E-PSO model. The mapping method, HyperMap, is described next.

\subsection{HyperMap}
\label{sec:hypermap}

HyperMap is based on Maximum Likelihood Estimation (MLE) and is fully specified in Fig.~\ref{fig:the_method}. On its input it takes the network adjacency matrix  $\alpha_{ij}$ ($\alpha_{ij}=\alpha_{ji}=1$ if there is a link between nodes $i$ and $j$, and $\alpha_{ij}=\alpha_{ji}=0$ otherwise), and the network parameters $m, L, \gamma, T, \zeta$. It then computes radial and angular coordinates $r_i(t), \theta_i$, for all nodes $i \leq t$ in the network.

HyperMap first estimates the MLE appearance (or birth) times of nodes $i=1,2,\ldots,t$. As shown in~\cite{hypermap_ton}, the higher the degree of a node in the E-PSO model the earlier its MLE appearance time. Therefore, HyperMap uses the following procedure for finding the MLE of the node appearance times in a given network with $t$ nodes. It sorts all nodes in the decreasing order of their degrees $k_1>k_2>\ldots>k_t$, with ties broken arbitrarily, and sets their MLE appearance times $i=1,2,\ldots,t$ in the same order. That is, the node with the largest degree $k_1$ is expected to appear first, $i=1$, the second largest degree node $k_2$ appeared second, $i=2$, and so on. The node born at time $i$ is called node $i$.

Having a sequence of MLE node birth times, HyperMap replays the hyperbolic growth of the network in accordance with the E-PSO model as follows. When a node is born at time $1 \leq i \leq t$, it is assigned an initial radial coordinate $r_i=\frac{2}{\zeta}\ln{i}$, and every existing node $j < i$ moves increasing its radial coordinate according to $r_j(i)=\beta r_j +(1-\beta)r_i$. The method assigns to a new node $i > 1$ the angular coordinate $\theta_i$ that maximizes its local likelihood
\begin{equation}
\label{eq:local_likelihood_links}
\mathcal L_\textnormal{L}^{i}=\prod_{1 \leq j < i} p(x_{ij})^{\alpha_{ij}}\left[1-p(x_{ij})\right]^{1-\alpha_{ij}}.
\end{equation}
This likelihood is a function of $\theta_i$, since $x_{ij}$ depends on $\theta_i$, see Eq.~(\ref{eq:x_ji}), $p(x_{ij})$ depends on $x_{ij}$, see Eq.~(\ref{eq:p_x_ji}), and $\mathcal L_\textnormal{L}^{i}$ depends on $p(x_{ij})$.
The product in Eq.~(\ref{eq:local_likelihood_links}) goes over all the old nodes $j < i$. The likelihood $\mathcal L_\textnormal{L}^{i}$ is called \emph{local} as it depends only on the connections (and disconnections) between new node $i$ and existing nodes $j < i$. For example, if new node $i=4$ is connected to nodes $1, 2$ but not to node $3$, i.e., $\alpha_{41}=1, \alpha_{42}=1, \alpha_{43}=0$, then $\mathcal L_\textnormal{L}^{4}$ would be $\mathcal L_\textnormal{L}^{4}=p(x_{41})p(x_{42})(1-p(x_{43}))$. We use the subscript L to emphasize that $\mathcal L_\textnormal{L}^{i}$ depends on the links between new node $i$ and existing nodes $j < i$, i.e., it is a \emph{link-based} approach. In the next section, we will derive an alternative local likelihood, $\mathcal L_\textnormal{CN}^{i}$, which depends on the number of common neighbors between new node $i$ and existing nodes $j < i$.
\begin{figure}[!ht]
\begin{center}
\begin{minipage}{3.4in}
\begin{algorithm}[H]
\begin{algorithmic}[1]
\STATE Sort node degrees in decreasing order $k_1> k_2>\ldots>k_t$ with ties broken arbitrarily.
\STATE Call node $i$, $i=1,2,\ldots,t$, the node with degree $k_i$.
\STATE Node $i=1$ is born, assign to it initial radial coordinate $r_1=0$ and random angular coordinate $\theta_1 \in [0, 2\pi]$.
\FOR{$i=2$ to $t$}
    \STATE ~~~Node $i$ is born, assign to it initial radial coordinate\\~~~$r_i=\frac{2}{\zeta}\ln{i}$.
    \STATE ~~~Increase the radial coordinate of every existing node\\~~~$j < i$ according to $r_j(i)=\beta r_j +(1-\beta)r_i$.
    \STATE ~~~Assign to node $i$ angular coordinate $\theta_i$ maximizing $\mathcal L_\textnormal{L}^{i}$\\~~~given by Equation (\ref{eq:local_likelihood_links}).
\ENDFOR
\end{algorithmic}
\end{algorithm}
\end{minipage}
\end{center}
\caption{The HyperMap Embedding Algorithm.
\label{fig:the_method}}
\end{figure}

The maximization of $\mathcal L_\textnormal{L}^{i}$ is performed numerically, by sampling the likelihood $\mathcal L_\textnormal{L}^{i}$ at different values of $\theta$ in $[0, 2\pi]$ separated by intervals $\Delta\theta=\frac{1}{i}$, and then setting $\theta_i$ to the value of $\theta$ that yields the largest value of $\mathcal L_\textnormal{L}^{i}$. Since, to compute $\mathcal L_\textnormal{L}^{i}$ for a given $\theta$ we need to compute the connection probability between node $i$ and all existing nodes $j <i$, we need a total of $O(i^2)$ steps to perform the maximization. If there are $t$ nodes in total, HyperMap needs $O(t^3)$ running time to map the full network.

\emph{Specifying input parameters.}  Parameter $\zeta > 0$ can be set to any value, e.g., $\zeta=1$. As mentioned, changing the value of this parameter corresponds to radial coordinate rescaling. Specifically, the radial coordinates of nodes will be rescaled by the factor $\zeta$, since as can be seen by steps 5 and 6 in Fig.~\ref{fig:the_method}, at the final time $i=t$, $r_j(t)=\beta r_j +(1-\beta)r_t=\frac{2 \beta}{\zeta}\ln{j} + \frac{2(1-\beta)}{\zeta}\ln{t},~j \leq t$. Furthermore, the likelihood $\mathcal L_\textnormal{L}^{i}$ in Eq.~(\ref{eq:local_likelihood_links}) does not depend on $\zeta$, as it cancels out in the connection probability  $p(x_{ij})$ in Eq.~(\ref{eq:p_x_ji}). That is, different values of $\zeta$ will yield exactly the same angular coordinates. Parameter $m$ can be obtained from historical data of the evolution of the network. If such data is available, then $m$ is the average number of connections that nodes have once they first appear in the data. If no historical data are available, $m$ can be set, as an approximation, to the minimum observed node degree in the network. Given the average node degree $\bar{k}$ in the network, and knowing $m$ and $\bar{k}$, we get $L=\frac{\bar{k}-2m}{2}$. The power law exponent $\gamma$ can be obtained from the degree distribution of the network, while parameter $T$ is found experimentally~\cite{hypermap_ton}. We emphasize that the parameters for HyperMap come directly from the observation of the real network. With these 5 parameters ($m, L, \gamma, T, \zeta$), and the network adjacency matrix $\alpha_{ij}$, HyperMap infers $2 t$ hyperbolic node coordinates in a network of $t$ nodes (a radial and angular coordinate for each node), and consequently, $O(t^2)$ hyperbolic distances between nodes.

\section{Inferring node similarity coordinates using the number of common neighbors}
\label{sec:common_neighbors_mapping}

We now show how the angular (similarity) coordinates of nodes can be inferred using the number of common neighbors between new and old nodes, instead of the connections and disconnections between them. Specifically, we first derive an alternative local likelihood, $\mathcal L_\textnormal{CN}^{i}$, which uses the observed number of common neighbors between each new node $i$ and each existing node $j < i$ at final time $t$. Then, we use this likelihood in place of $\mathcal L_\textnormal{L}^{i}$ in Equation~(\ref{eq:local_likelihood_links}) in order to infer the angular coordinate of each node.

In Section~\ref{sec:validation}, we show that for small $i$'s, i.e., for nodes that appear at early MLE times, which are the high degree nodes, $\mathcal L_\textnormal{CN}^{i}$ yields a more accurate angular coordinate inference than $\mathcal L_\textnormal{L}^{i}$. This is because, for all node pairs $i, j$, $j < i$, $\mathcal L_\textnormal{CN}^{i}$ utilizes more information, since it uses the final number of common neighbors between the pairs. That is, it considers the \emph{full} network adjacency matrix, i.e., the network adjacency matrix at the final time $t$, and uses the number of common neighbors between the node pairs at that time. In contrast, $\mathcal L_\textnormal{L}^{i}$ in Eq.~(\ref{eq:local_likelihood_links}) uses less information, since at each time $i \leq t$ it considers only the connections and disconnections between node $i$ and old nodes $j < i$. To derive $\mathcal L_\textnormal{CN}^{i}$ we first need to compute the distribution of the number of common neighbors between node pairs in the E-PSO model, which is the task we perform next.

\subsection{Distribution of the number of common neighbors}

Consider a network that has grown up to $t$ nodes according to E-PSO (Section~\ref{sec:e_pso}), where nodes are numbered according to the order they appear. Consider two nodes $i, j$ with $j < i$ and a third node $k$. The initial radial coordinates of these nodes are $r_i=\frac{2}{\zeta}\ln{i}, r_j=\frac{2}{\zeta}\ln{j}$ and $r_k=\frac{2}{\zeta}\ln{k}$. We first need to find $p(i,j, \theta_i, \theta_j ; k)$, which is the probability that $i$ and $j$ are both connected to $k$ given their angular coordinates $\theta_i, \theta_j$. Below, we distinguish three cases and compute corresponding probabilities $p_1(i, j, \theta_i, \theta_j ; k)$, $p_2(i, j, \theta_i, \theta_j ; k)$ and $p_3(i, j, \theta_i, \theta_j ; k)$.

\underline{Case 1: $i > j > k$.} In this case, the connections to $k$ happen when $j$ and $i$ first appear, i.e., at times $j$ and $i$ respectively.
Therefore,
\begin{widetext}
\begin{eqnarray}
\label{eq:p_case_1}
p_1(i, j, \theta_i, \theta_j ; k)=\frac{1}{2\pi}\int_{0}^{2\pi} \frac{1}{1+e^{\frac{\zeta}{2T}(x_{jk}-R_j)}}\times\frac{1}{1+e^{\frac{\zeta}{2T}(x_{ik}-R_i)}} d\theta_k,\text{ where}\\
\nonumber x_{jk}=\frac{1}{\zeta}\mathrm{arccosh}\left[\cosh{\zeta r_j}\cosh{\zeta r_k(j)}-\sinh{\zeta r_j} \sinh {\zeta r_k(j)} \cos{\theta_{jk}}\right],\\
\nonumber x_{ik}=\frac{1}{\zeta}\mathrm{arccosh}\left[\cosh{\zeta r_i}\cosh{\zeta r_k(i)}-\sinh{\zeta r_i} \sinh {\zeta r_k(i)} \cos{\theta_{ik}}\right],\\
\nonumber R_j=r_j-\frac{2}{\zeta}\ln\left[\frac{2T}{\sin{T\pi}}\frac{I_j}{\bar{m}_j(t)}\right],~~R_i=r_i-\frac{2}{\zeta}\ln\left[\frac{2T}{\sin{T\pi}}\frac{I_i}{\bar{m}_i(t)}\right],\\
\nonumber r_k(j)=\beta r_k+(1-\beta)r_j,~~r_k(i)=\beta r_k+(1-\beta)r_i.
\end{eqnarray}
\end{widetext}
\underline{Case 2: $i > k > j$.} Here the connection between $i$ and $k$ happens when $i$ first appears, i.e., at time $i$, and the connection between $j$ and $k$ happens when $k$ first appears, i.e., at time $k$. Thus,
\begin{widetext}
\begin{eqnarray}
\label{eq:p_case_2}
p_2(i, j, \theta_i, \theta_j ; k) =\frac{1}{2\pi}\int_{0}^{2\pi} \frac{1}{1+e^{\frac{\zeta}{2T}(x_{kj}-R_k)}}\times\frac{1}{1+e^{\frac{\zeta}{2T}(x_{ik}-R_i)}} d\theta_k,\text{ where}\\
\nonumber x_{kj}=\frac{1}{\zeta}\mathrm{arccosh}\left[\cosh{\zeta r_k}\cosh{\zeta r_j(k)}-\sinh{\zeta r_k} \sinh {\zeta r_j(k)} \cos{\theta_{jk}}\right],\\
\nonumber x_{ik}=\frac{1}{\zeta}\mathrm{arccosh}\left[\cosh{\zeta r_i}\cosh{\zeta r_k(i)}-\sinh{\zeta r_i} \sinh {\zeta r_k(i)} \cos{\theta_{ik}}\right],\\
\nonumber R_k=r_k-\frac{2}{\zeta}\ln\left[\frac{2T}{\sin{T\pi}}\frac{I_k}{\bar{m}_k(t)}\right],~~R_i=r_i-\frac{2}{\zeta}\ln\left[\frac{2T}{\sin{T\pi}}\frac{I_i}{\bar{m}_i(t)}\right],\\
\nonumber r_j(k)=\beta r_j+(1-\beta)r_k,~~r_k(i)=\beta r_k+(1-\beta)r_i.
\end{eqnarray}
\end{widetext}
\underline{Case 3: $k > i > j$.} In this final case, both connections with $k$ happen when $k$ appears, i.e., at time $k$. Therefore,
\begin{widetext}
\begin{eqnarray}
\label{eq:p_case_3}
p_3(i, j, \theta_i, \theta_j ; k)=\frac{1}{2\pi}\int_{0}^{2\pi} \frac{1}{1+e^{\frac{\zeta}{2T}(x_{kj}-R_k)}}\times\frac{1}{1+e^{\frac{\zeta}{2T}(x_{ki}-R_k)}} d\theta_k,\text{ where}\\
\nonumber x_{kj}=\frac{1}{\zeta}\mathrm{arccosh}\left[\cosh{\zeta r_k}\cosh{\zeta r_j(k)}-\sinh{\zeta r_k} \sinh {\zeta r_j(k)} \cos{\theta_{jk}}\right],\\
\nonumber x_{ki}=\frac{1}{\zeta}\mathrm{arccosh}\left[\cosh{\zeta r_k}\cosh{\zeta r_i(k)}-\sinh{\zeta r_k} \sinh {\zeta r_i(k)}
\cos{\theta_{ik}}\right],\\
\nonumber R_k=r_k-\frac{2}{\zeta}\ln\left[\frac{2T}{\sin{T\pi}}\frac{I_k}{\bar{m}_k(t)}\right],~~r_j(k)=\beta r_j+(1-\beta)r_k,~~r_i(k)=\beta r_i+(1-\beta)r_k.
\end{eqnarray}
\end{widetext}
The integrals in Equations~(\ref{eq:p_case_1})--(\ref{eq:p_case_3}) can be only computed numerically. Since the connection events are statistically independent, the number of common neighbors between nodes $i$ and $j$, $ j < i$, given their angles $\theta_i, \theta_j$, is a sum of independent Bernoulli trials with different success probabilities, given by Equations~(\ref{eq:p_case_1})--(\ref{eq:p_case_3}). Therefore, by the Central Limit Theorem~\cite{probability-and-measure}, for sufficiently large network sizes $t$, the distribution of the number of common neighbors $n_{ij}$ between $i$ and $j$ is approximately normally distributed, i.e., its probability density is approximately
\begin{equation}
\label{eq:normal_cn}
f(n_{ij} | \theta_i, \theta_j)=\frac{1}{\sigma(i, j, \theta_i, \theta_j)\sqrt{2\pi}}e^{-\frac{\left(n_{ij}-\mu(i, j, \theta_i, \theta_j)\right)^2}{2\sigma^2(i, j, \theta_i, \theta_j)}},
\end{equation}
where its mean $\mu(i, j, \theta_i, \theta_j)$ and variance $\sigma^2(i, j, \theta_i, \theta_j)$ are
\begin{eqnarray}
\label{eq:mu}
\nonumber \mu(i, j, \theta_i, \theta_j)&=&\sum_{k=1}^{j-1} p_1(i, j, \theta_i, \theta_j ; k)+\sum_{k=j+1}^{i-1} p_2(i, j, \theta_i, \theta_j ; k)\\
&+& \sum_{k=i+1}^{t} p_3(i, j, \theta_i, \theta_j ; k),
\end{eqnarray}
\begin{eqnarray}
\label{eq:sigma}
\nonumber \sigma^2(i, j, \theta_i, \theta_j)&=&\sum_{k=1}^{j-1} p_1(i, j, \theta_i, \theta_j ; k)(1-p_1(i, j, \theta_i, \theta_j ; k))\\
\nonumber &+&\sum_{k=j+1}^{i-1} p_2(i, j, \theta_i, \theta_j ; k)(1-p_2(i, j, \theta_i, \theta_j ; k))\\
\nonumber &+&\sum_{k=i+1}^{t} p_3(i, j, \theta_i, \theta_j ; k)(1-p_3(i, j, \theta_i, \theta_j ; k)).\\
\end{eqnarray}
To compute $\mu(i, j, \theta_i, \theta_j)$ and $\sigma(i, j, \theta_i, \theta_j)$ we use the fact that the mean of a Bernoulli random variable with success probability $p$ is $p$, and its variance is $p(1-p)$. The computation of  $\mu(i, j, \theta_i, \theta_j)$ and $\sigma(i, j, \theta_i, \theta_j)$ for each $i, j$ pair requires $O(t)$ steps.

\subsection{Likelihood and likelihood maximization}

We are now ready to derive the likelihood $\mathcal L_{\textnormal{CN}}^{i}$ that we can use in place of $\mathcal L_{L}^{i}$ in Equation~(\ref{eq:local_likelihood_links}), in order to infer the node angular coordinates.

Consider new node $i \leq t$ in a network that grows according to E-PSO up to time $t$. We denote by $\mathcal L_1^{i} \equiv \mathcal L(\theta_i |r_i, \{r_j(i), \theta_j\}, \{n_{ij}^t\}, m, L, \gamma, T, \zeta)_{j<i}$ the likelihood that $i$'s angular coordinate takes value $\theta_i$, given its $r_i$, the coordinates of the old nodes $\{r_j(i), \theta_j\}\equiv\{r_1(i), \theta_1, r_2(i), \theta_2,\ldots,r_{i-1}(i), \theta_{i-1}\}$, the number of common neighbors between $i$ and each old node $j < i$ at the final time $t$, $\{n_{ij}^t\} \equiv \{n_{i1}^t, n_{i2}^t, \ldots, n_{ii-1}^t\}$, and the network parameters $m, L, \gamma, T, \zeta$. Since the distribution of the angular coordinates is uniform on $[0, 2\pi]$, we can rewrite $\mathcal L_1^{i}$ using Bayes' rule as
\begin{equation}
\label{eq:local_likelihood_1}
\mathcal L_1^{i}=\frac{1}{2\pi}\frac{\mathcal L_2^{i}}{\mathcal L_3^{i}},
\end{equation}
where $\mathcal L_2^{i} \equiv \mathcal L(\{n_{ij}^{t}\} | r_i, \theta_i, \{r_j(i), \theta_j\}, m, L, \gamma, T, \zeta)_{j<i}$ is the likelihood to have the numbers of common neighbors $\{n_{ij}^t\}$, if the angular coordinate of node $i$ has value $\theta_i$, conditioned on its radial coordinate, the coordinates of the old nodes, and the network parameters. Likelihood $\mathcal L_3^{i} \equiv \mathcal L(\{n_{ij}^t\} | r_i, \{r_j(i), \theta_j\}, m, L, \gamma, T, \zeta)_{j<i}$, independent of $\theta_i$, is the probability that $i$ has the numbers of common neighbors with old nodes specified by $\{n_{ij}^t\}$, conditioned as shown by notation.

We are looking for the angle $\theta_i^{*}$ that maximizes the likelihood $\mathcal L_1^{i}$ in Equation~(\ref{eq:local_likelihood_1}), or equivalently, $\mathcal{L}_{2}^i$. We can compute $\mathcal L_2^{i}$ using Equation~(\ref{eq:normal_cn})
\begin{equation}
\label{eq:local_likelihood_CN}
\mathcal L_2^{i}=\prod_{1 \leq j < i} f(n_{ij}^t | \theta_i, \theta_j)\equiv \mathcal{L}_{\textnormal{CN}}^i.
\end{equation}
The product goes over all the old nodes $j < i$. Equation~(\ref{eq:local_likelihood_CN}) gives the likelihood $\mathcal L_{\textnormal{CN}}^{i}$ that we can use in HyperMap in place of $\mathcal L_{L}^{i}$ in Eq.~(\ref{eq:local_likelihood_links}), where $n_{ij}^t$ is the observed number of common neighbors between nodes appearing at MLE times $i, j$, computed from the given network adjacency matrix $\alpha_{ij}$. Note that maximizing  $\mathcal{L}_{\textnormal{CN}}^i$ is equivalent to maximizing its logarithm $\ln{\mathcal{L}_{\textnormal{CN}}^i}$,
\begin{equation}
\label{eq:log_likelihood}
\ln{\mathcal{L}_{\textnormal{CN}}^i}=C-\sum_{j=1}^{i-1}\ln{\sigma(i, j, \theta_i, \theta_j)}-\sum_{j=1}^{i-1}\frac{\left(n_{ij}^t-\mu(i, j, \theta_i, \theta_j)\right)^2}{2\sigma^2(i, j, \theta_i, \theta_j)},
\end{equation}
where $C=(i-1)\ln{\frac{1}{\sqrt{2\pi}}}$, independent of $\theta_i$.

\subsection{$\mathcal L_{\textnormal{CN}}^{i}$ versus $\mathcal L_{\textnormal{L}}^{i}$, and the hybrid method}
\label{sec:comparison_and_hybrid_approach}

As with $\mathcal{L}_{\textnormal{L}}^i$, the maximization of $\mathcal L_{\textnormal{CN}}^{i}$ can be only performed numerically. This can be done in the same way as with $\mathcal{L}_{\textnormal{L}}^i$ (Section~\ref{sec:hypermap}), i.e., by sampling the likelihood $\mathcal L_{\textnormal{CN}}^{i}$ at different values of $\theta$ in $[0, 2\pi]$ separated by intervals $\Delta\theta=\frac{1}{i}$, and then setting $\theta_i^{*}$ to the value of $\theta$ that yields the largest value of $\mathcal L_{\textnormal{CN}}^{i}$. (To be more precise, we will be using sampling intervals $\Delta\theta=\min\{0.01, \frac{1}{i}\}$.) Since, to compute $\mathcal L_{\textnormal{CN}}^{i}$ for a given $\theta$ we need to compute $\mu(i, j, \theta_i, \theta_j)$ and $\sigma(i, j, \theta_i, \theta_j)$ between node $i$ and every existing node $j < i$, we need a total of $O(i^2 t)$ steps to perform the maximization of $\mathcal L_{\textnormal{CN}}^{i}$. Therefore, if there are $t$ nodes in total, HyperMap with $\mathcal L_{\textnormal{CN}}^{i}$ requires $O(t^4)$ running time to map the full network, versus $O(t^3)$ with  $\mathcal L_{\textnormal{L}}^{i}$.

Likelihoods $\mathcal L_{\textnormal{CN}}^{i}$ and $\mathcal L_{\textnormal{L}}^{i}$ yield different results for the first few nodes appearing at early MLE times. Specifically, all nodes $i$ for which  their average number of connections to previous nodes in Eq.~(\ref{eq:m_i_t}) is $\bar{m}_i(t) \geq i-1$, are expected to be connected to all previous nodes $j \leq i-1$ with a high probability. This condition holds for high degree nodes appearing at early MLE times, rendering their exact angular coordinate inference with $\mathcal L_{\textnormal{L}}^{i}$ infeasible. This is because $\mathcal L_{\textnormal{L}}^{i}$ uses the connections and disconnections between new and old nodes in order to place the nodes at the right angles; if new node $i$ is connected to all previous nodes $j < i$ with high probability then large zones of different angular coordinates are all quite likely with $\mathcal L_{\textnormal{L}}^{i}$. This effect was noted in~\cite{hypermap_ton}. In contrast, $\mathcal L_{\textnormal{CN}}^{i}$ can accurately infer the angular coordinates of nodes appearing early because it effectively utilizes ``future" connectivity information as well, i.e., the number of common neighbors between the nodes at the \emph{final} time $t$. This important difference between  $\mathcal L_{\textnormal{CN}}^{i}$ and $\mathcal L_{\textnormal{L}}^{i}$ is illustrated in Section~\ref{sec:validation}. We note that since the inference of  the angular coordinates of new nodes appearing at later MLE times depends on the inferred angles of high degree nodes appearing early, then if the latter are not accurately inferred, the former will not be accurately inferred either.

Given the angular coordinates of high degree nodes appearing at early MLE times, the inference of the angular coordinates of nodes appearing at later MLE times, e.g., of nodes $i$ for which $\bar{m}_i(t) < i-1$, using either $\mathcal L_{\textnormal{CN}}^{i}$ or $\mathcal L_{\textnormal{L}}^{i}$ yields similar results, i.e., the two likelihoods infer approximately the same angular coordinates for later nodes. This effect is also illustrated in Section~\ref{sec:validation}, and it means that one can use the following \emph{hybrid approach}:  use $\mathcal L_{\textnormal{CN}}^{i}$ for the first nodes $i$ for which $\bar{m}_i(t) \geq i-1$, and then use $\mathcal L_{\textnormal{L}}^{i}$ for the rest of the nodes for which $\bar{m}_i(t) < i-1$. The benefit of this approach is running time, as the number of nodes for which $\bar{m}_i(t) \geq  i-1$ is usually quite small, e.g., in the order of few tens of nodes. Therefore, HyperMap with this hybrid approach will still have $O(t^3)$ running time. In the next section, we describe a simple heuristic to reduce this running time to $O(t^2)$.

\subsection{Hybrid method versus $\mathcal L_{\textnormal{L}}^{i}$ with correction steps}
\label{sec:correction_steps}

It was shown in~\cite{hypermap_ton} that the accuracy of HyperMap can be improved by occasionally running ``correction steps''  right after step $7$ in Fig.~\ref{fig:the_method}. Specifically, at some predefined set of times $i$, we visit each existing node $j \leq i$, and having the coordinates of the rest of the nodes $l \leq i$, $l \neq j$, we update $j$'s angle to the value $\theta_j'$ that maximizes
\begin{equation}
\label{eq:local_likelihood_links_correction}
\widetilde{\mathcal L_\textnormal{L}^{j}}=\prod_{1 \leq l \leq i} p(x_{jl})^{\alpha_{jl}}\left[1-p(x_{jl})\right]^{1-\alpha_{jl}},~~l\ne j,
\end{equation}
where $x_{jl}$ is the hyperbolic distance between $j$ and $l$ when the youngest of the two nodes appeared, and $p(x_{jl})$ is given by Eq.~(\ref{eq:p_x_ji}), using in it $R_j$ if $j > l$ or $R_l$ if $j < l$. It has been observed in~\cite{hypermap_ton} that these correction steps are beneficial when run at relatively small times $i$. This observation fully agrees with our results in this paper.

These correction steps are a heuristic that tries to effectively recompute improved angles for the first (high degree) nodes, by considering not only the connections to their previous nodes, but also connections to nodes that appear later, i.e., future connectivity information, as in the common-neighbors approach. In Section~\ref{sec:validation} we show that HyperMap with $\mathcal L_{\textnormal{L}}^{i}$ and correction steps yields similar results to the hybrid method that does not use correction steps.

\section{Speeding up the method}
\label{sec:embedding_speedup}

As explained in Section~\ref{sec:comparison_and_hybrid_approach}, the running time of HyperMap with either the hybrid or link-based approaches is
$O(t^3)$. Here we introduce a simple heuristic that reduces this running time to $O(t^2)$ without significantly sacrificing embedding accuracy, as we verify in the next section. We first observe that connected nodes are attracted to each other, and are expected to be placed close to each other in the angular space~\cite{BoPa10}. This means that for each node $i$, we can get an \emph{initial estimate} for its angular coordinate, $\theta_i^{\textnormal{init}}$, by considering only the previous nodes $ j < i$ in $\mathcal L_{\textnormal{L}}^{i}$ (Eq.~(\ref{eq:local_likelihood_links})) that are its neighbors. This requires only $O(k_{i})$ steps, where $k_{i}$ is $i$'s degree, and $k_i=O(\bar{k})$ for sufficiently large $i$. That is, we can estimate $\theta_i^{\textnormal{init}}$ by maximizing the likelihood
\begin{equation}
\label{eq:local_likelihood_init}
\mathcal L_\textnormal{L-init}^{i}=\prod_{1 \leq j < i, \alpha_{ij}=1} p(x_{ij}),
\end{equation}
where the product goes over all previous nodes $j < i$ that are $i$'s neighbors. The maximization of Eq.~(\ref{eq:local_likelihood_init}) can be performed numerically by sampling the likelihood at intervals $\Delta\theta=\frac{1}{i}$ as before, yielding a total running time of $O(\bar{k}i)=O(i)$ to find $\theta_i^{\textnormal{init}}$.

Once we estimate $\theta_i^{\textnormal{init}}$, we can consider a region around it, $[\theta_i^{\textnormal{init}}-\frac{C}{i}, \theta_i^{\textnormal{init}}+\frac{C}{i}]$, where $0 < C \ll t$ is a constant, and set the angular coordinate of node $i$, $\theta_i$, to the value of $\theta$ that yields the largest value of $\mathcal L_{\textnormal{L}}^{i}$ (Eq.~(\ref{eq:local_likelihood_links})) in this region. Since we sample the likelihood at intervals $\Delta\theta=\frac{1}{i}$, we need $O(C)$ steps to perform this maximization. Taken altogether, at sufficiently large times $i \gg C$ we need $O(i C)=O(i)$ steps to find $\theta_i$. Therefore, if we have $t \gg C$ nodes in total, the total running time to find their angles following this procedure is $O(t^2)$. The larger the value of $C$ the better the results are expected to be in general, as we are searching for the optimal value of $\theta_i$ over a larger region, but the procedure will also be slower.  We validate this speedup heuristic in the next section, where we set $C=200$, and show that it produces good results.

\section{Validation}
\label{sec:validation}

In this section we validate the new mapping method and its variations. To do so, we first grow synthetic networks according to E-PSO up to $t=5000$ nodes, with $m=1.5$, $L=2.5$, $\gamma=2.1$, $\zeta=1$, and $T=0.05, 0.4, 0.7$. Similar results hold for other parameter values. Then, we pass these synthetic networks to HyperMap, using their corresponding $m, L, \gamma, T, \zeta$ values, and compute radial and angular coordinates for the nodes, using either  $\mathcal L_{\textnormal{CN}}^{i}$ (common-neighbors method), $\mathcal L_{\textnormal{L}}^{i}$ (link-based method), or the hybrid method. We consider the real Internet in the next section.

\begin{figure*}
\centerline{
\subfigure[~$T=0.05$, $\mathcal L_{\textnormal{CN}}^{i}$.]{\includegraphics[width=1.9in, height=1.45in]{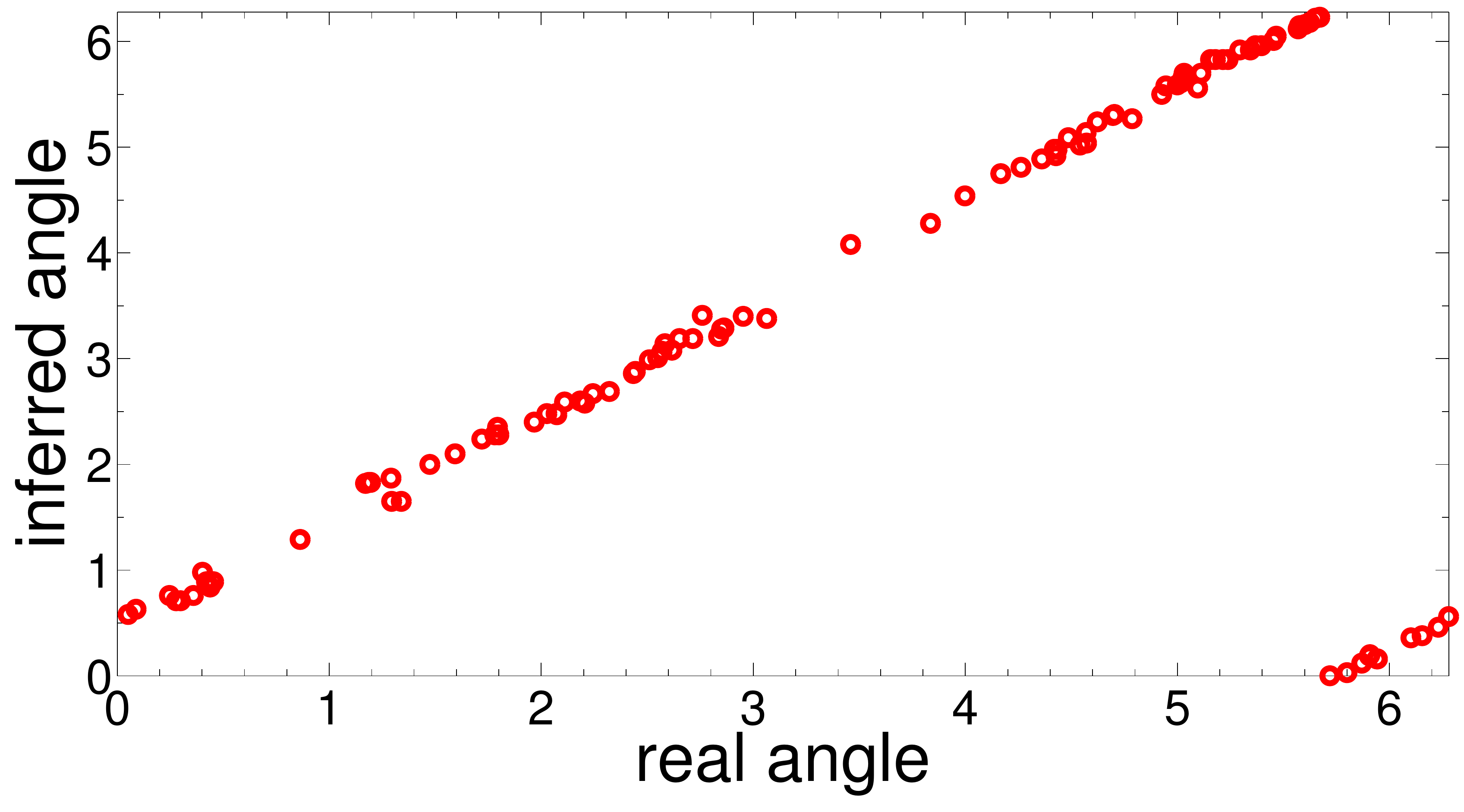}}
\subfigure[~$T=0.4$, $\mathcal L_{\textnormal{CN}}^{i}$.]{\includegraphics[width=1.9in, height=1.45in]{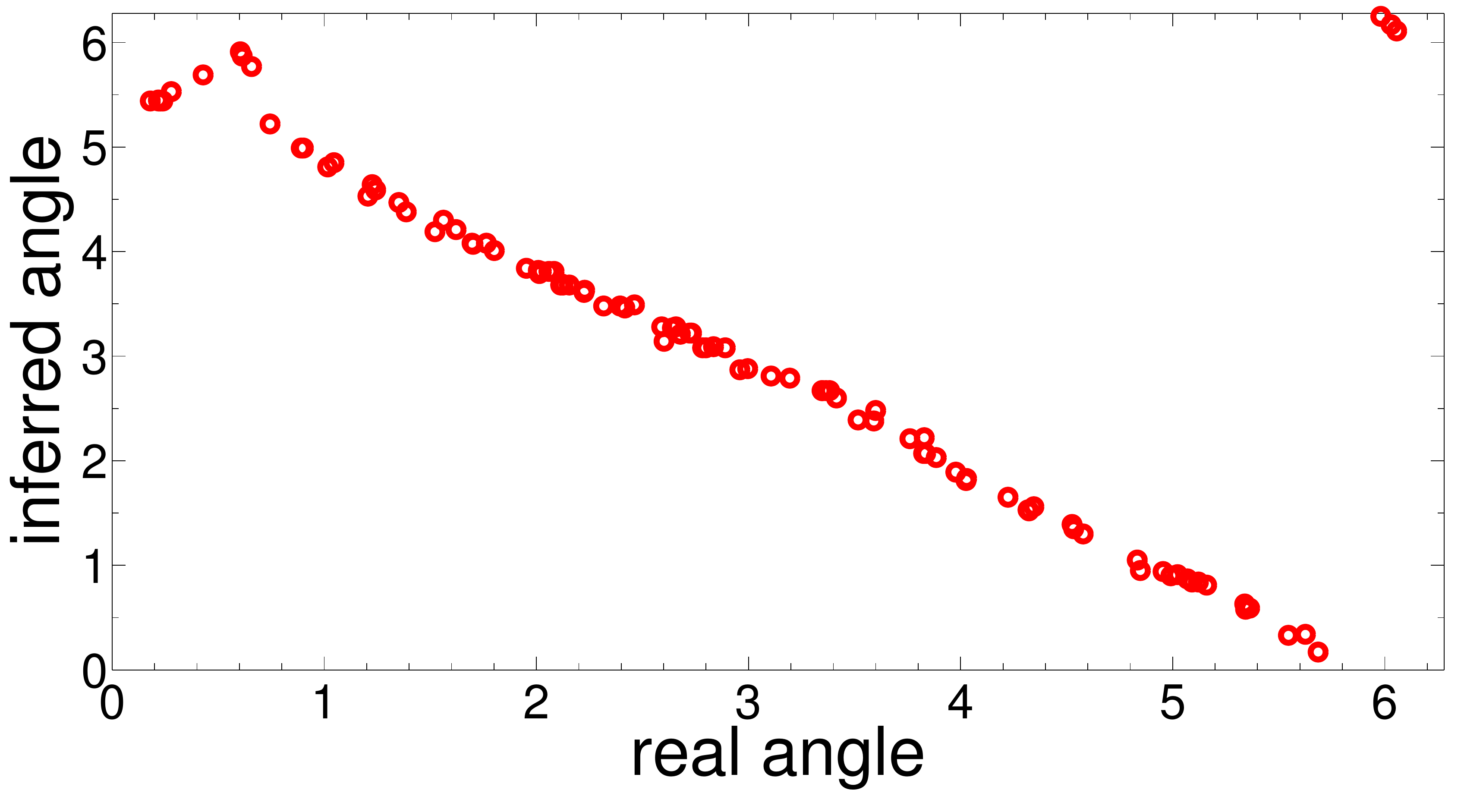}}
\subfigure[~$T=0.7$, $\mathcal L_{\textnormal{CN}}^{i}$.]{\includegraphics[width=1.9in, height=1.45in]{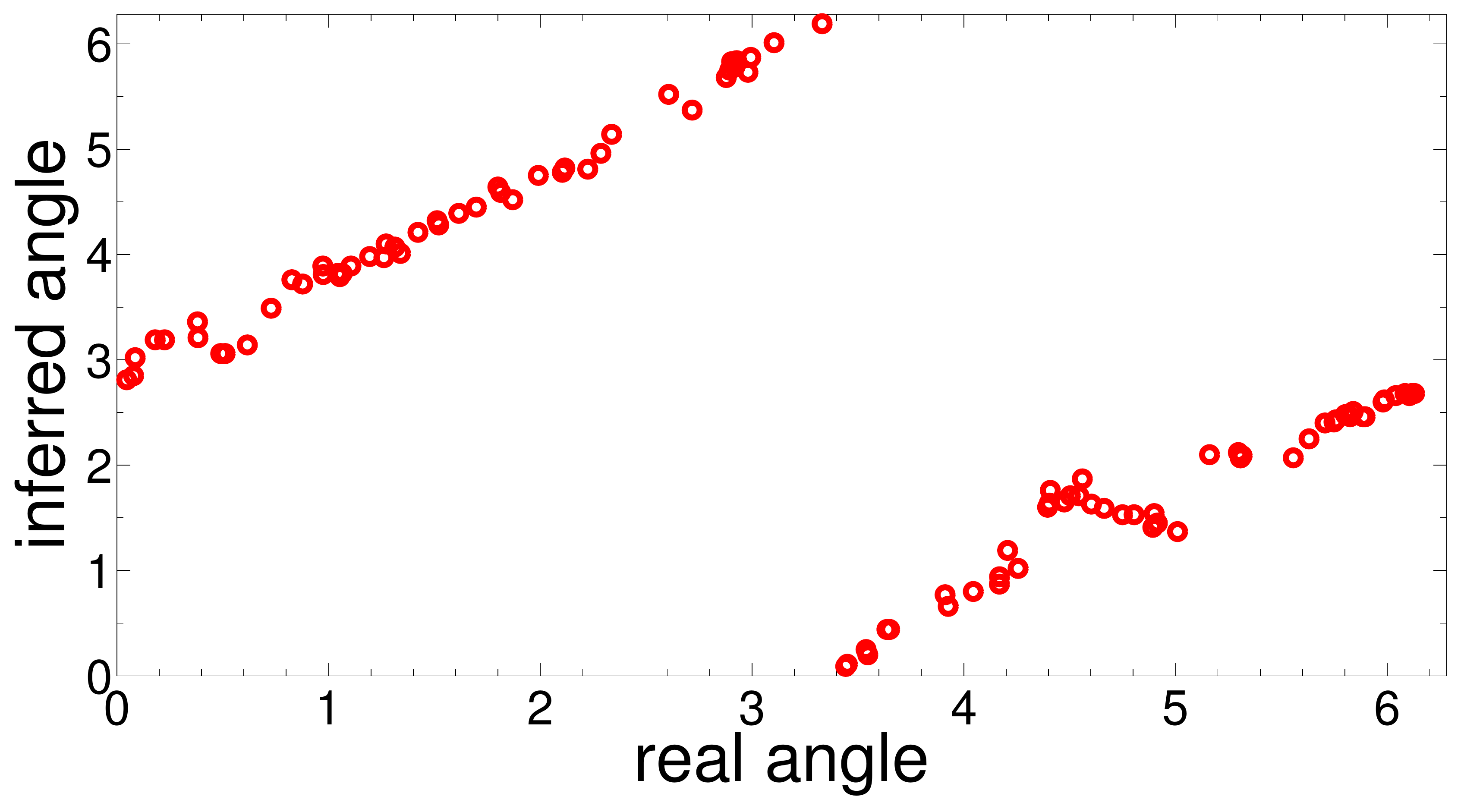}}
}
\centerline{
\subfigure[~$T=0.05$, $\mathcal L_{\textnormal{L}}^{i}$.]{\includegraphics[width=1.9in, height=1.45in]{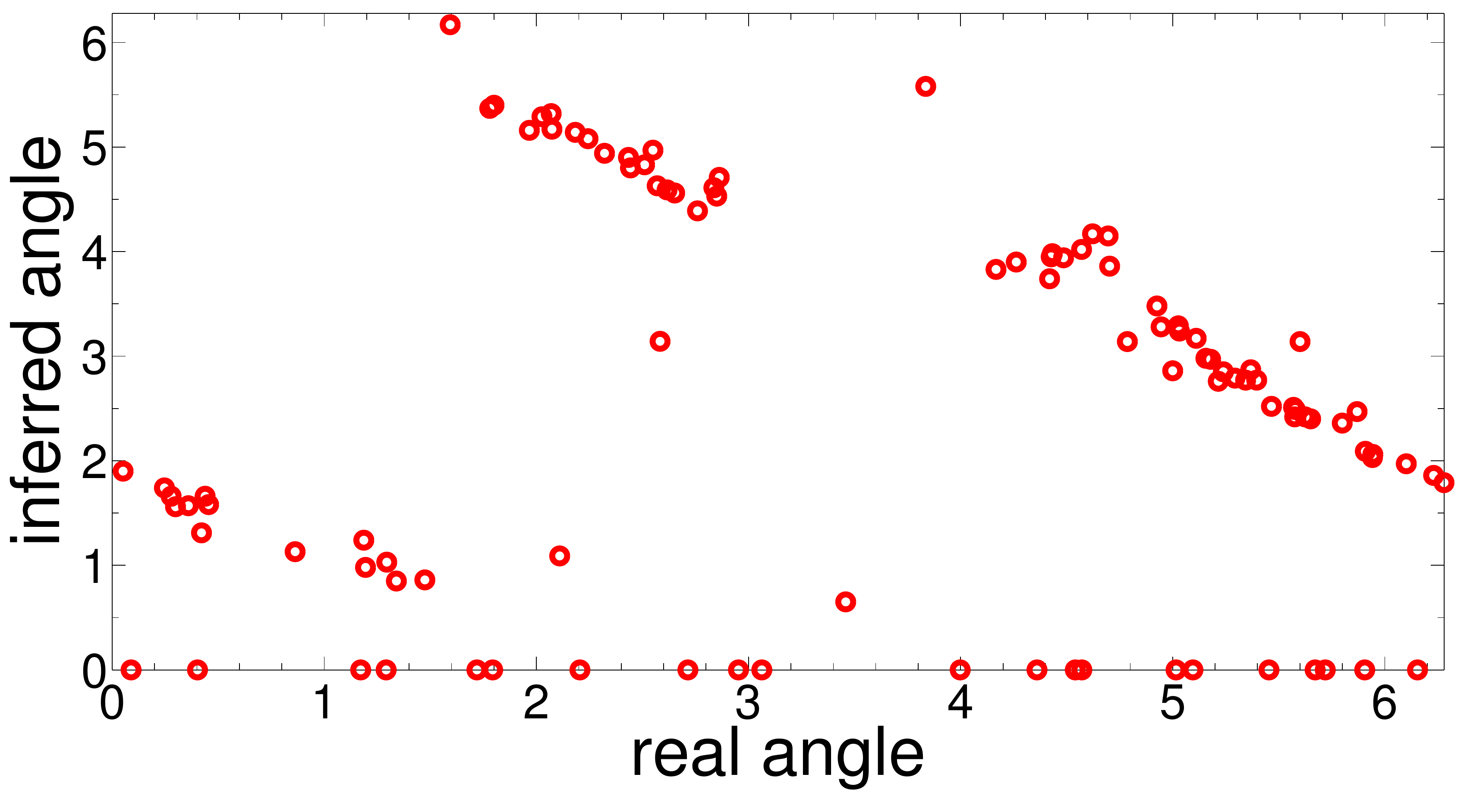}}
\subfigure[~$T=0.4$, $\mathcal L_{\textnormal{L}}^{i}$.]{\includegraphics[width=1.9in, height=1.45in]{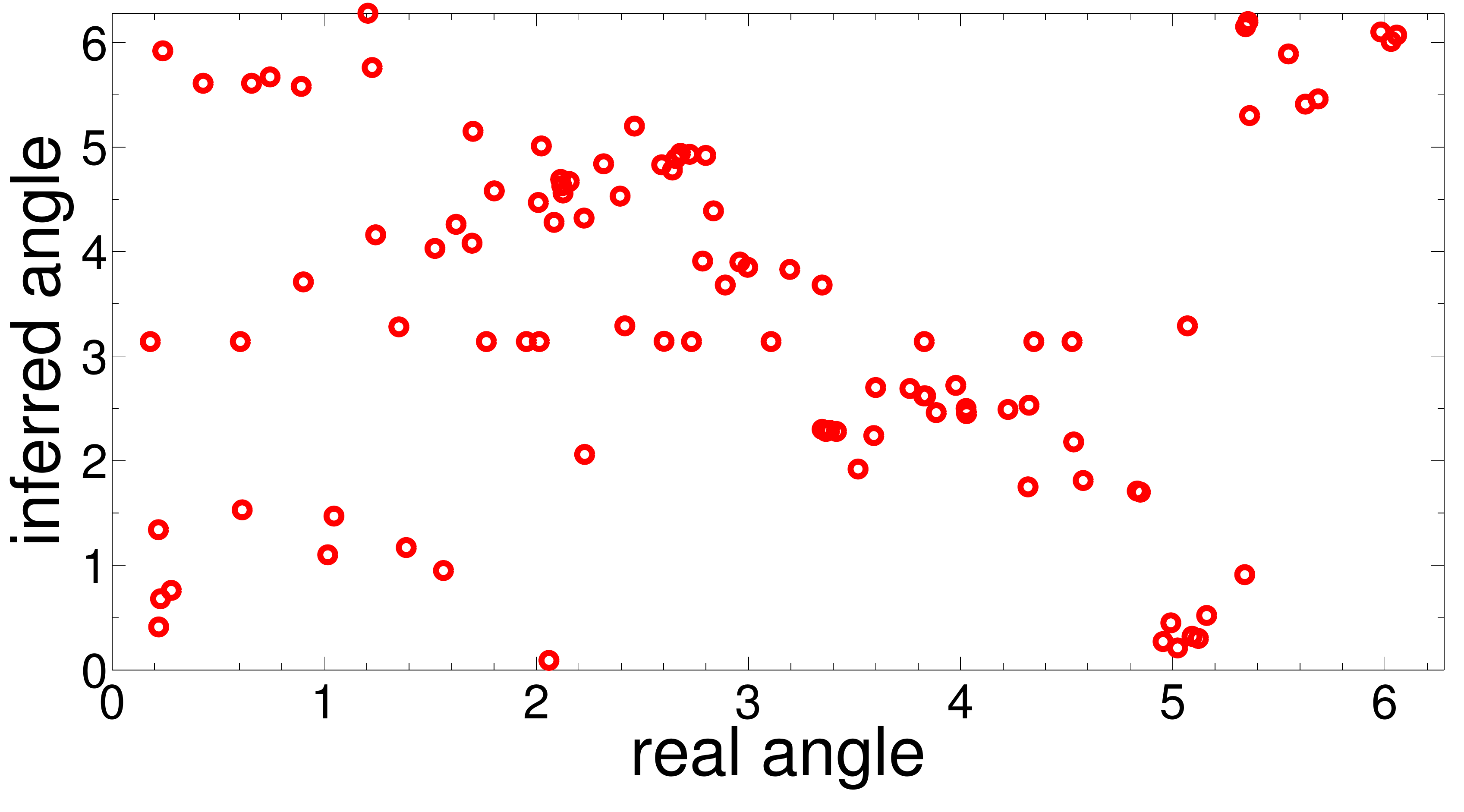}}
\subfigure[~$T=0.7$, $\mathcal L_{\textnormal{L}}^{i}$.]{\includegraphics[width=1.9in, height=1.45in]{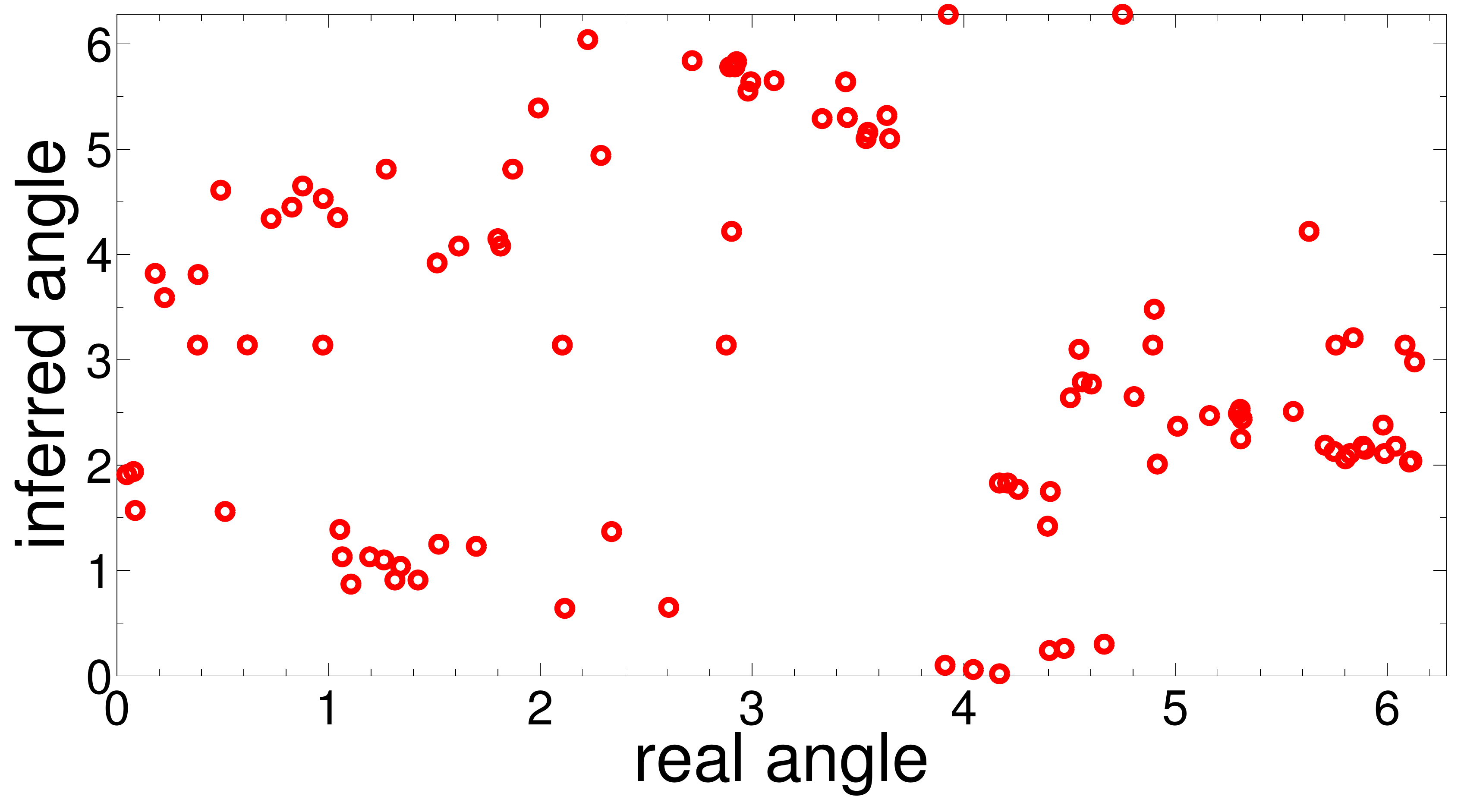}}
}
\caption{Inferred~vs.~real angles (in radians) for synthetic networks with $t=5000$ nodes and parameters $m=1.5, L=2.5, \gamma=2.1$, and $T$ as shown in the captions.  The plots juxtapose the inferred against the real angles for the first $100$ nodes, i.e., the nodes that appear at MLE times $1 \leq i \leq 100$. In (a-c) the common-neighbors method is used, while in (d-f) the link-based method is used.
\label{fig:inferred_vs_real_first_100}}
\end{figure*}

\textbf{Inferred versus real angles for nodes appearing at early MLE times.} Fig.~\ref{fig:inferred_vs_real_first_100} juxtaposes the inferred against the real angles for the first $100$ nodes, i.e., for the nodes that appear at MLE times $1 \leq i \leq 100$ for each considered network, when $\mathcal L_{\textnormal{CN}}^{i}$ or $\mathcal L_{\textnormal{L}}^{i}$ is used. We observe that the common-neighbors method is more accurate at inferring the angles of these first nodes. The reason for this was explained in Section~\ref{sec:comparison_and_hybrid_approach}. Specifically, we see in Figs.~\ref{fig:inferred_vs_real_first_100}(a-c) that $\mathcal L_{\textnormal{CN}}^{i}$ can infer the real angles of the nodes quite accurately, subject only to a \emph{global} phase shift. This phase shift can take any value in $[0, 2\pi]$, and it is due to the rotational symmetry of the model. The exact value of this shift is not important, and it depends on the initialization of the angle of the first node in HyperMap, which can be any random value in $[0, 2\pi]$ (cf. Step 3 in Fig.~\ref{fig:the_method}).

\begin{figure*}
\centerline{
\subfigure{\includegraphics[width=1.7in, height=1.25in]{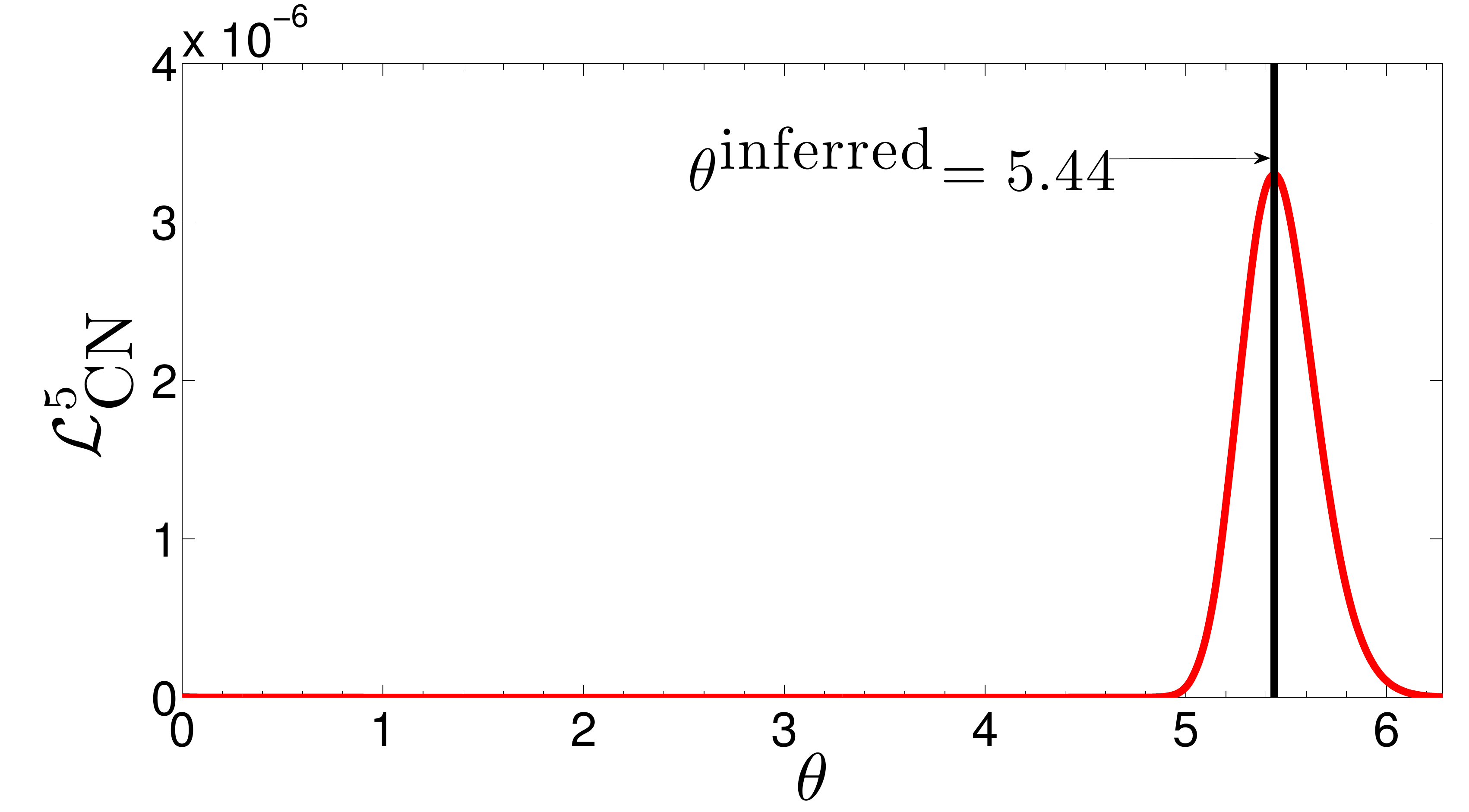}}
\subfigure{\includegraphics[width=1.7in, height=1.2in]{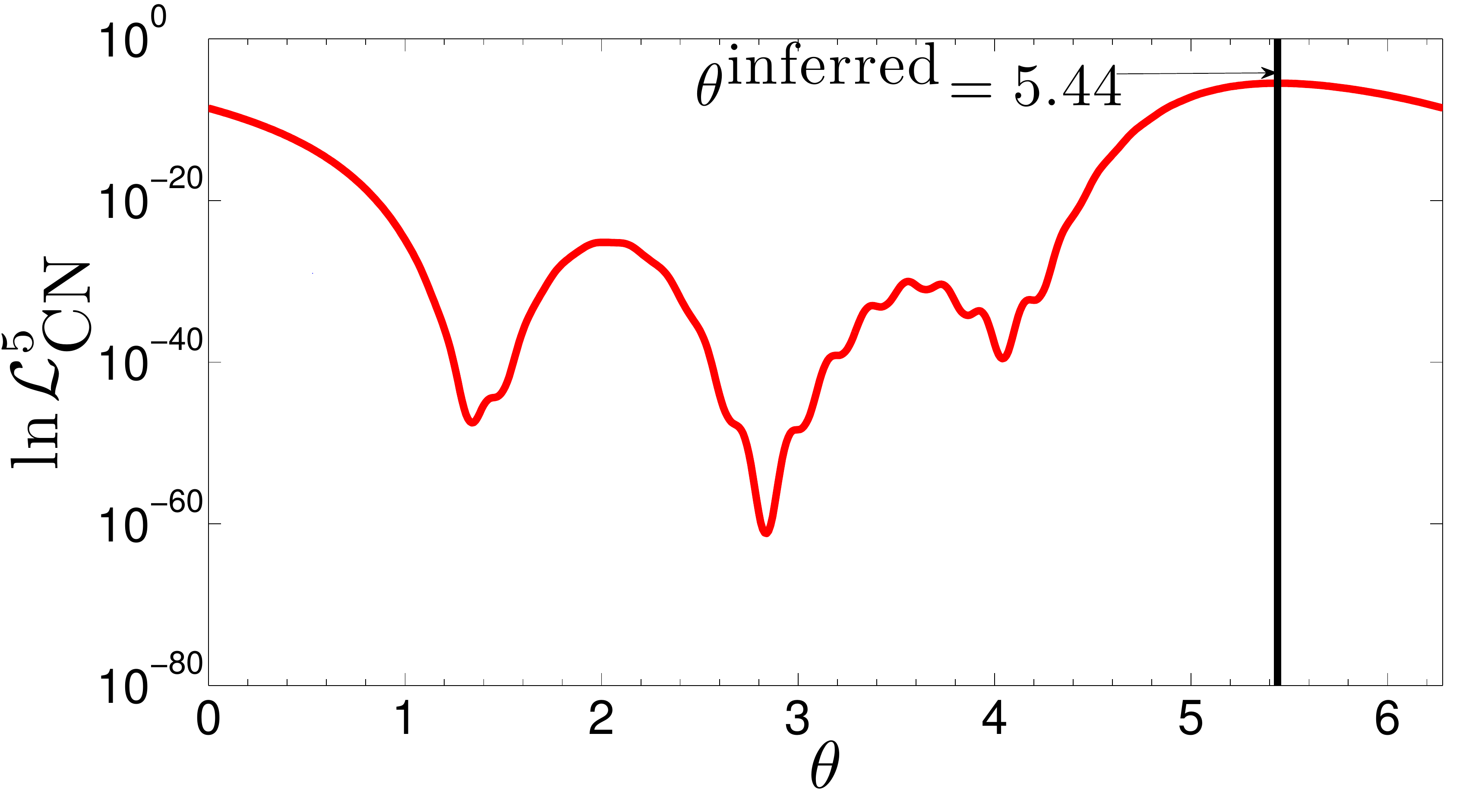}}
\subfigure{\includegraphics[width=1.7in, height=1.22in]{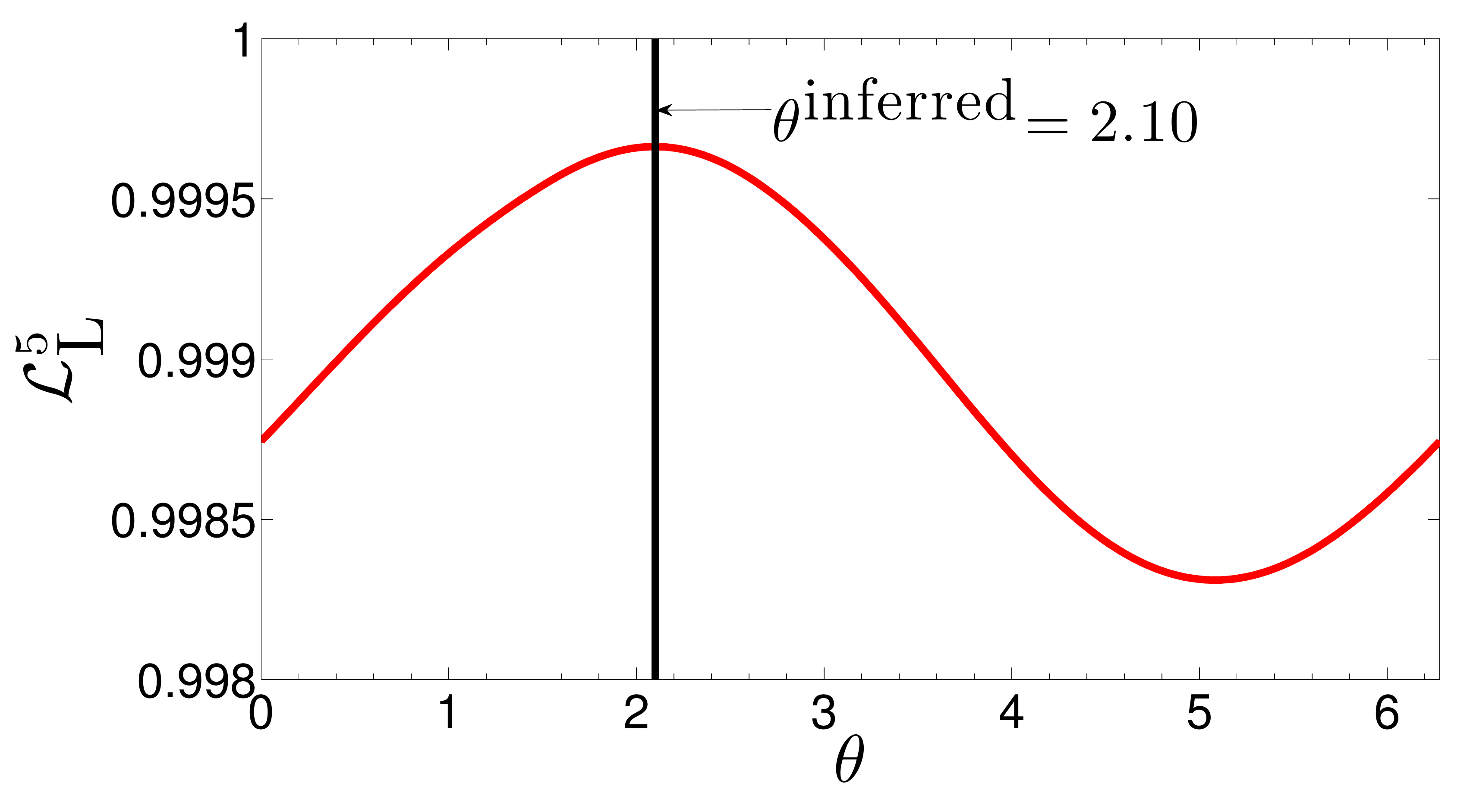}}
\subfigure{\includegraphics[width=1.7in, height=1.22in]{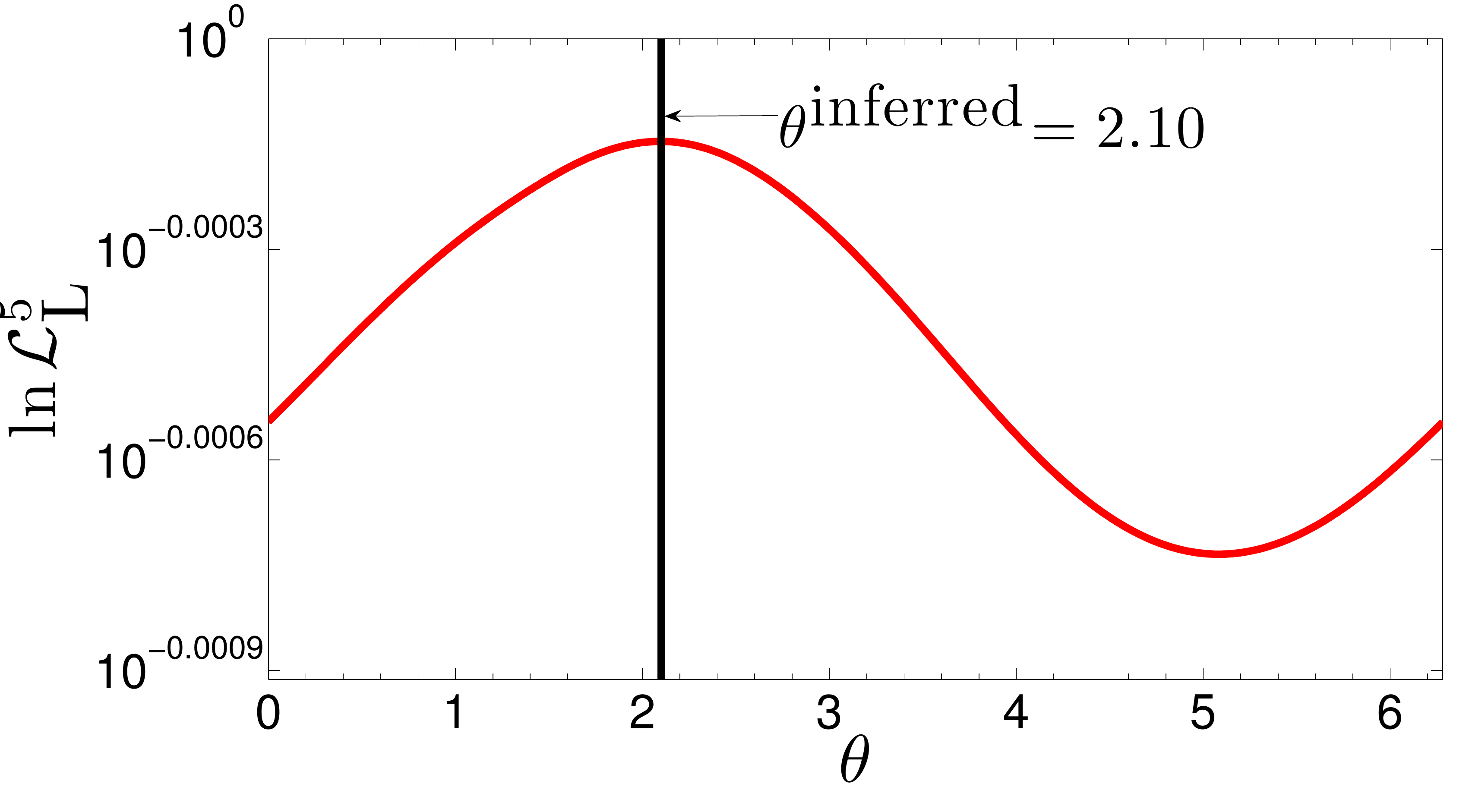}}
}
\centerline{
\subfigure{\includegraphics[width=1.7in, height=1.25in]{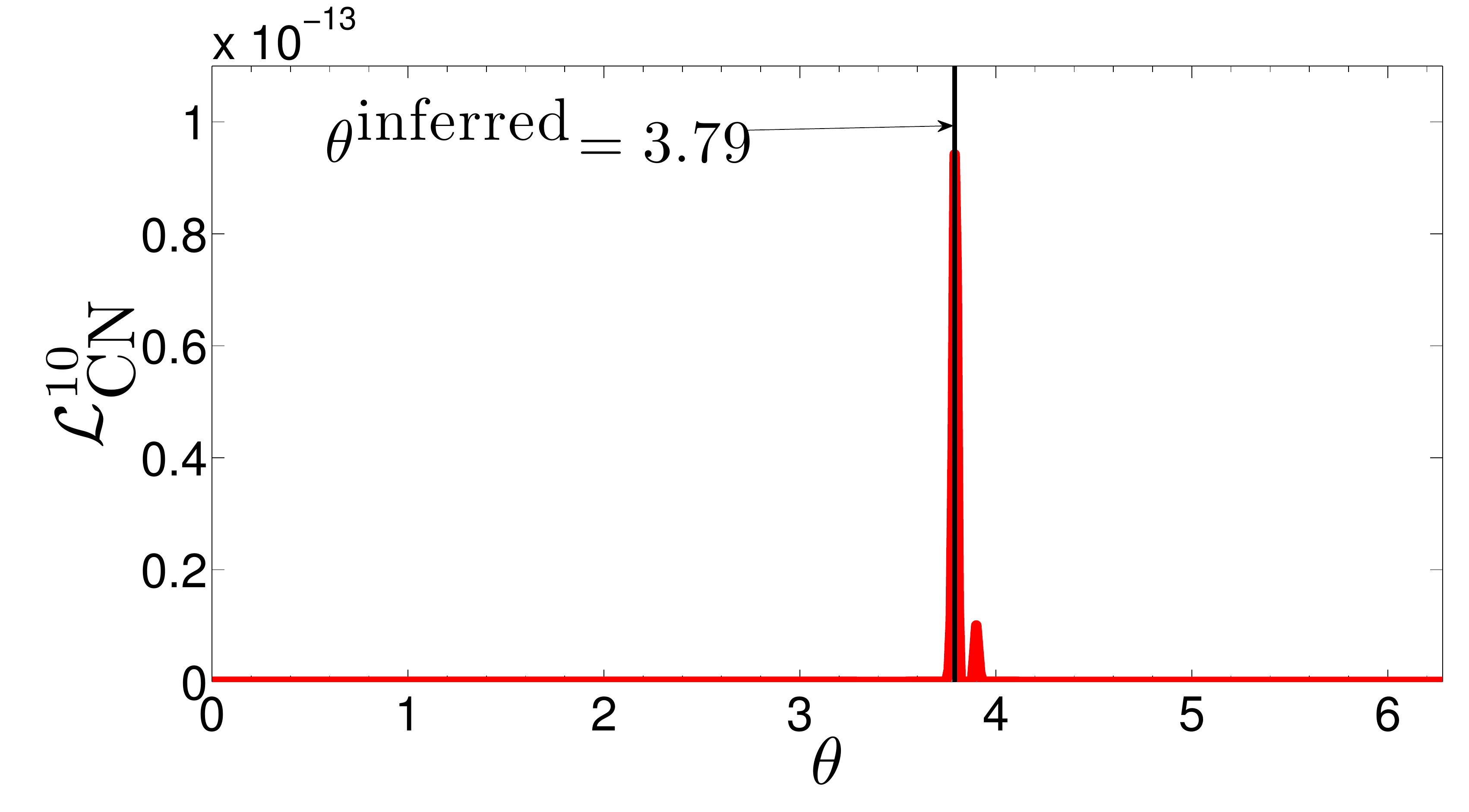}}
\subfigure{\includegraphics[width=1.7in, height=1.2in]{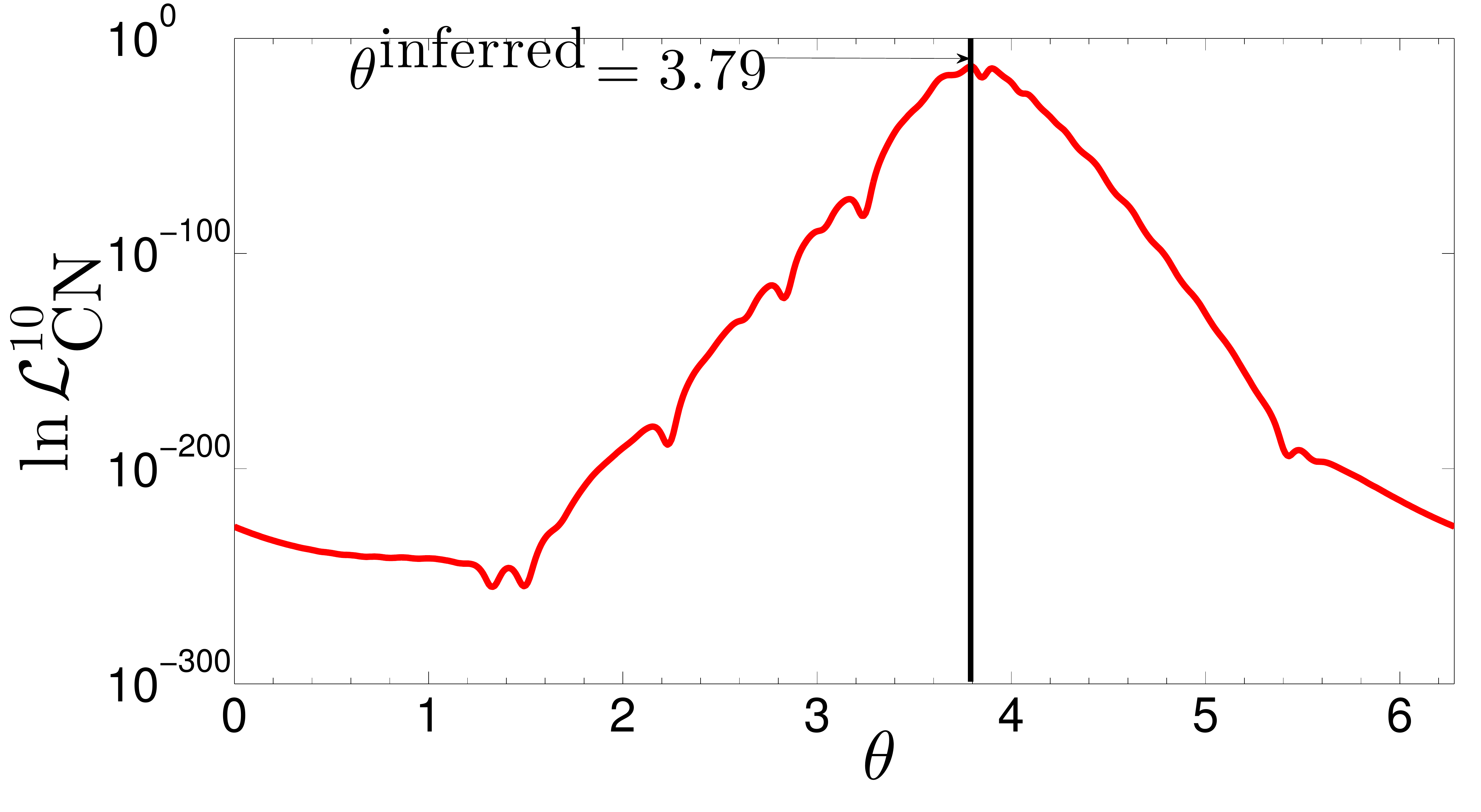}}
\subfigure{\includegraphics[width=1.7in, height=1.22in]{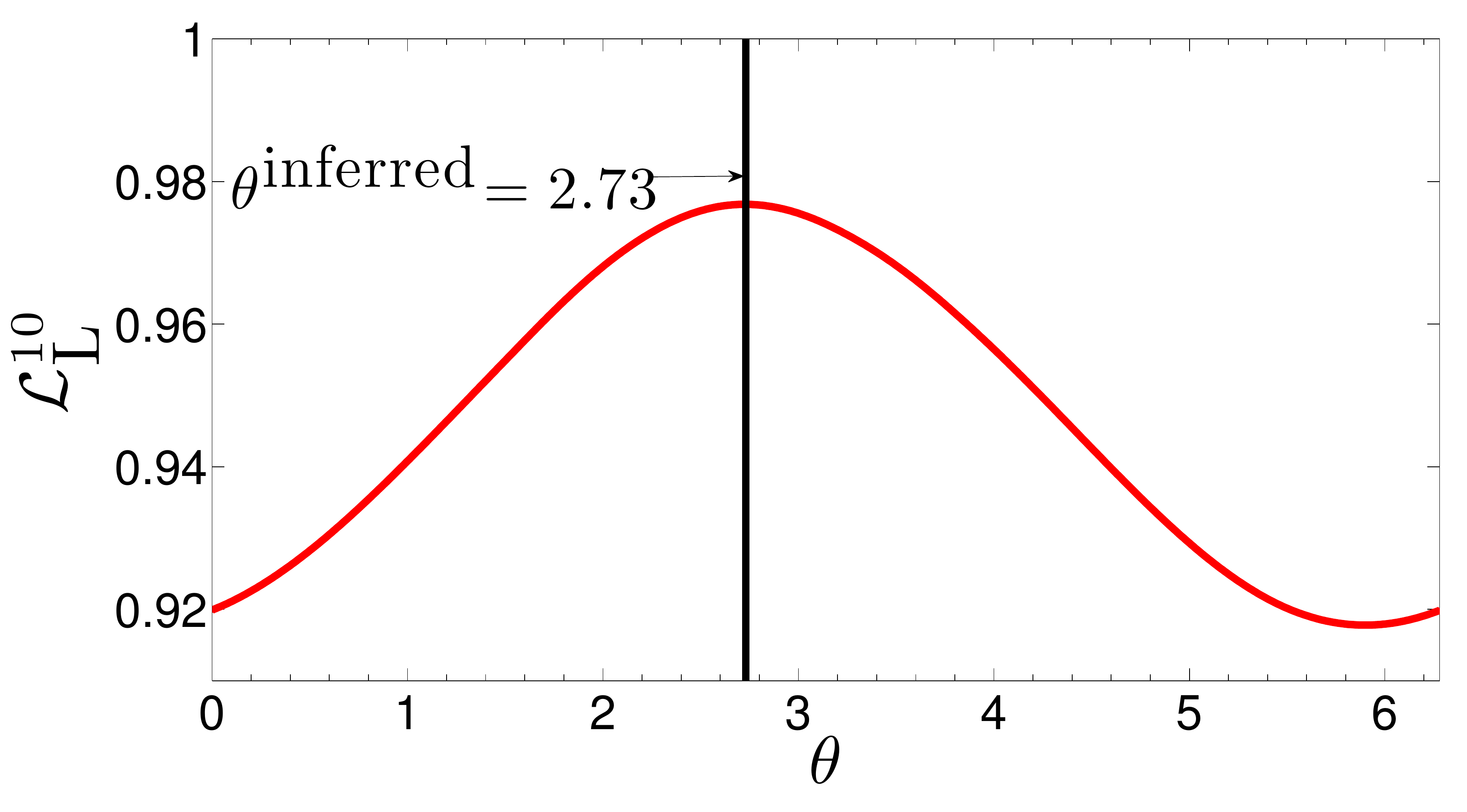}}
\subfigure{\includegraphics[width=1.7in, height=1.22in]{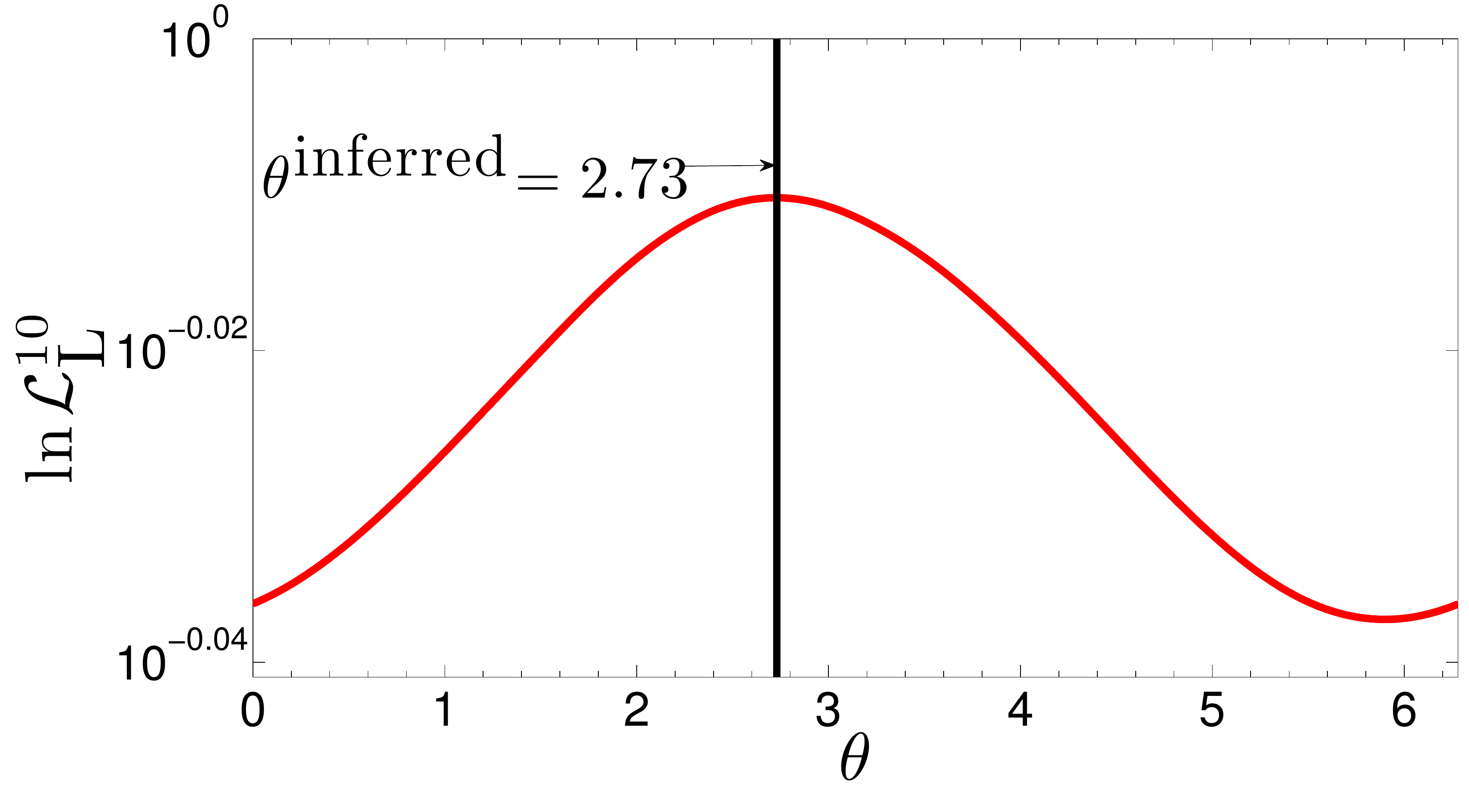}}
}
\centerline{
\subfigure{\includegraphics[width=1.7in, height=1.25in]{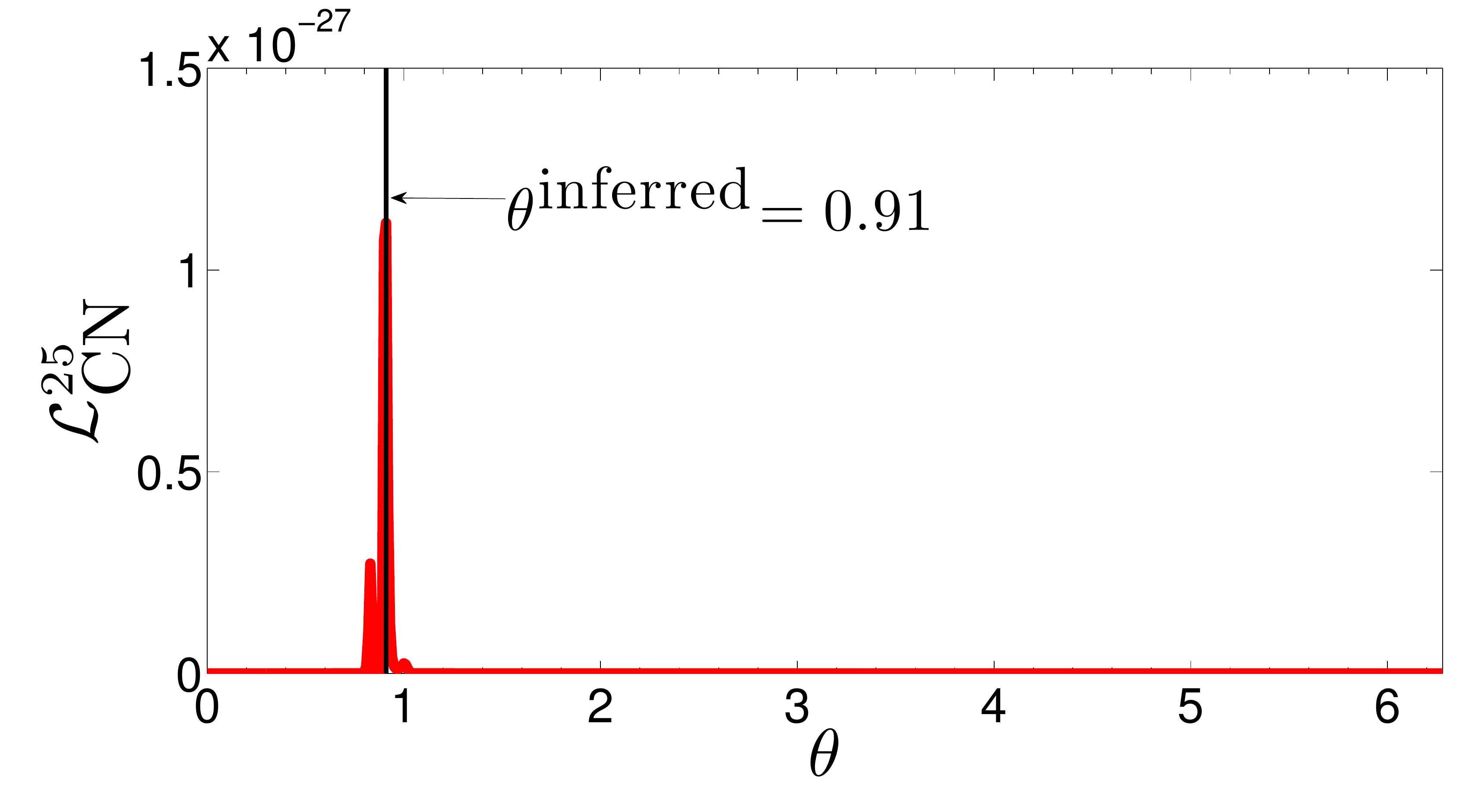}}
\subfigure{\includegraphics[width=1.7in, height=1.2in]{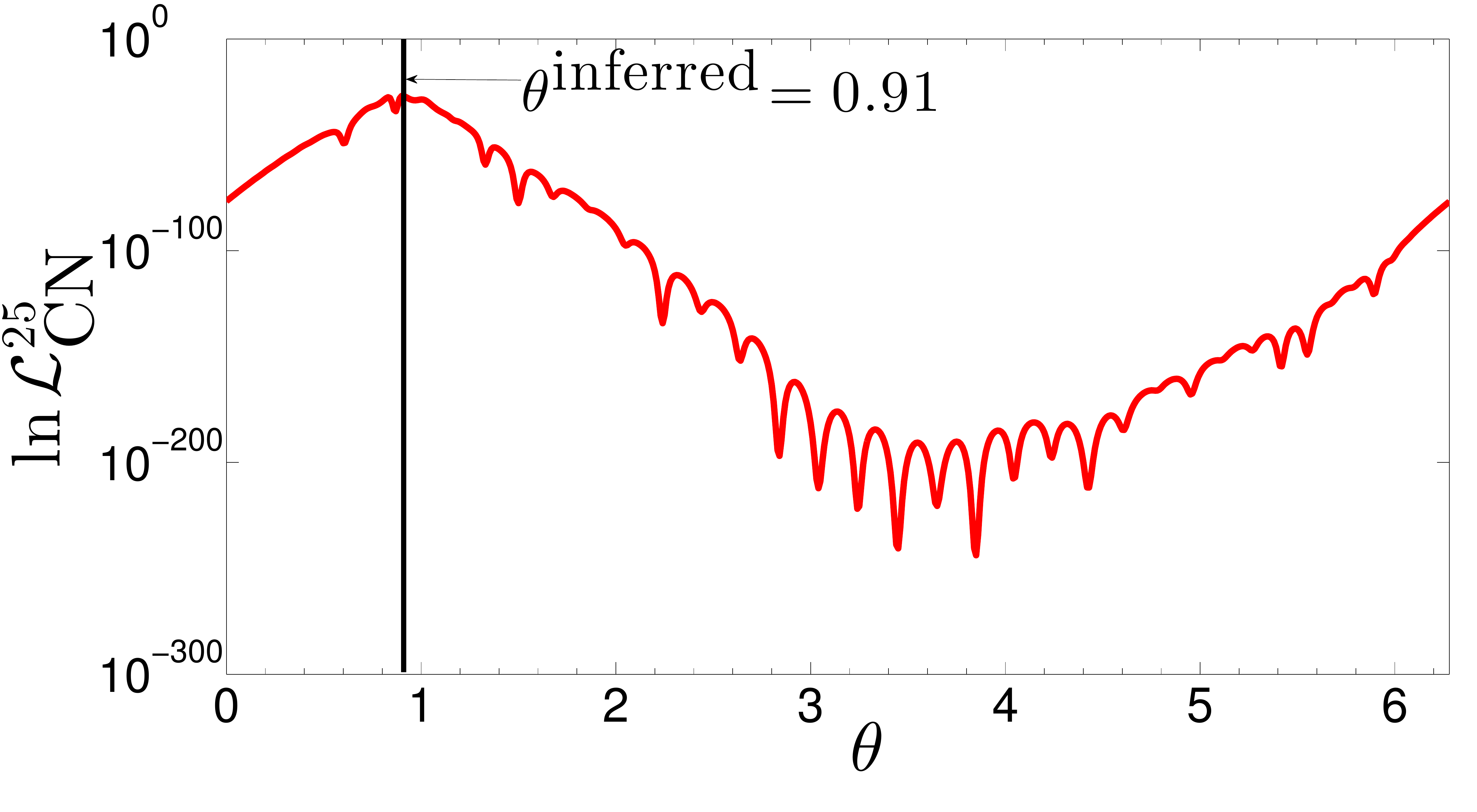}}
\subfigure{\includegraphics[width=1.7in, height=1.25in]{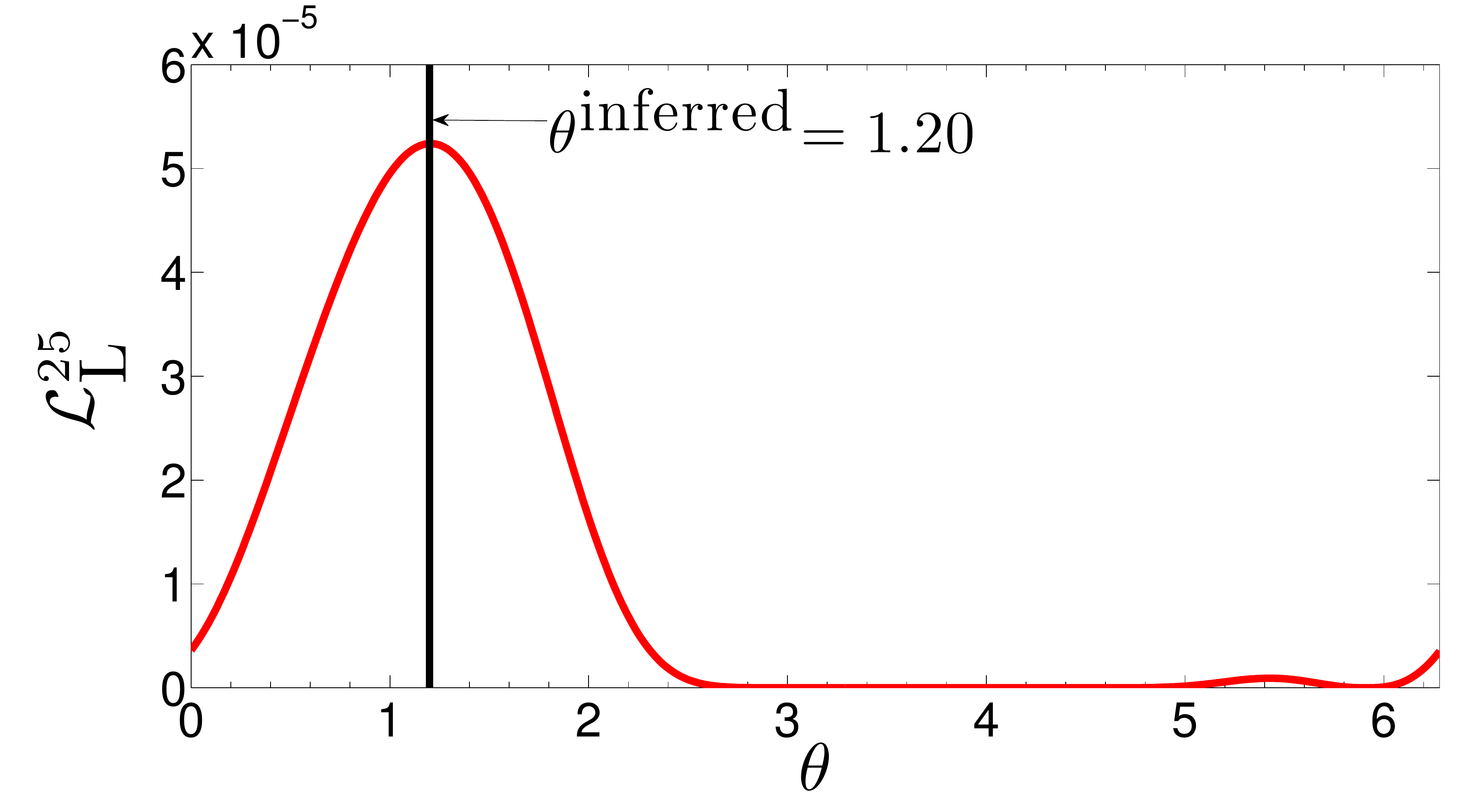}}
\subfigure{\includegraphics[width=1.7in, height=1.22in]{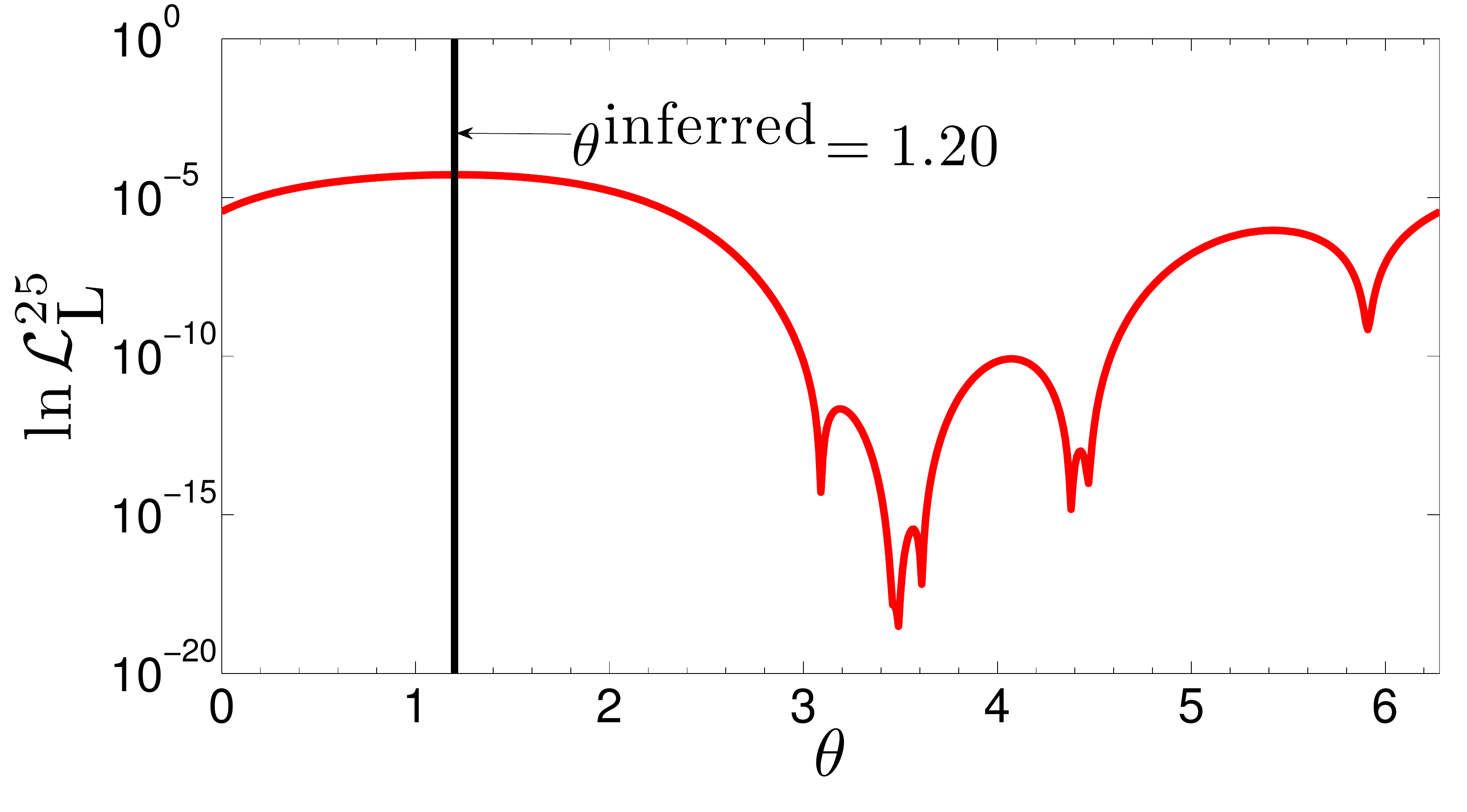}}
}
\centerline{
\subfigure{\includegraphics[width=1.7in, height=1.25in]{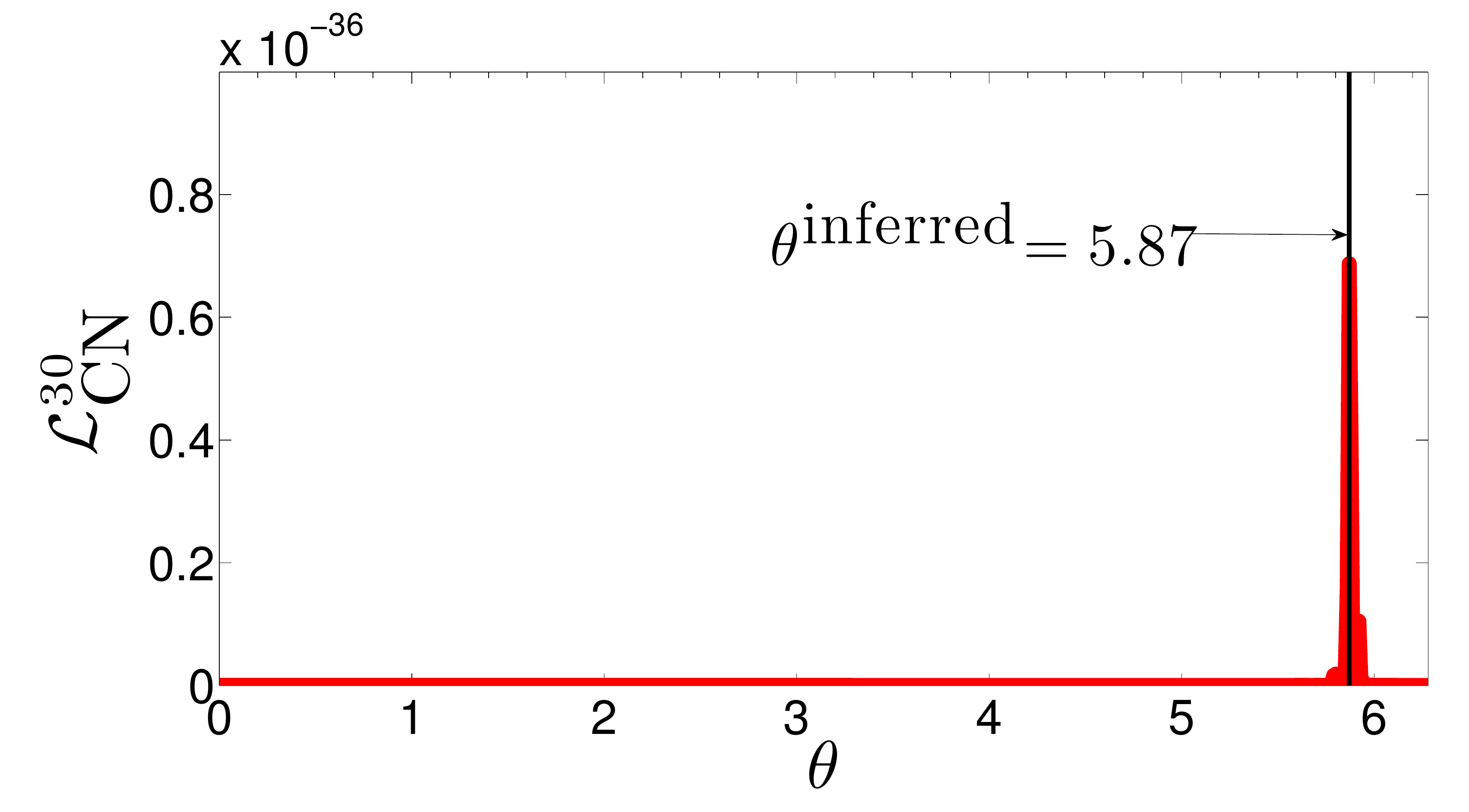}}
\subfigure{\includegraphics[width=1.7in, height=1.2in]{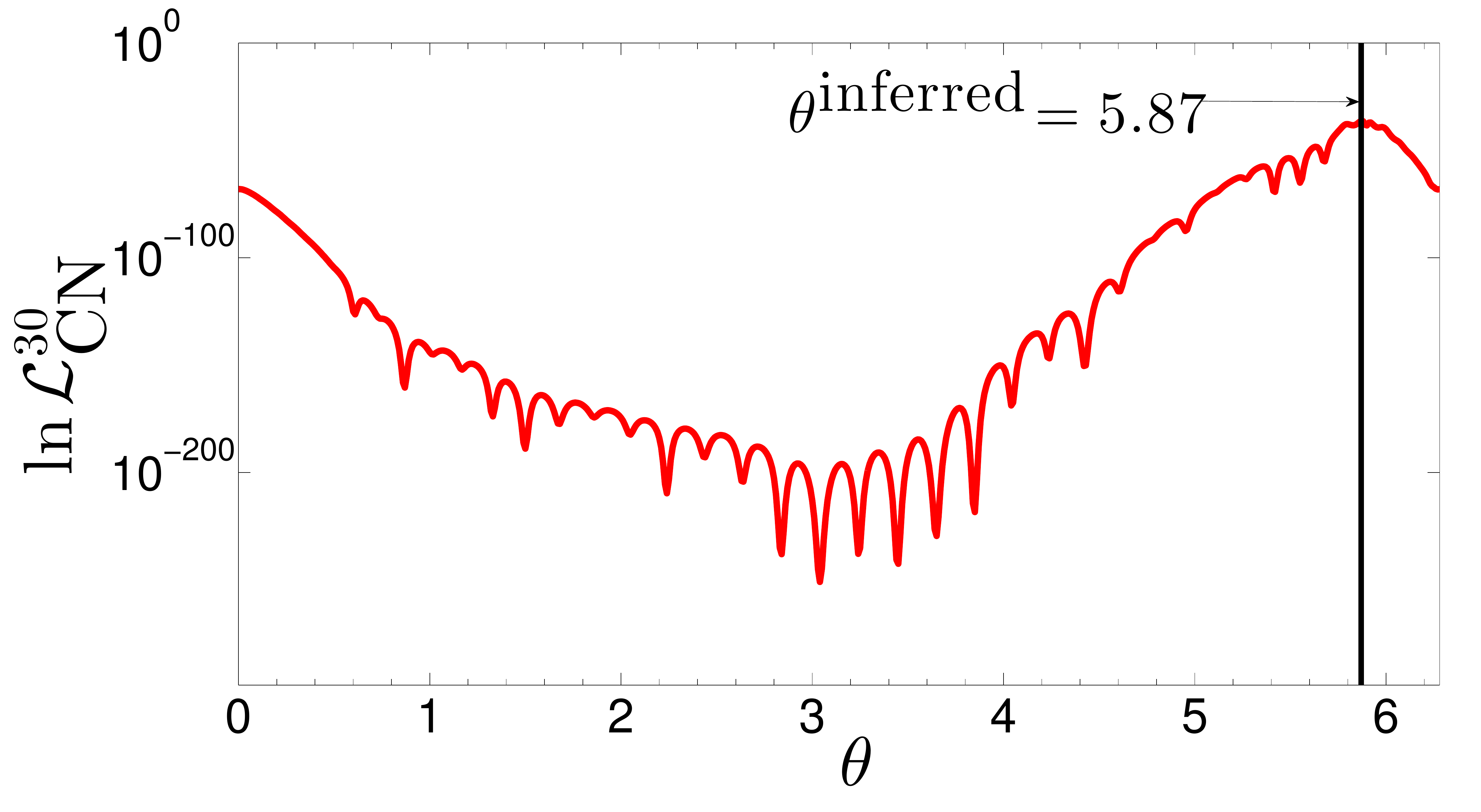}}
\subfigure{\includegraphics[width=1.7in, height=1.25in]{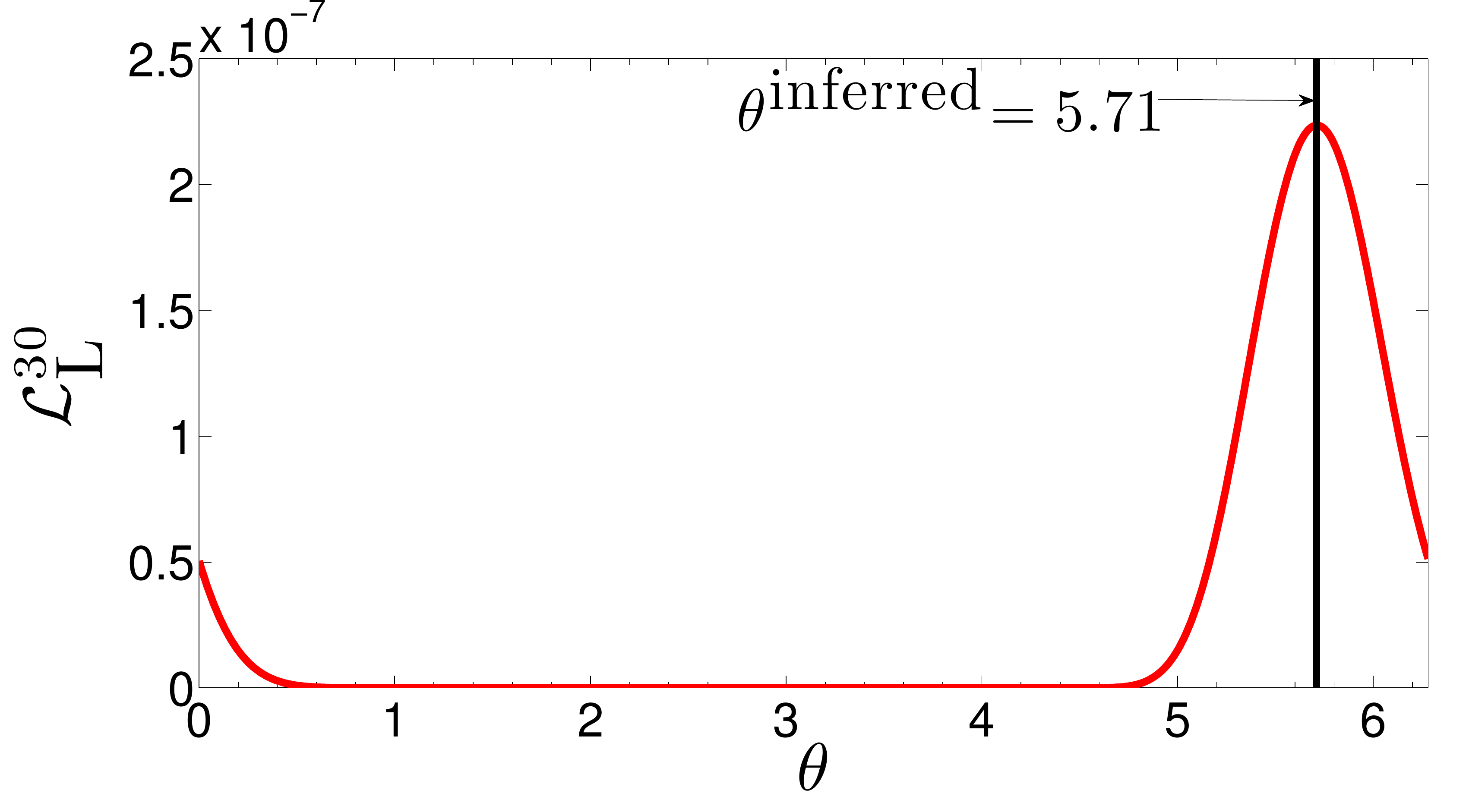}}
\subfigure{\includegraphics[width=1.7in, height=1.22in]{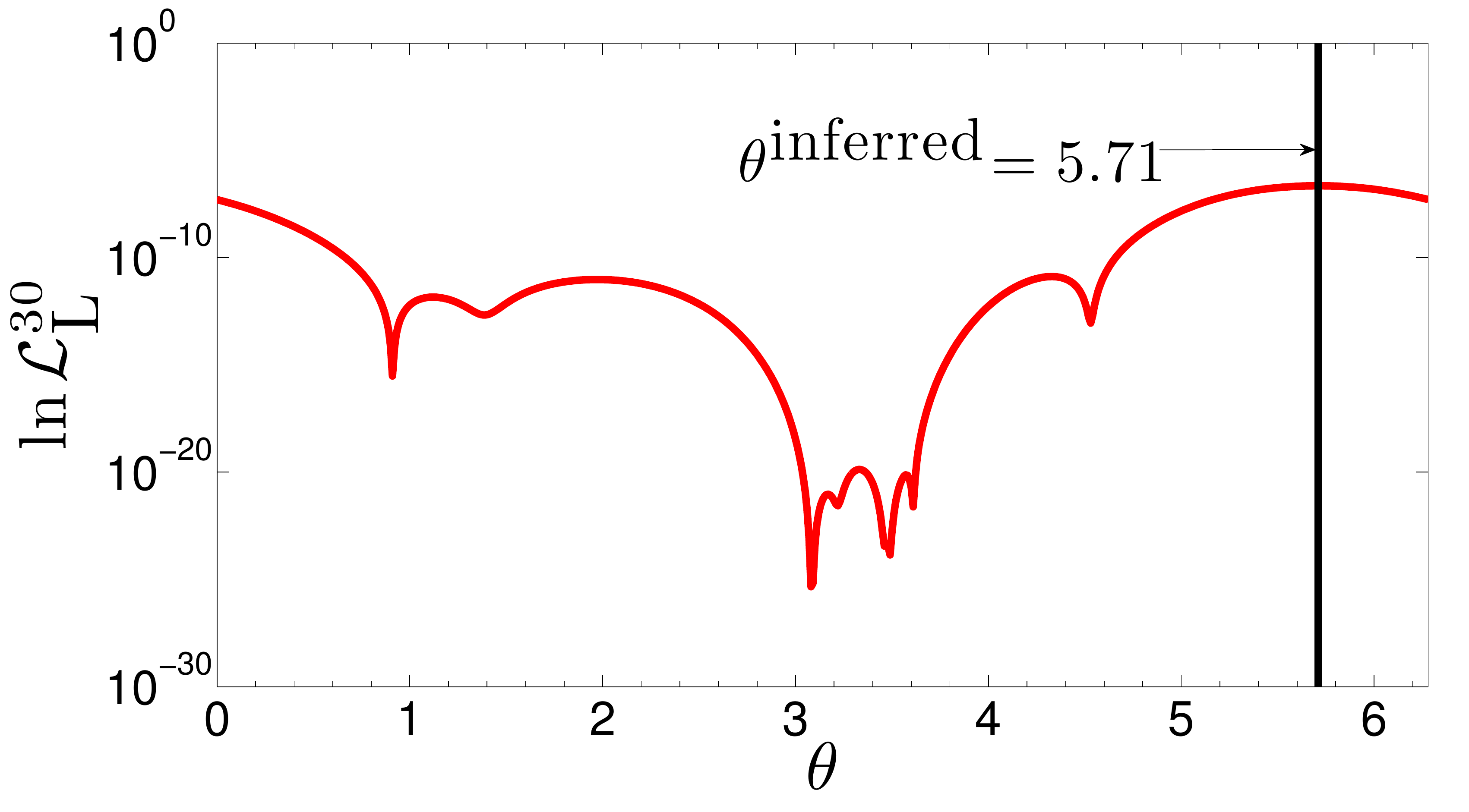}}
}
\centerline{
\subfigure{\includegraphics[width=1.7in, height=1.25in]{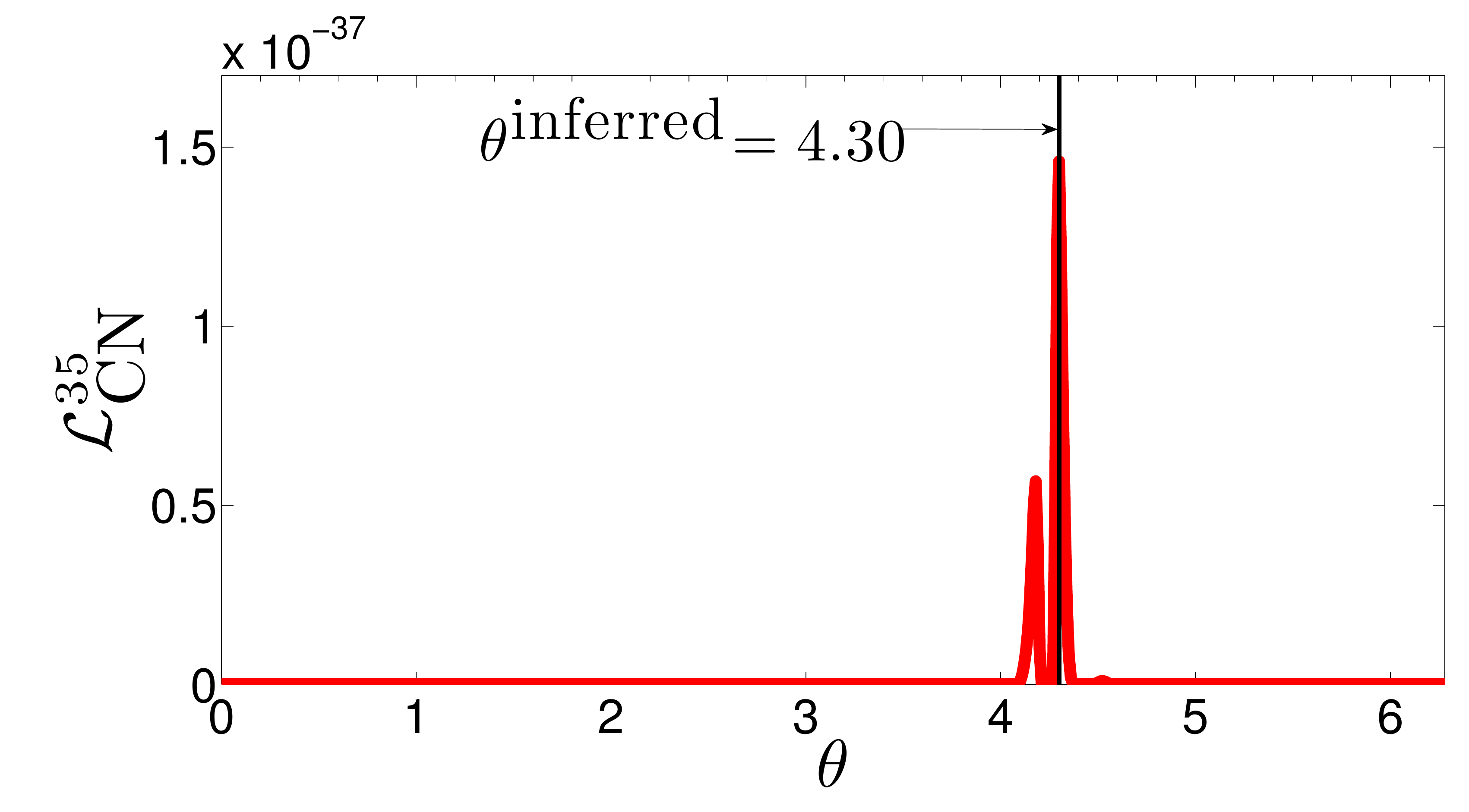}}
\subfigure{\includegraphics[width=1.7in, height=1.2in]{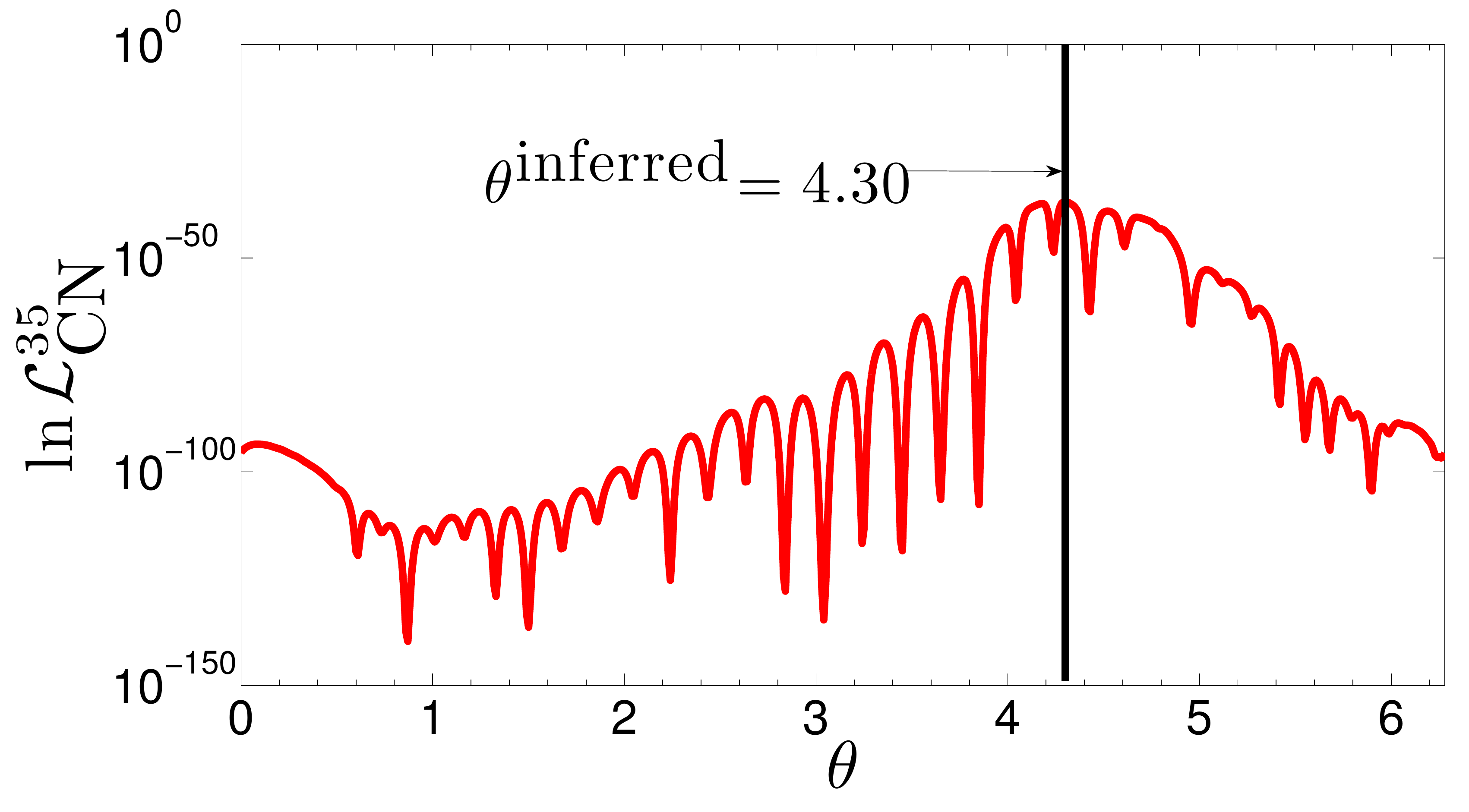}}
\subfigure{\includegraphics[width=1.7in, height=1.25in]{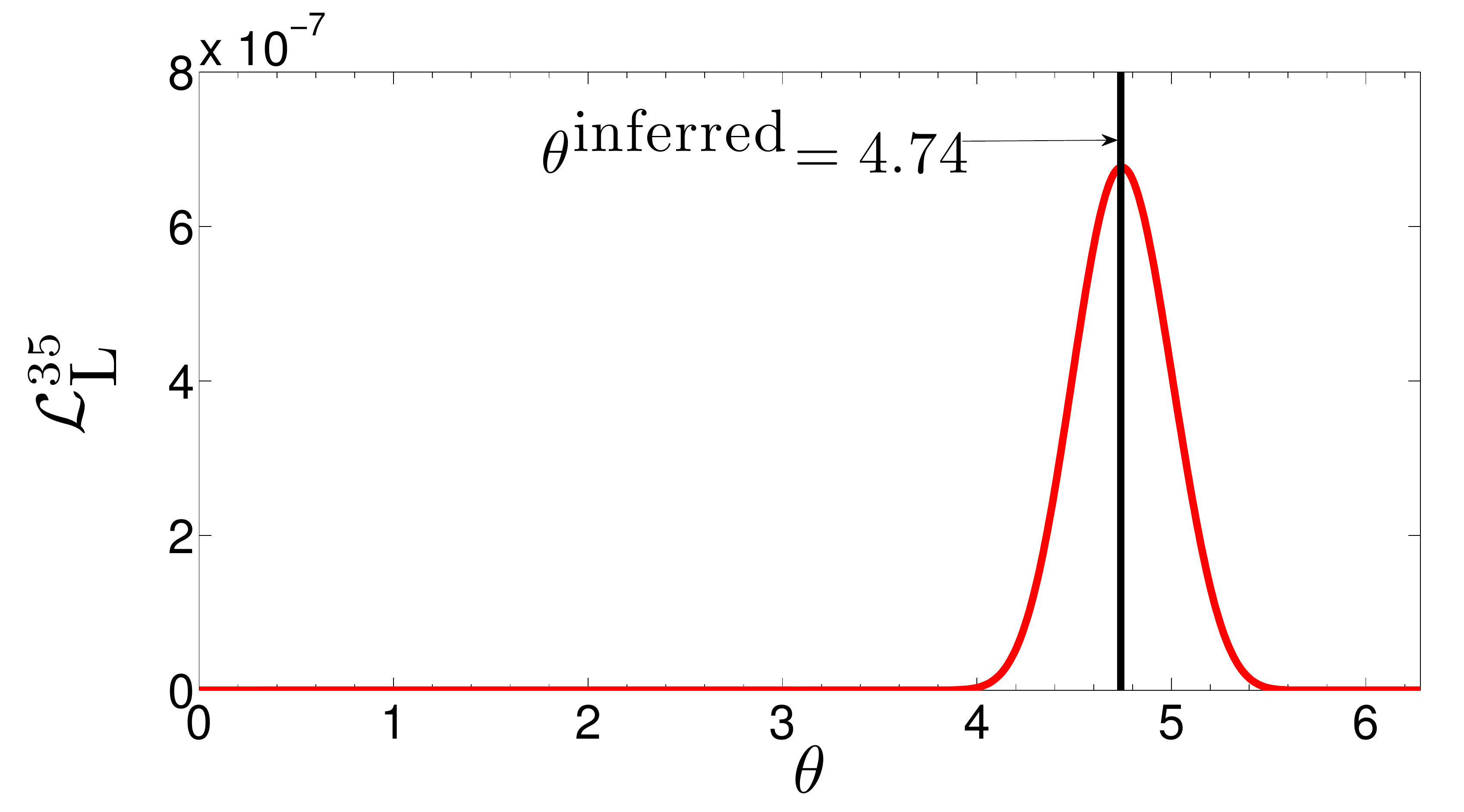}}
\subfigure{\includegraphics[width=1.7in, height=1.22in]{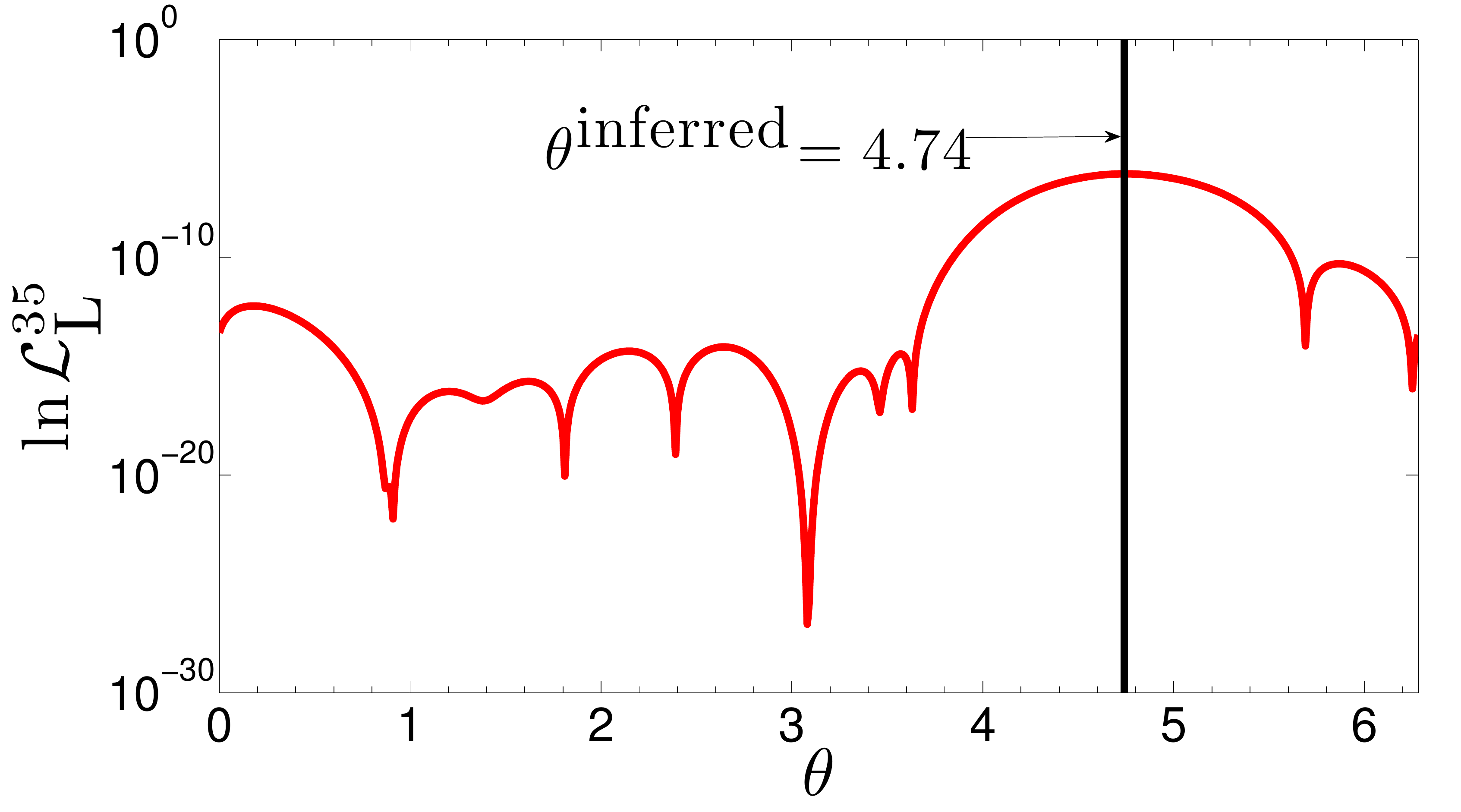}}
}
\centerline{
\subfigure{\includegraphics[width=1.7in, height=1.25in]{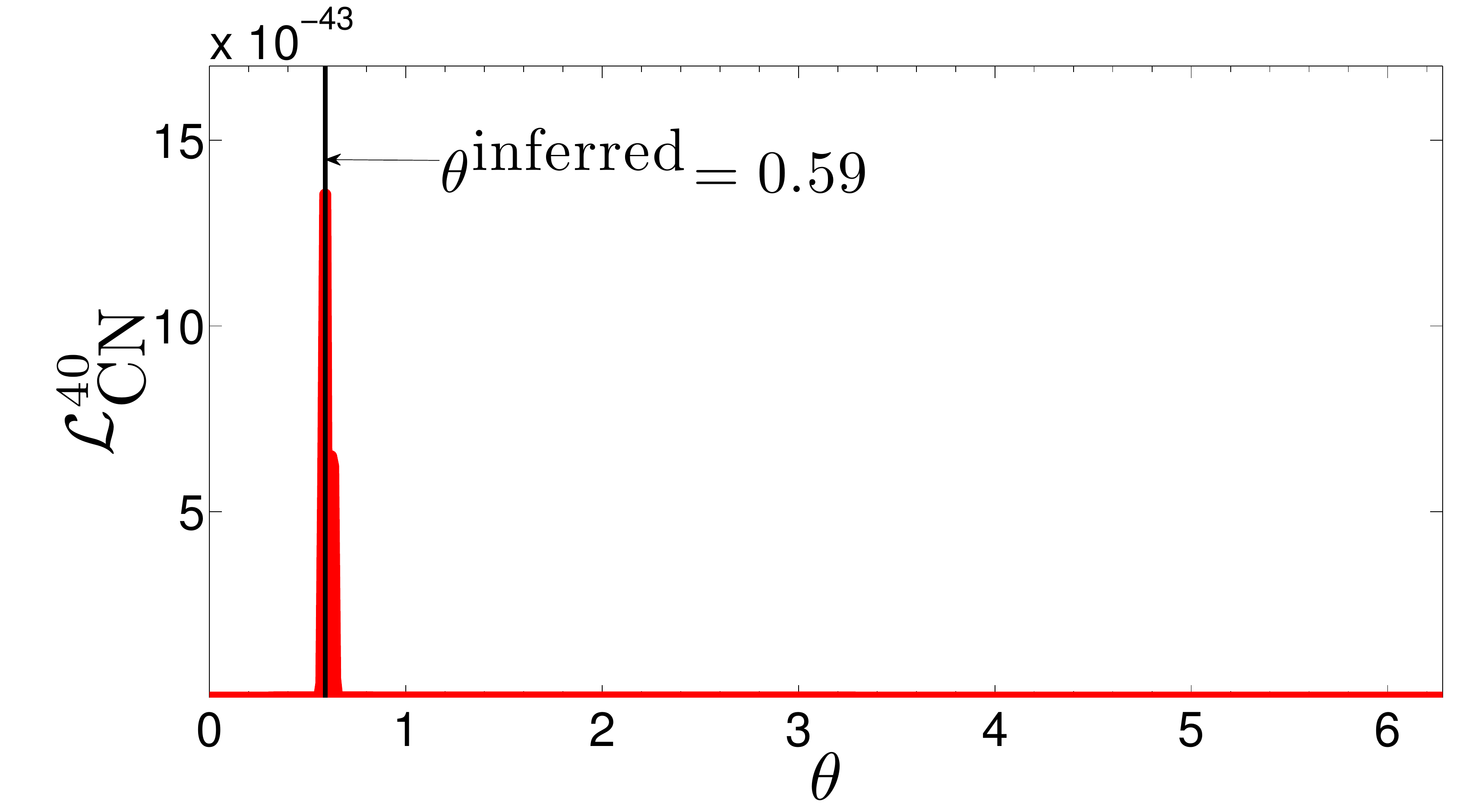}}
\subfigure{\includegraphics[width=1.7in, height=1.2in]{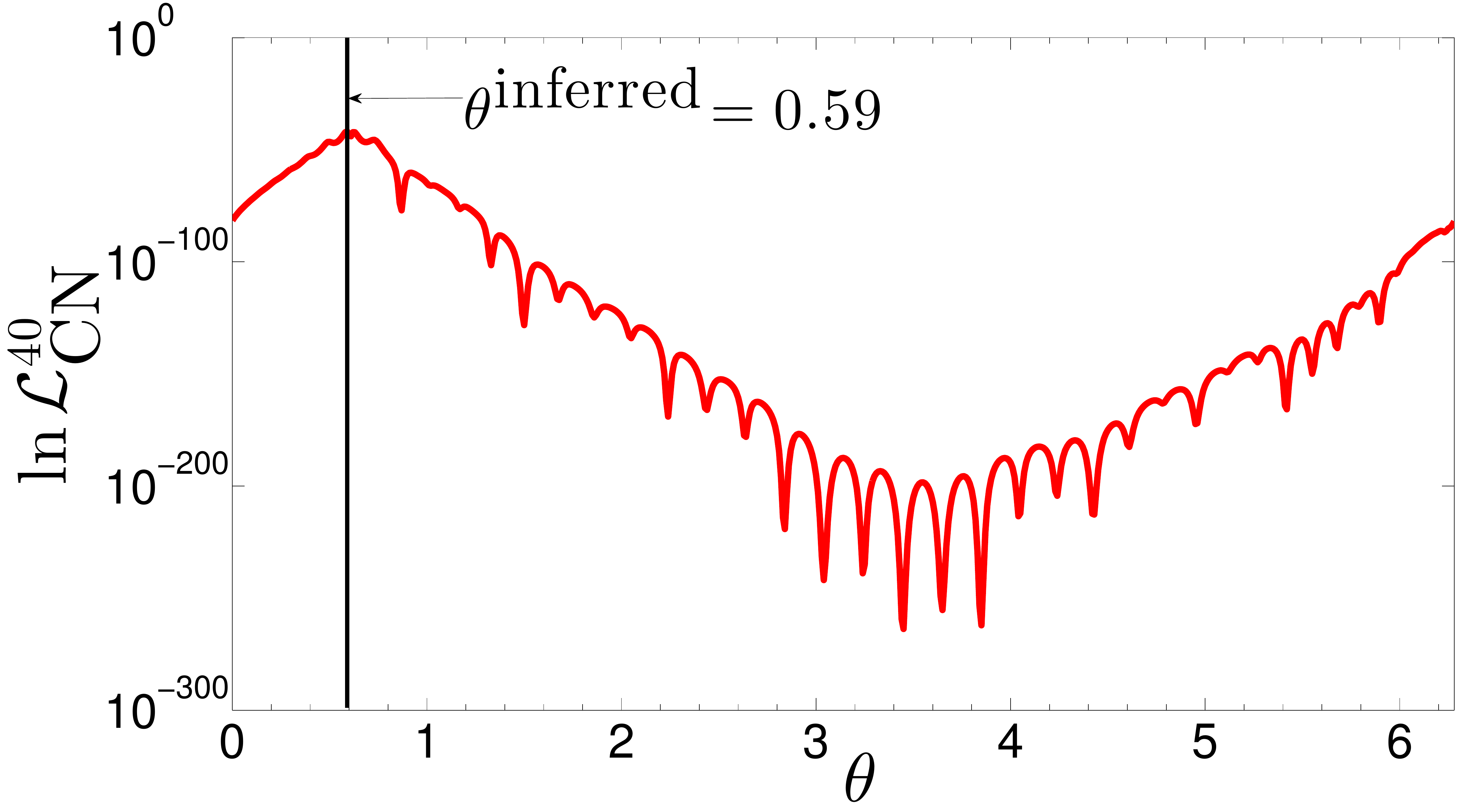}}
\subfigure{\includegraphics[width=1.7in, height=1.26in]{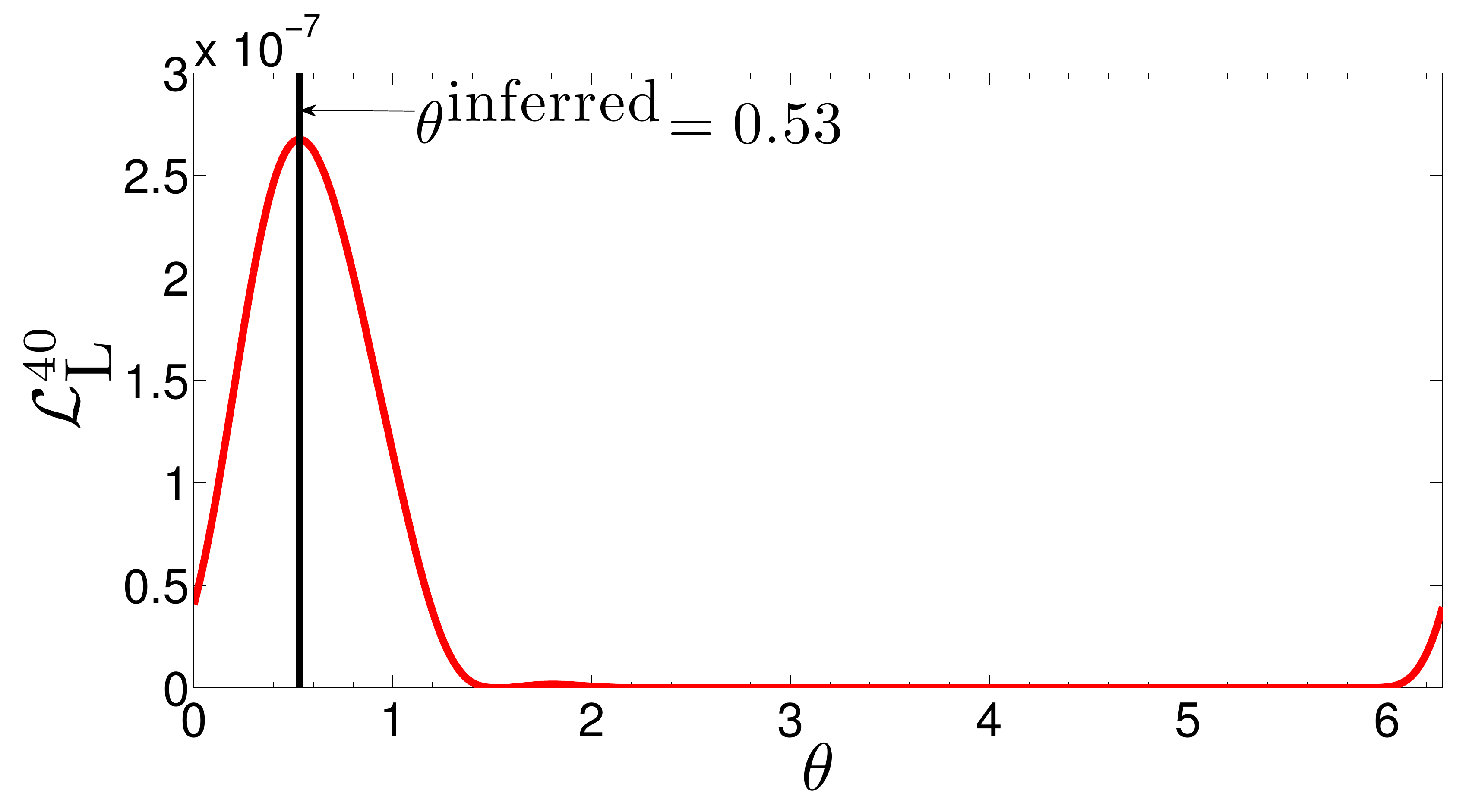}}
\subfigure{\includegraphics[width=1.7in, height=1.22in]{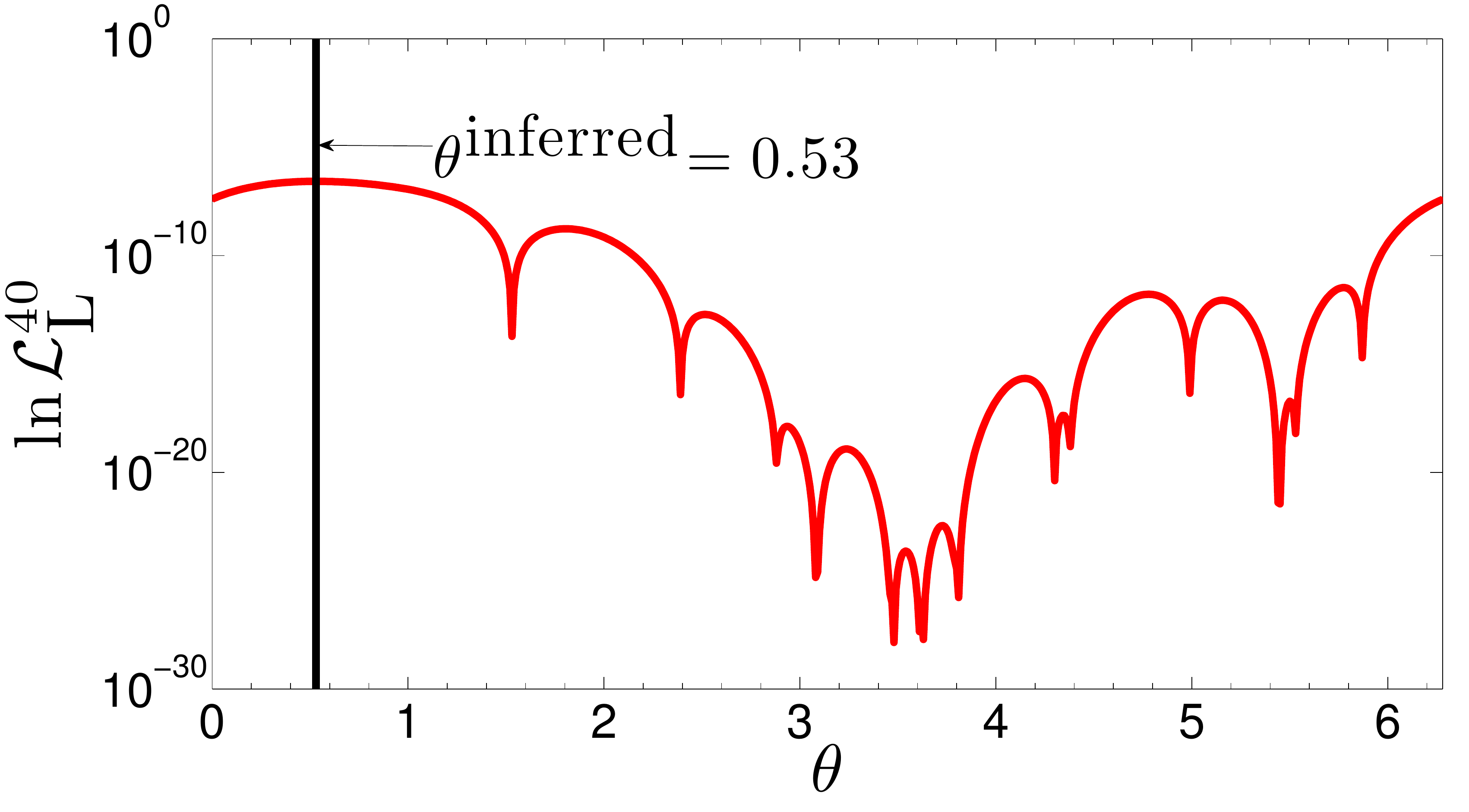}}
}
\caption{Likelihood landscapes for different nodes in a synthetic network with $t=5000$ nodes and parameters $m=1.5, L=2.5, \gamma=2.1, T=0.4$. The plots show the likelihoods $\mathcal{L}^{i}_{\textnormal{CN}}$, $\mathcal{L}^{i}_{\textnormal{L}}$ (Eqs.~(\ref{eq:local_likelihood_CN}),(\ref{eq:local_likelihood_links})) and the log-likelihoods  $\ln{\mathcal{L}^{i}_{\textnormal{CN}}}$, $\ln{\mathcal{L}^{i}_{\textnormal{L}}}$, for nodes appearing at MLE times $i=5, 10, 25, 30, 35, 40$, as a function of the angular coordinate $\theta$ (in radians). The vertical line in each plot shows the inferred angle $\theta^{\textnormal{inferred}}$ (in radians), which always  corresponds to the global maximum of the likelihood.
\label{fig:likelihood_landscapes}}
\end{figure*}

\textbf{Likelihood landscapes.} To gain a deeper understanding on the behavior of $\mathcal L_{\textnormal{CN}}^{i}$ and $\mathcal L_{\textnormal{L}}^{i}$, we show in Fig.~\ref{fig:likelihood_landscapes} the corresponding likelihood landscapes for different nodes that appear at early MLE times, $i=5, 10, 25, 30, 35, 40$. To enable comparison between the two methods, the link-based likelihood $\mathcal{L}^{i}_{\textnormal{L}}$ is computed after fixing the angles of the old nodes $j < i$ to the angles inferred by the common-neighbors method. We observe that at small $i$, $i=5,10$, $\mathcal L_{\textnormal{CN}}^{i}$ and $\mathcal L_{\textnormal{L}}^{i}$ behave quite differently, achieving their maximum at different values of $\theta$. As discussed in Section~\ref{sec:comparison_and_hybrid_approach}, nodes appearing at early MLE times are connected to all previous nodes with high probability. Therefore large zones of angular coordinates are nearly equally likely according to $\mathcal L_{\textnormal{L}}^{i}$, which is not the case with $\mathcal L_{\textnormal{CN}}^{i}$. This difference is evident in the first two rows of Fig.~\ref{fig:likelihood_landscapes}, showing the landscapes of $\mathcal L_{\textnormal{CN}}^{5}$, $\mathcal L_{\textnormal{CN}}^{10}$ and $\mathcal L_{\textnormal{L}}^{5}$, $\mathcal L_{\textnormal{L}}^{10}$. We also observe that $\mathcal L_{\textnormal{L}}^{i}$ of all possible angular coordinates is quite high for early nodes: $\mathcal L_{\textnormal{L}}^{5}$ of any angle is above $99\%$, and $\mathcal L_{\textnormal{L}}^{10}$ is above $92\%$ for all angles.

At larger times $i$, $i \geq 25$, the two likelihoods achieve their maximum around the same angle, while their landscapes vary in a somewhat similar manner. This justifies the hybrid approach of Section~\ref{sec:comparison_and_hybrid_approach}, which uses $\mathcal L_{\textnormal{CN}}^{i}$ to infer the angles of the first $i$ nodes for which $\bar{m}_i(t) \geq i-1$, and then $\mathcal L_{\textnormal{L}}^{i}$ to infer the angles of the rest of the nodes. For the considered networks, relation $\bar{m}_i(t) \geq i-1$ holds only for the first $33$ nodes, while for the AS Internet snapshots in the next section it holds only for the first $36$-$40$ nodes.

\begin{figure*}
\centerline{
\subfigure[~$T=0.05$, hybrid.]{\includegraphics[width=1.9in, height=1.45in]{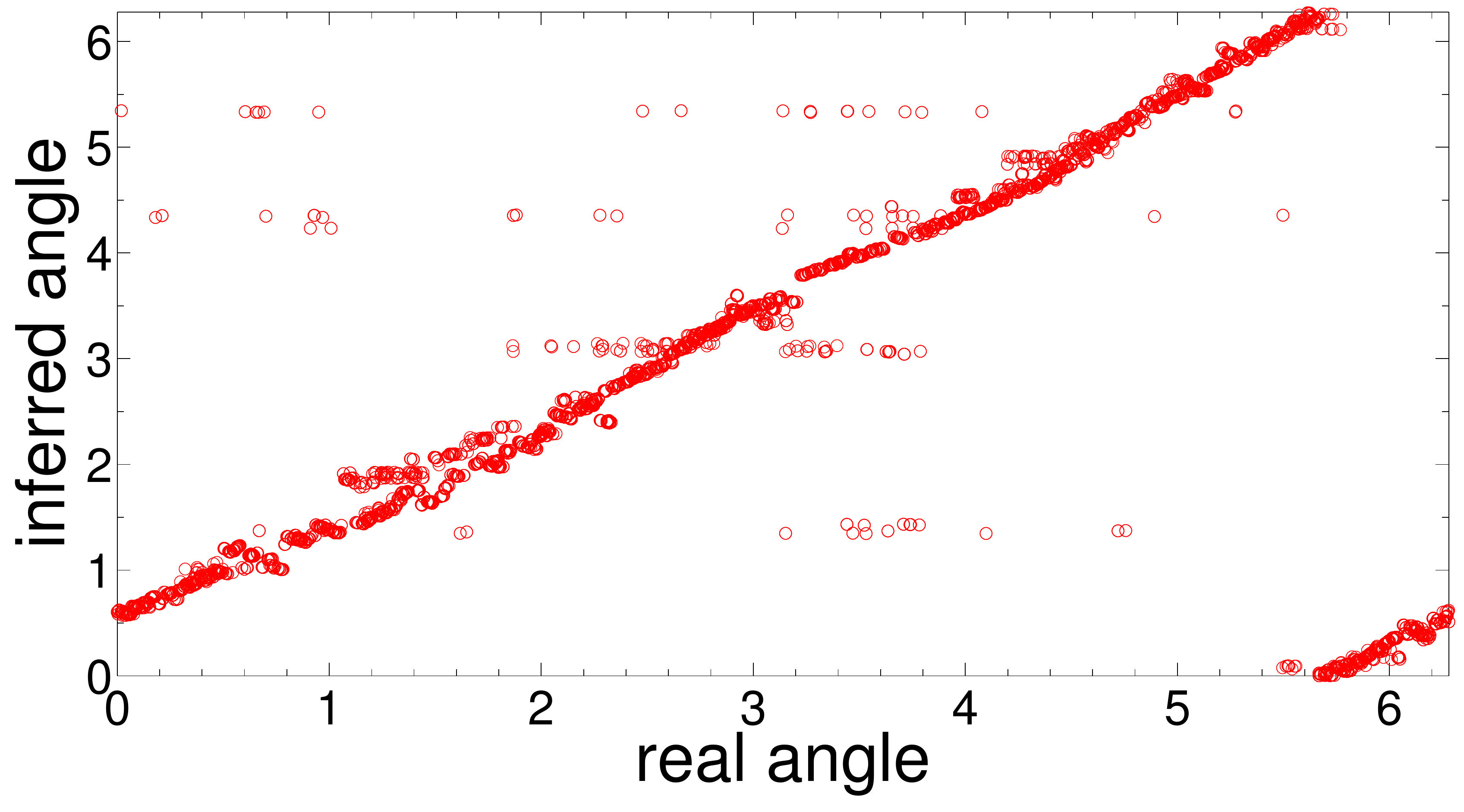}}
\subfigure[~$T=0.4$, hybrid.]{\includegraphics[width=1.9in, height=1.45in]{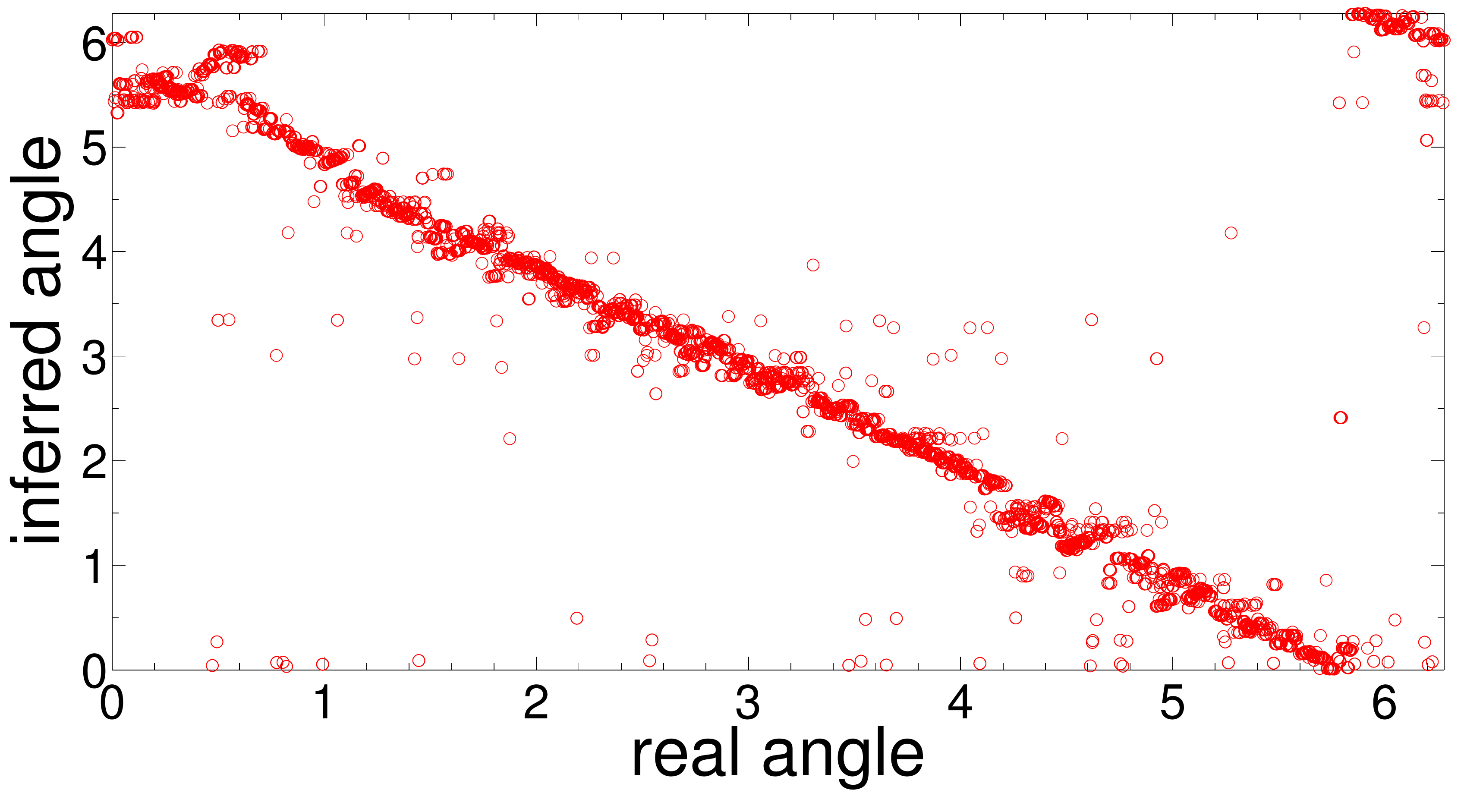}}
\subfigure[~$T=0.7$, hybrid.]{\includegraphics[width=1.9in, height=1.45in]{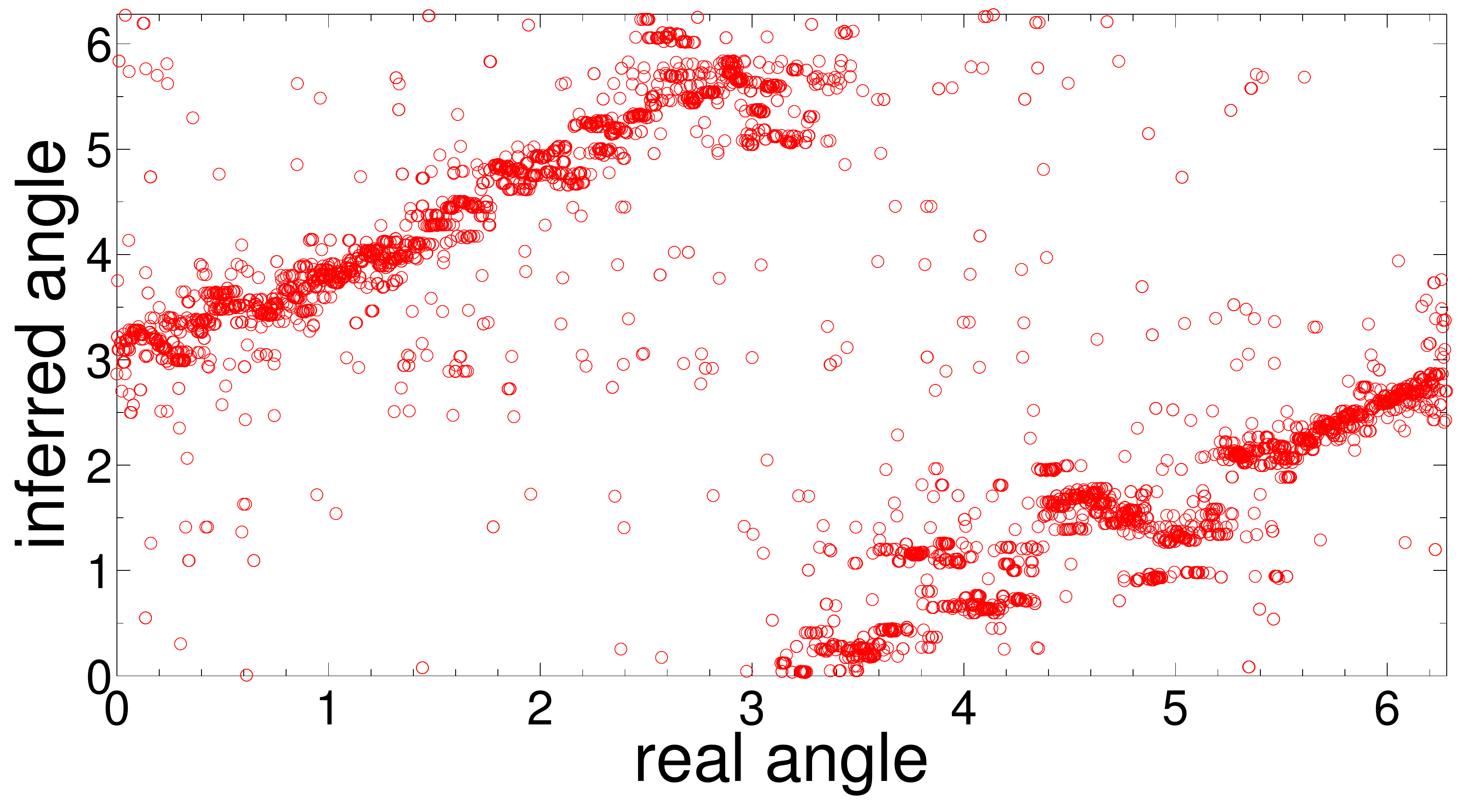}}
}
\centerline{
\subfigure[~$T=0.05$, $\mathcal L_{\textnormal{L}}^{i}$.]{\includegraphics[width=1.9in, height=1.45in]{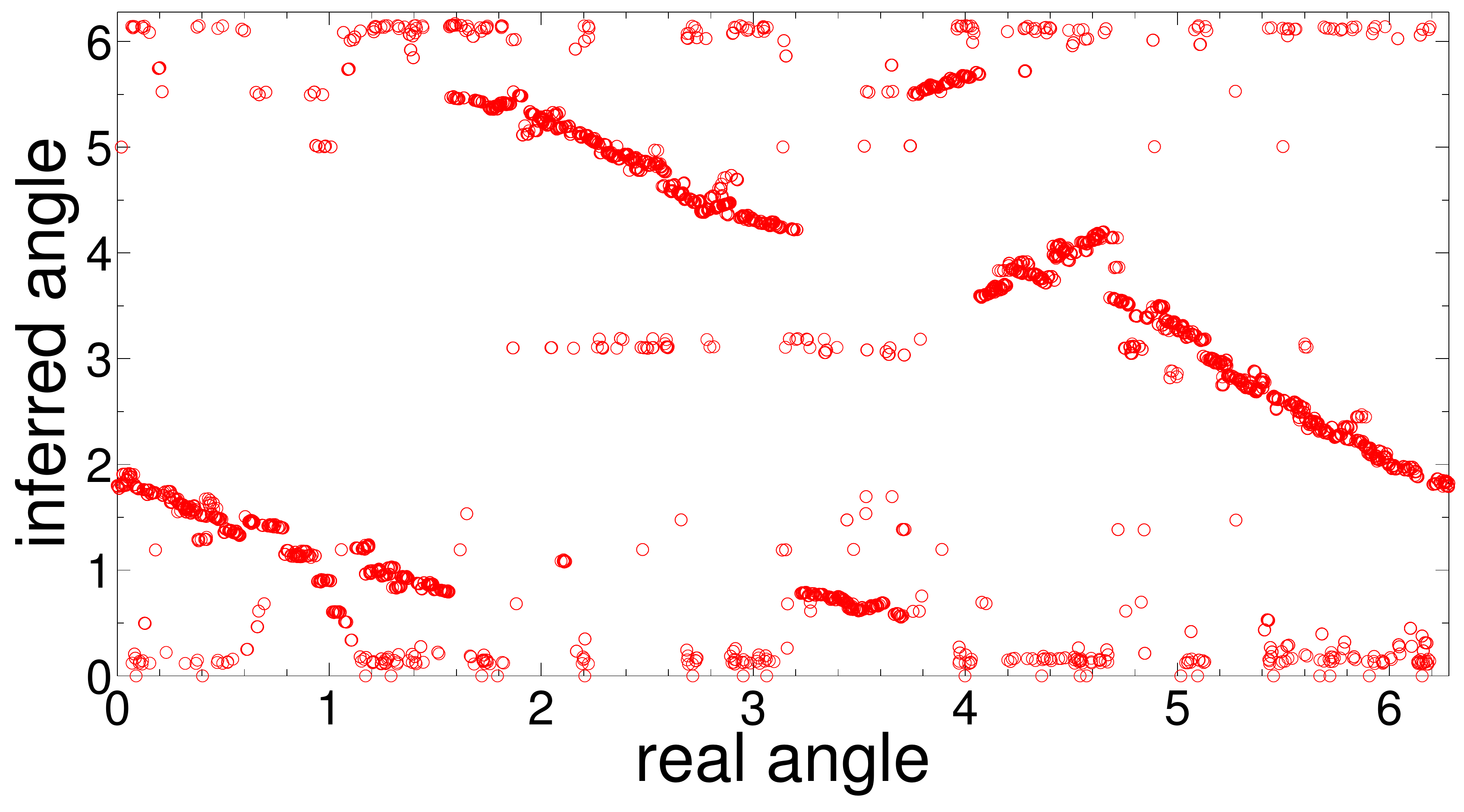}}
\subfigure[~$T=0.4$, $\mathcal L_{\textnormal{L}}^{i}$.]{\includegraphics[width=1.9in, height=1.45in]{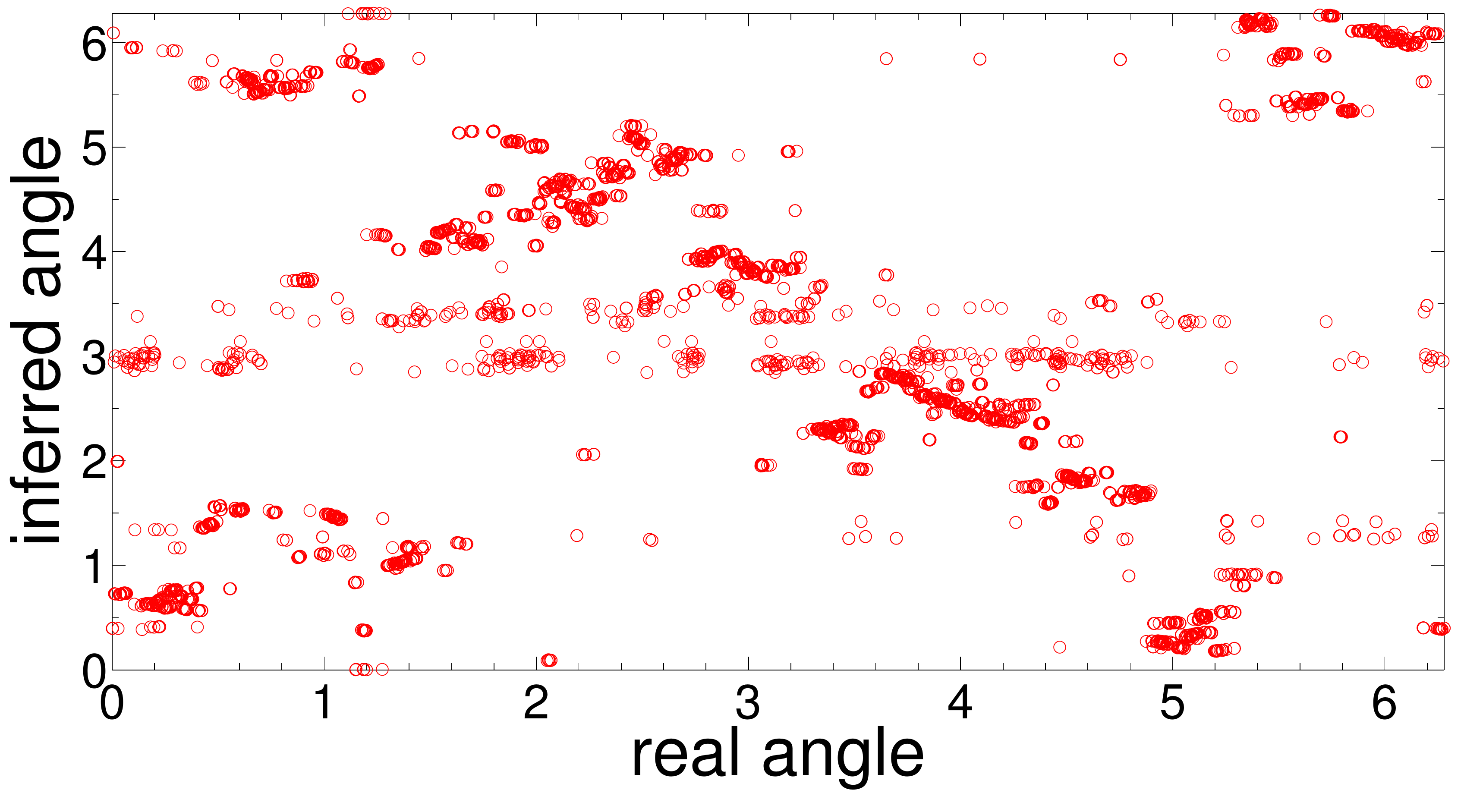}}
\subfigure[~$T=0.7$, $\mathcal L_{\textnormal{L}}^{i}$.]{\includegraphics[width=1.9in, height=1.45in]{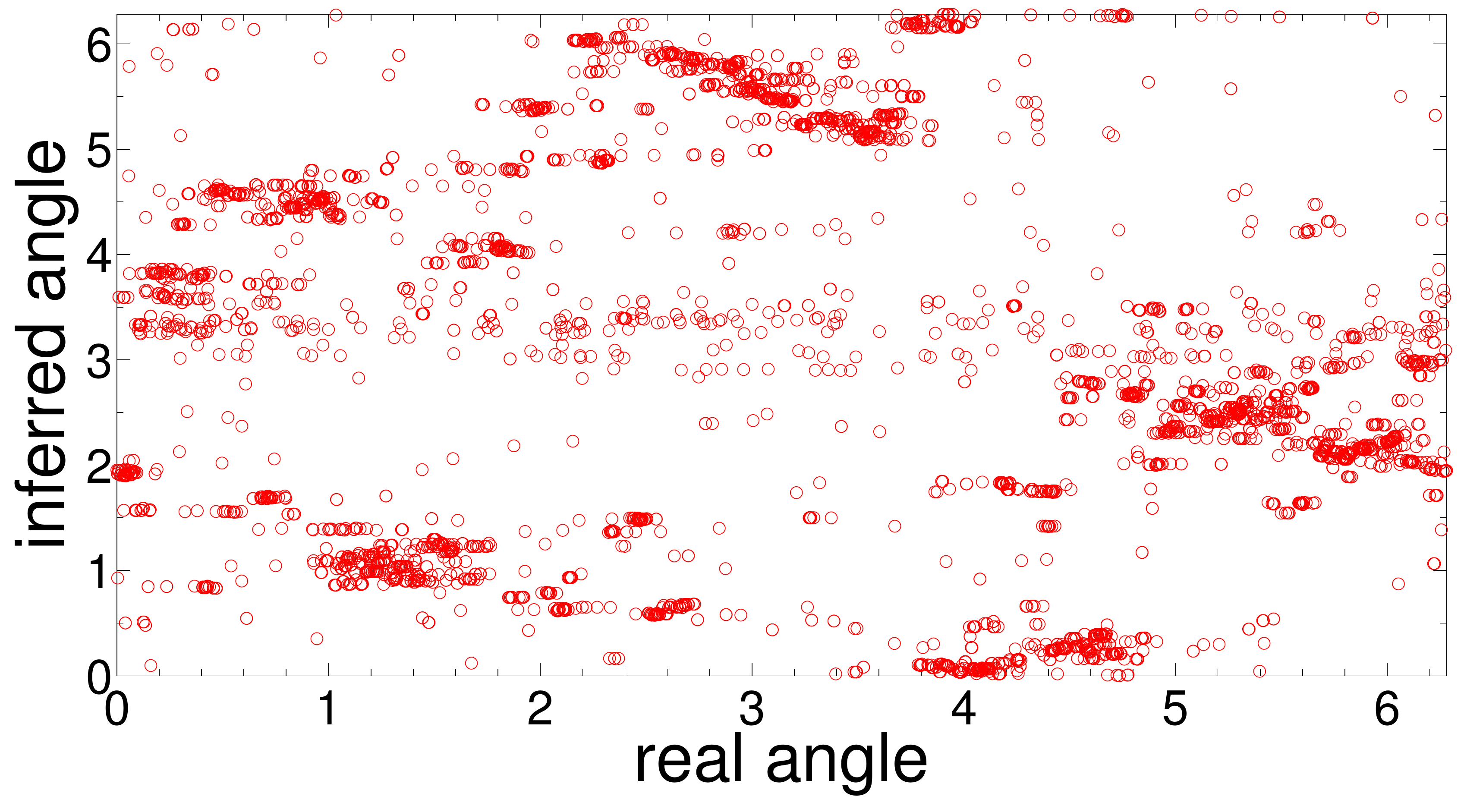}}
}
\caption{Inferred~vs.~real angles (in radians) for all the nodes in the synthetic networks of Fig.~\ref{fig:inferred_vs_real_first_100}. In (a-c) the hybrid method is used, while in (d-f) the link-based method is used.
\label{fig:inferred_vs_real_all_no_corrections}}
\end{figure*}

\textbf{Inferred versus real angles for all the nodes.} Fig.~\ref{fig:inferred_vs_real_all_no_corrections} juxtaposes the inferred against the real angles for all nodes in each considered network, when the hybrid and link-based methods are used. We observe that: (i) the hybrid method is more accurate than the link-based method, as expected; (ii) Figs.~\ref{fig:inferred_vs_real_all_no_corrections}(a-c) are similar to Figs.~\ref{fig:inferred_vs_real_first_100}(a-c), meaning that as long as the angular coordinates of the first few nodes are accurately inferred, then the angular coordinates of the rest of the nodes will also be accurately inferred; (iii) the inference is in general better at lower temperatures $T$; and (iv) that the inference is in general better for higher degree nodes appearing at early MLE times, cf. Figs.~\ref{fig:inferred_vs_real_first_100}(a-c) and \ref{fig:inferred_vs_real_all_no_corrections}(a-c).

\begin{figure*}
\centerline{
\subfigure[~$T=0.05$.]{\includegraphics[width=1.9in, height=1.3in]{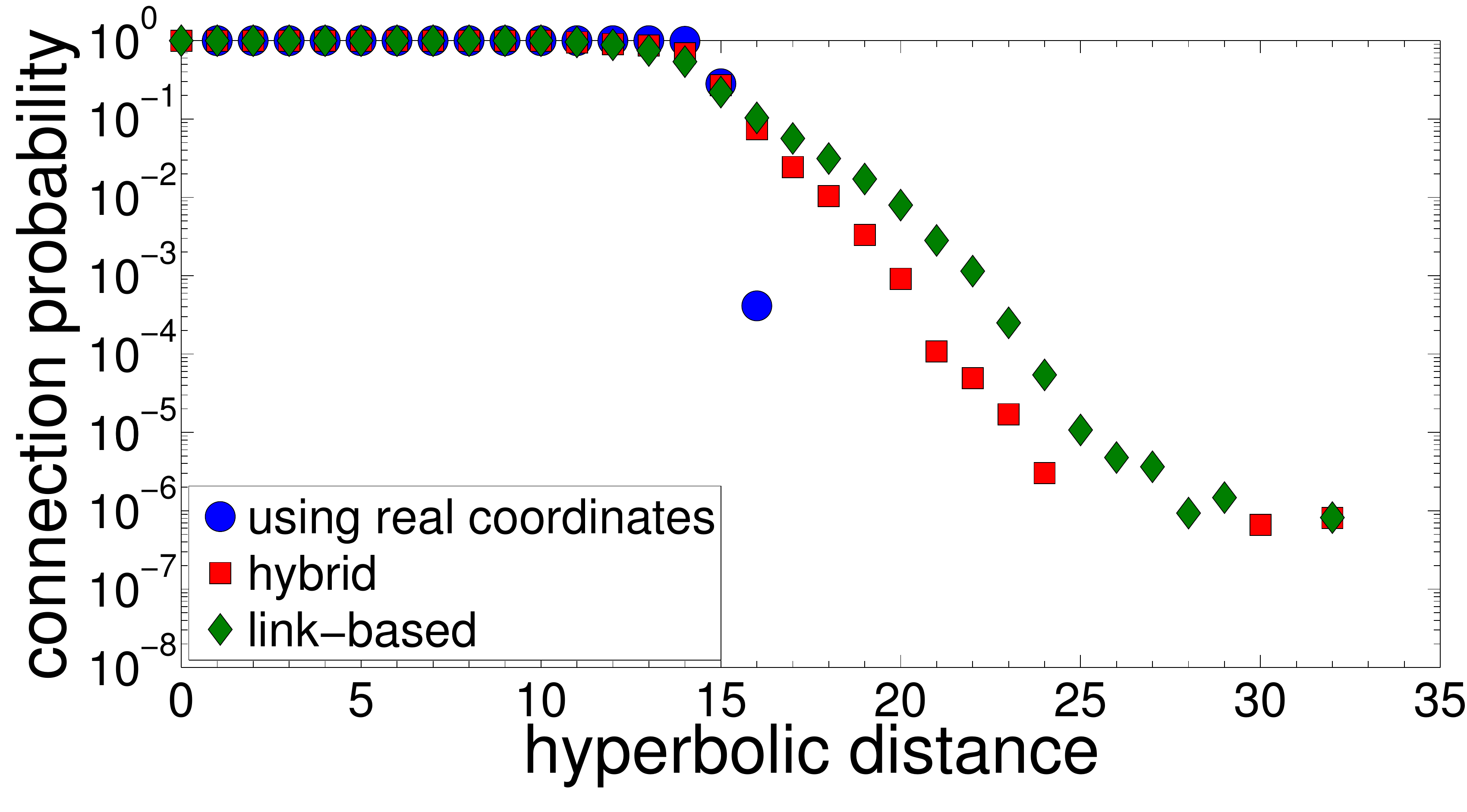}}
\subfigure[~$T=0.4$.]{\includegraphics[width=1.9in, height=1.3in]{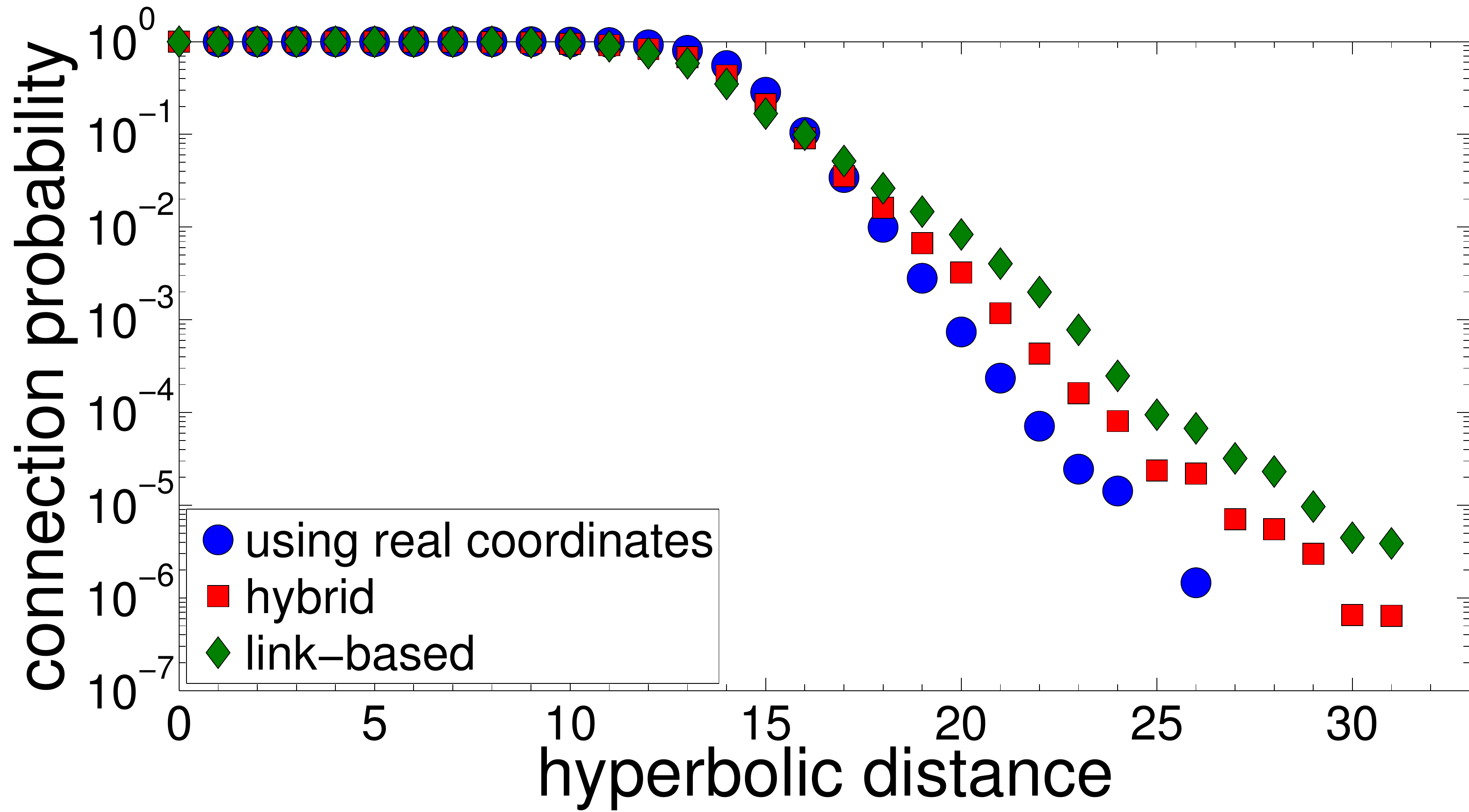}}
\subfigure[~$T=0.7$.]{\includegraphics[width=1.9in, height=1.3in]{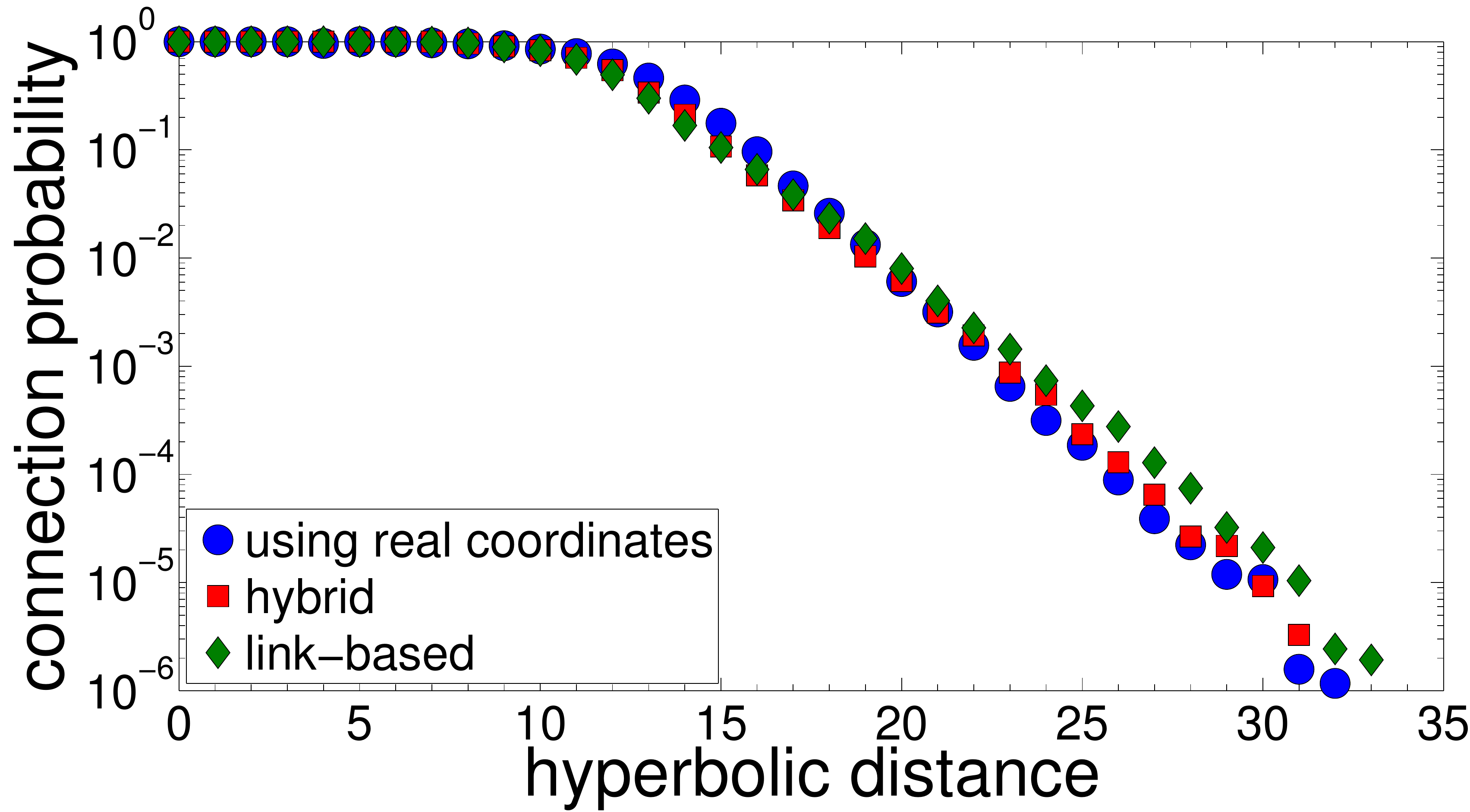}}
}
\caption{Connection probabilities with inferred (radial and angular) node coordinates obtained by the hybrid and link-based methods, and with real node coordinates. The results correspond to the mappings of Fig.~\ref{fig:inferred_vs_real_all_no_corrections}.
\label{fig:connection_probability_no_corrections}}
\end{figure*}

\textbf{Connection probability.} In Fig.~\ref{fig:connection_probability_no_corrections} we report the connection probability, which is the probability that there is a link between a pair of nodes located at hyperbolic distance $x$, using real and inferred node coordinates. This probability is computed as the ratio of the number of connected node pairs to the total number of pairs of nodes located at distance $x$. From the figure, we observe that all inferred connection probabilities are close to the real ones, except from some discrepancies at their tails, which are more pronounced at lower  $T$'s. Furthermore, we see that the results with the hybrid method are only slightly better compared to the link-based method in terms of the connection probability. This suggests that the link-based method also produces relatively good mappings, even though it cannot infer as well the real angular coordinates.

We also quantify the quality of the obtained mappings using two other metrics: (i)~the logarithmic loss, and (ii)~the performance of greedy routing.

\textbf{Logarithmic loss.}
The logarithmic loss is a quality metric for statistical inference defined as $LL=-{\ln{\mathcal L}}$, where $\mathcal L$ in our case is the global likelihood
\begin{equation}
\label{eq:global_likelihood}
\mathcal L=\prod_{1 \leq j < i \leq t} p(x_{ij}(t))^{\alpha_{ij}}\left[1-p(x_{ij}(t))\right]^{1-\alpha_{ij}}.
\end{equation}
The product goes over all node pairs $i, j$ in the network, $x_{ij}(t)$ is the hyperbolic distance between pair $i, j$, and $p(x_{ij}(t))=1/(1+e^{\frac{\zeta}{2T}(x_{ij}(t)-R_t)})$ is the connection probability. We use $LL$ to quantify the quality of the inference of the node angular coordinates. Specifically, we first compute $LL$ using the inferred node coordinates $\{r_i(t), \theta_i\}$, and then compare the result to the case where $LL$ is computed using the inferred $r_i(t)$'s and \emph{random} $\theta_i$'s drawn uniformly from $[0, 2\pi]$. We denote the former by $LL^{\textnormal{inf}}$ and the latter by $LL^{\textnormal{rand}}$. The smaller the $LL^{\textnormal{inf}}$ compared to $LL^{\textnormal{rand}}$, the better the quality of the mapping.  In particular, the ratio $r_{LL}=e^{-LL^{\textnormal{inf}}}/e^{-LL^{\textnormal{rand}}}=e^{(LL^{\textnormal{rand}}-LL^{\textnormal{inf}})}$ is the ratio of the likelihood with the inferred angular coordinates to the likelihood with random angular coordinates. The higher this ratio, the better the mapping quality. Table~\ref{tab:logloss_no_corrections} reports the logarithmic losses $LL^{\textnormal{inf}}$, $LL^{\textnormal{rand}}$, the ratio $r_{LL}$, as well as $LL^{\textnormal{real}}$ that is the logarithmic loss if we use the real radial and angular coordinates of nodes. We observe that: (i) the hybrid method yields lower logarithmic losses compared to the link-based method, which is expected since it infers the node angular coordinates more accurately; and (ii) that the logarithmic losses for both methods are significantly lower than those obtained with random angular coordinates, and closer to the logarithmic losses obtained with the real coordinates. These results suggest that the link-based method yields relatively good results, but the hybrid approach is better, as expected.
\begin{table*}
\begin{center}
\caption{Logarithmic Losses in the mappings of Fig.~\ref{fig:inferred_vs_real_all_no_corrections}
\label{tab:logloss_no_corrections}}
\begin{tabular}{|c|c|c|c|c|c|c|}
\hline Network & $LL^{\textnormal{real}}$ &  $LL^{\textnormal{inf}}$, hybrid & $LL^{\textnormal{inf}}$, link-based & $LL^{\textnormal{rand}}$ & $r_{LL}$, hybrid & $r_{LL}$, link-based \\
\hline $T=0.05$ & $1.1 \times 10^4$ & $9.6 \times 10^4$ & $24.8 \times 10^4$& $123 \times 10^4$ &$e^{1134000}$&$e^{982000}$\\
\hline $T=0.4$ & $2.4 \times 10^4$ & $3.5 \times 10^4$ & $5.4 \times 10^4$& $17 \times 10^4$&$e^{135000}$&$e^{116000}$\\
\hline $T=0.7$ & $4.1 \times 10^4$ & $4.4 \times 10^4$ & $5.2\times10^4$& $11 \times 10^4$&$e^{66000}$&$e^{58000}$\\
\hline
\end{tabular}
\end{center}
\end{table*}

\textbf{Performance of greedy routing.}
One specific class of network functions that are impossible without underlying geometry are efficient targeted transport processes without global knowledge of the network structure. Many real networks have this routing or navigation function in common; in some networks, including the Internet, this function is their primary function~\cite{BoKrKc08}. Therefore navigability can be used as an alternative indirect metric of embedding quality. Navigability of an embedding is also of independent interest for some applications, such as Internet routing~\cite{BoPa10}. A network embedded in a geometric space is said \emph{navigable} if \emph{greedy routing (GR)} is efficient according to the metrics considered below. In GR, a node's address is its coordinates in the space, and each node knows only the addresses of its neighbors, and the destination node address of a ``packet''. Upon receipt of such a packet, the GR node, if it is not a destination, forwards the packet to its neighbor closest to the destination in the geometric space, and drops the packet if a local minimum loop is detected, i.e., if this neighbor is the same as the previous node visited by the packet.

We evaluate the efficiency of GR in the synthetic networks of Fig.~\ref{fig:inferred_vs_real_all_no_corrections}, using both the HyperMap-inferred (hybrid, link-based) and the real node coordinates. We consider the following two GR efficiency metrics~\cite{BoKrKc08}: (i) the percentage of successful paths, $p_s$, which is the proportion of paths that do not get looped and reach their destinations; and (ii) the average hop-length $\bar{h}$ of the successful paths. The results are shown in Table~\ref{tab:navigation_no_corrections}, where we see that: (i) both the hybrid and link-based methods yield mappings where GR is quite efficient, yielding high $p_s$'s and low path lengths $\bar{h}$, as it is the case with the real node coordinates; and (ii) that the hybrid method performs better, as expected, especially at lower $T$'s.
\begin{table*}
\begin{center}
\caption{Success ratio $p_s$ and average hop-length $\bar{h}$ of greedy paths in the mappings of Fig.~\ref{fig:inferred_vs_real_all_no_corrections}
\label{tab:navigation_no_corrections}}
\begin{tabular}{|c|c|c|c|}
\hline Network & using real coordinates & using inferred coordinates, hybrid & using inferred coordinates, link-based\\
\hline $T=0.05$ & $p_s=0.99, \bar{h}=3.0$ & $p_s=0.96, \bar{h}=3.1$ & $p_s=0.82, \bar{h}=3.3$\\
\hline $T=0.4$ & $p_s=0.94, \bar{h}=3.2$ & $p_s=0.95, \bar{h}=3.4$ & $p_s=0.87, \bar{h}=3.5$ \\
\hline $T=0.7$ & $p_s=0.77, \bar{h}=3.5$ & $p_s=0.92, \bar{h}=3.8$ & $p_s=0.89, \bar{h}=3.9$ \\
\hline
\end{tabular}
\end{center}
\end{table*}

\textbf{Correction steps.} We now repeat the same experiments applying the link-based and hybrid methods \emph{with} correction steps, in order to investigate the differences. Specifically, for each method we run $4$ correction steps as described in Section~\ref{sec:correction_steps}, right after all nodes with degrees $k \geq 60, 40, 20, 10$ appear in the network. Each of these correction steps is repeated $8$ times, which equals the average degree $\bar{k}$ in each network. In the hybrid method, the node angular coordinates that were inferred using the common-neighbors approach are not altered by the correction steps.

\begin{figure*}
\centerline{
\subfigure[~$T=0.05$, hybrid with correction steps.]{\includegraphics[width=1.9in, height=1.45in]{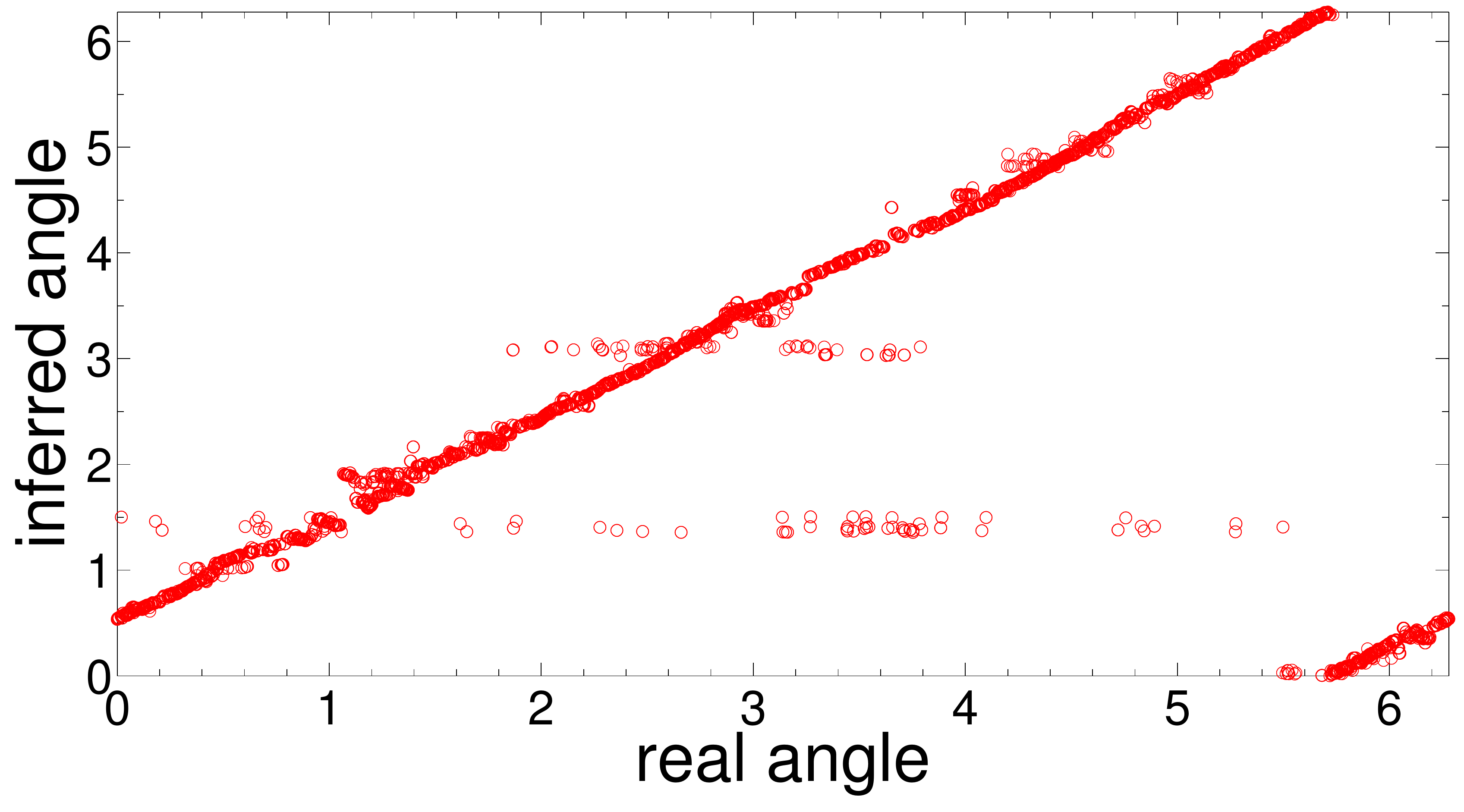}}
\subfigure[~$T=0.4$, hybrid with correction steps.]{\includegraphics[width=1.9in, height=1.45in]{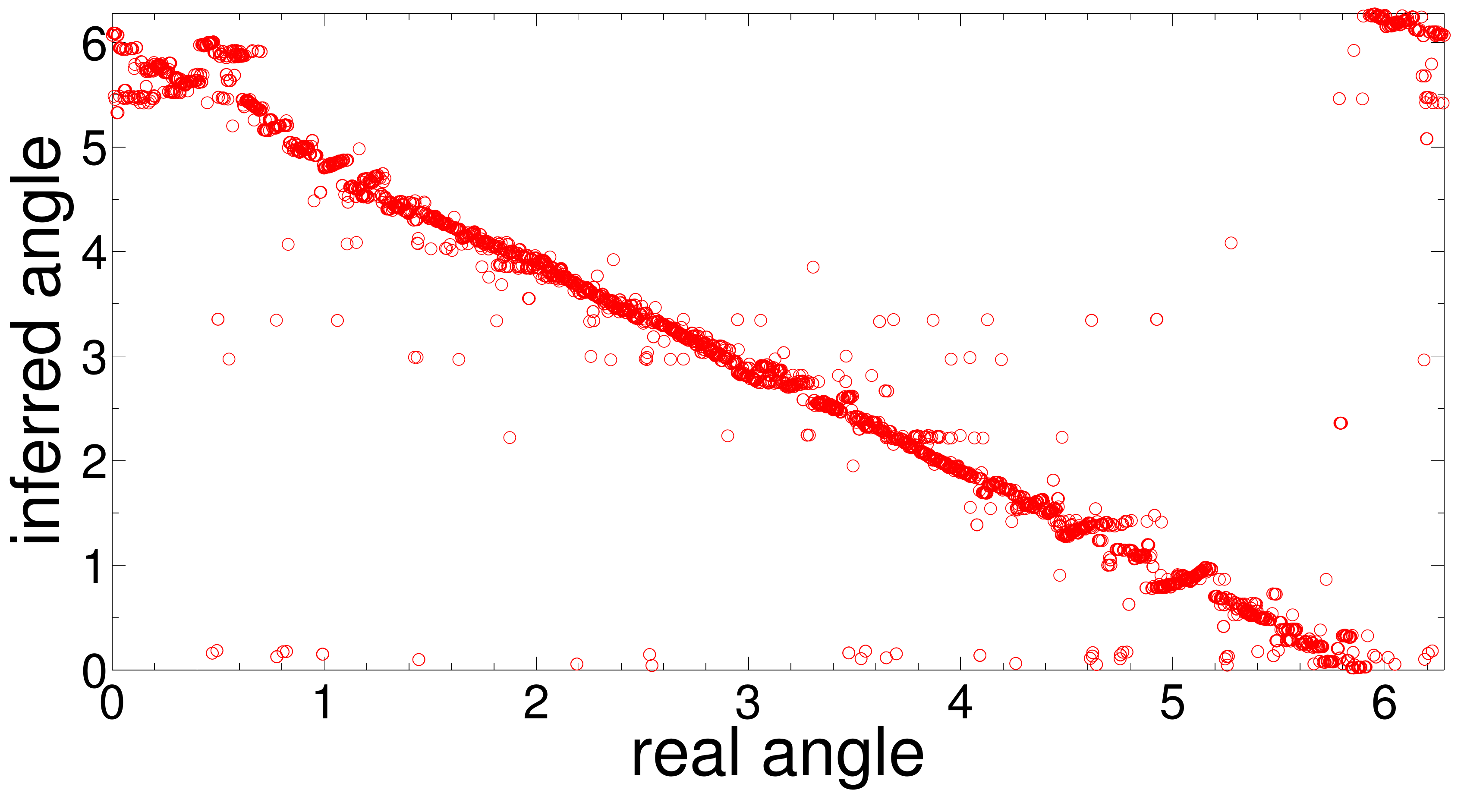}}
\subfigure[~$T=0.7$, hybrid with correction steps.]{\includegraphics[width=1.9in, height=1.45in]{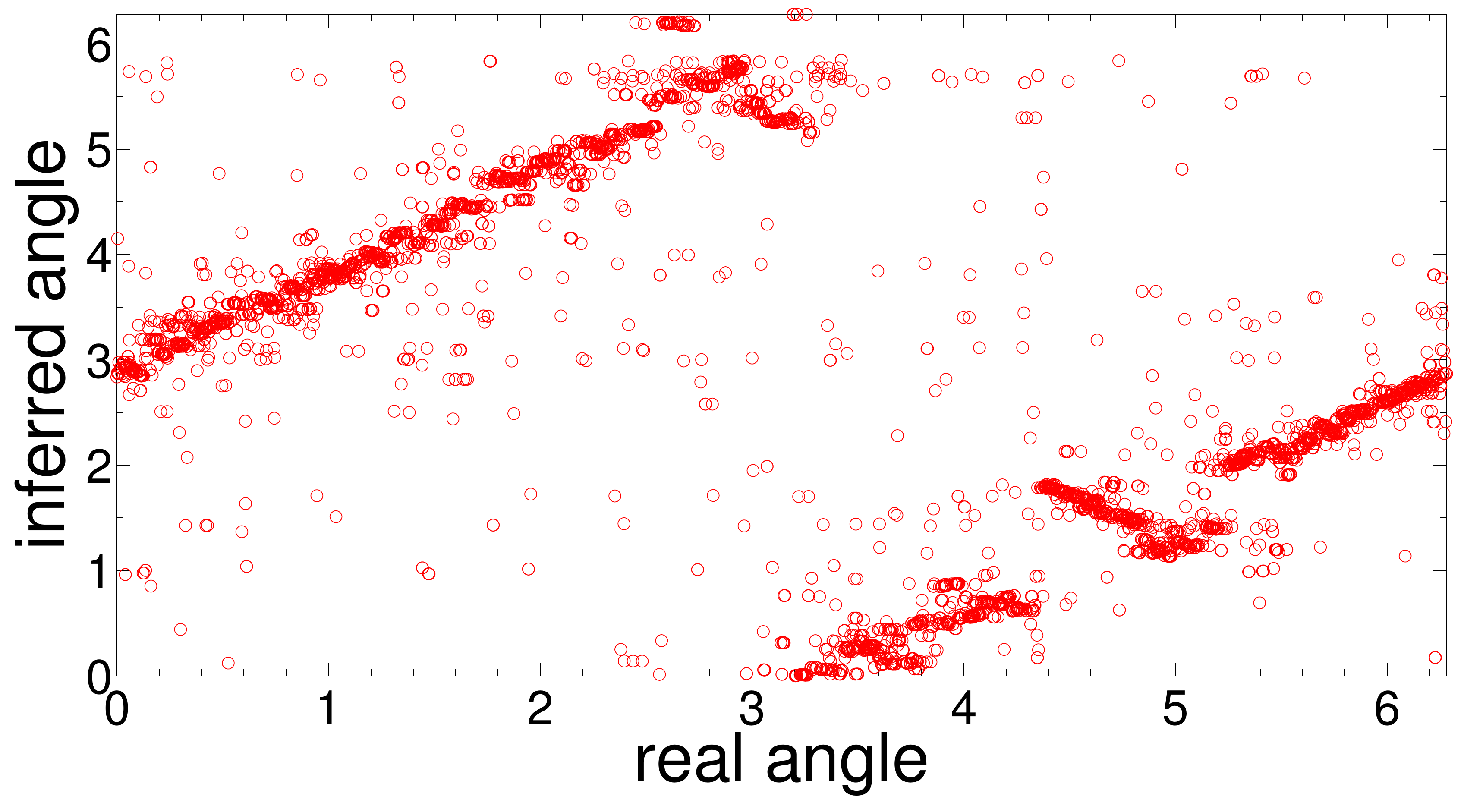}}
}
\centerline{
\subfigure[~$T=0.05$, link-based with correction steps.]{\includegraphics[width=1.9in, height=1.45in]{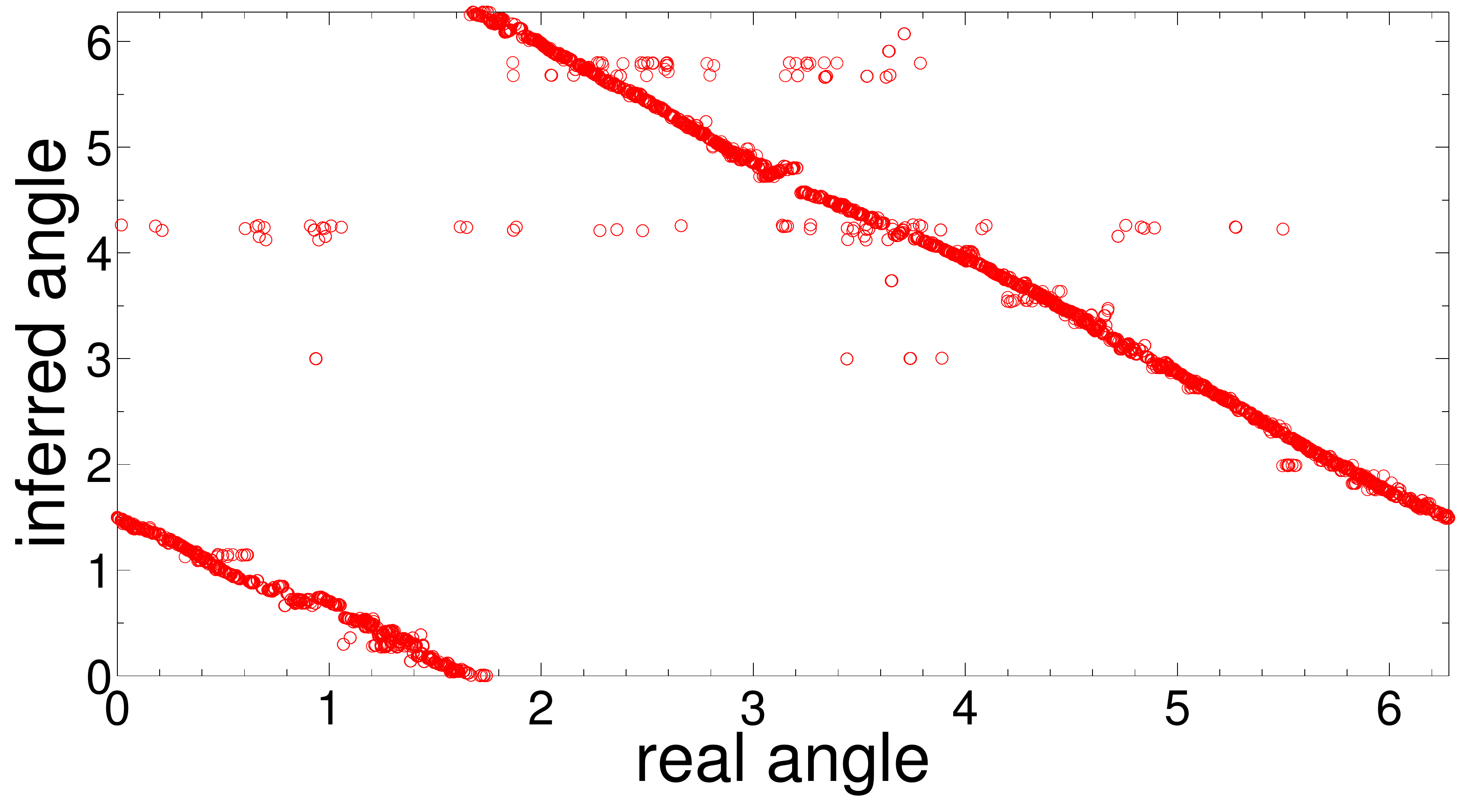}}
\subfigure[~$T=0.4$, link-based with correction steps.]{\includegraphics[width=1.9in, height=1.45in]{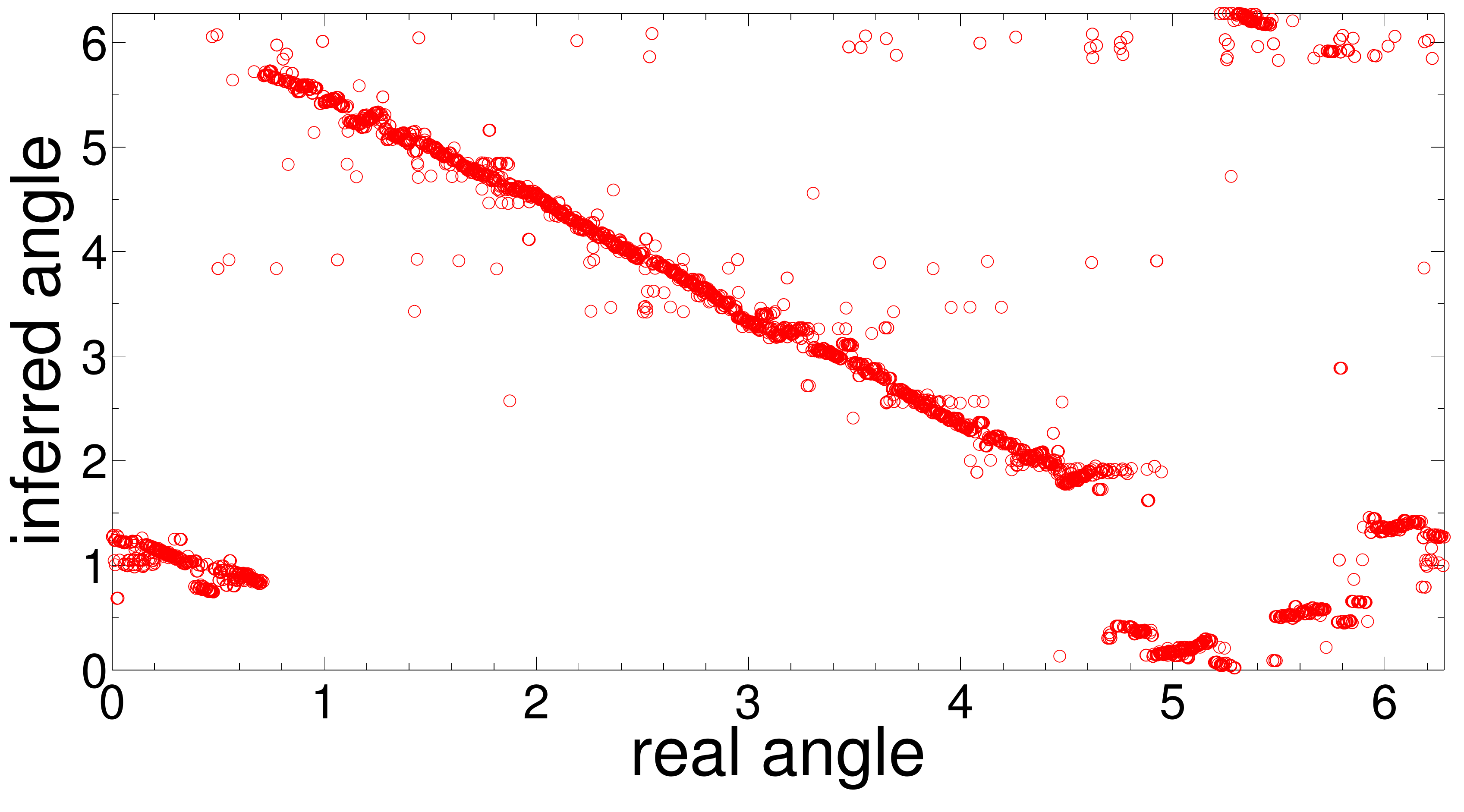}}
\subfigure[~$T=0.7$, link-based with correction steps.]{\includegraphics[width=1.9in, height=1.45in]{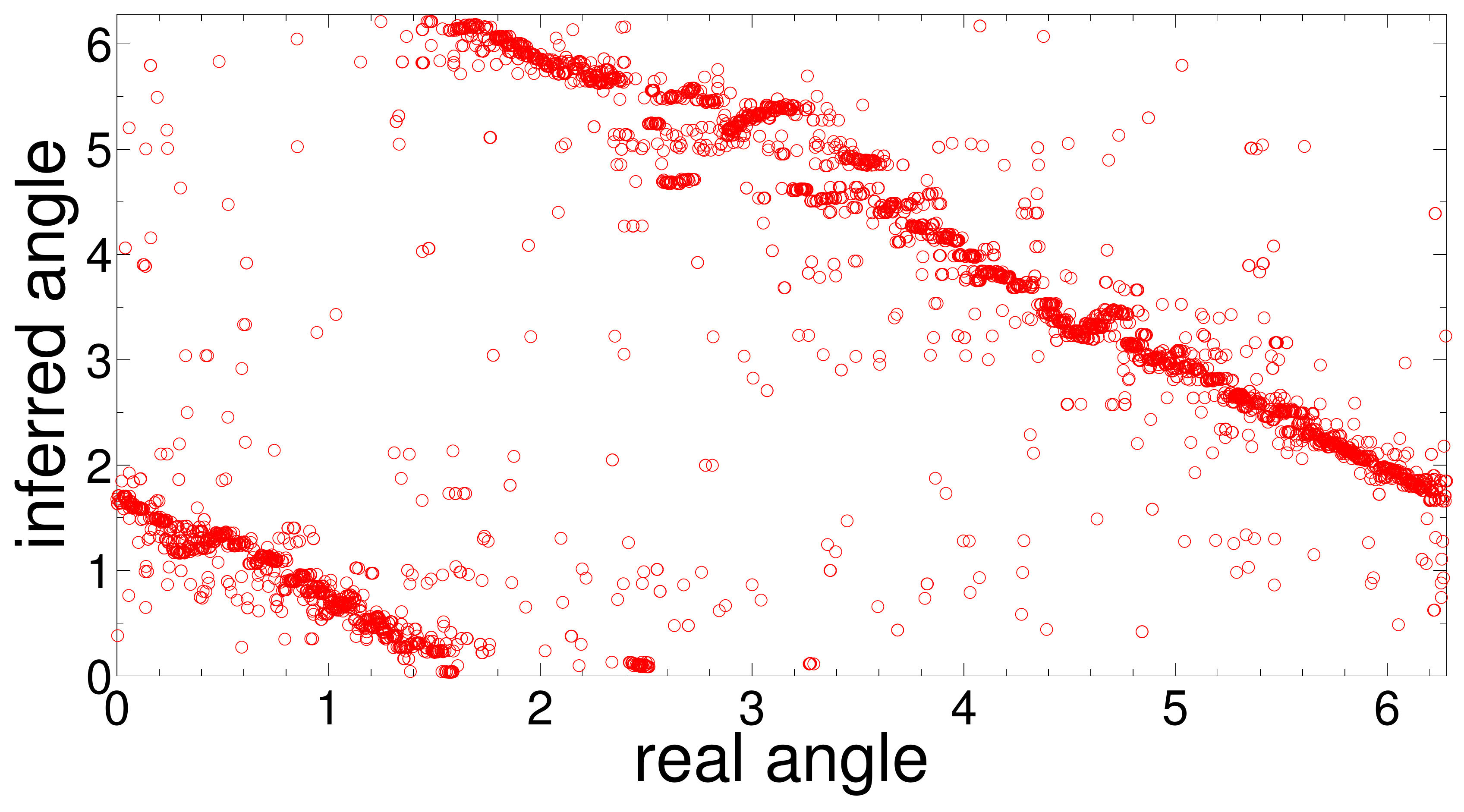}}
}
\caption{Inferred~vs.~real angles (in radians) for all nodes in the networks of Fig.~\ref{fig:inferred_vs_real_all_no_corrections}. In (a-c) the hybrid method is used, while in (d-f) the link-based method is used. In both methods the correction steps are run as described in the text.
\label{fig:inferred_vs_real_all_with_corrections}}
\end{figure*}

Figures~\ref{fig:inferred_vs_real_all_with_corrections}(d-f) show that the node angular coordinates are now inferred quite accurately with the link-based method. Figures~\ref{fig:inferred_vs_real_all_with_corrections}(a-c) show the results for the hybrid method, which look similar to Figs.~\ref{fig:inferred_vs_real_all_no_corrections}(a-c); this means that the effect of correction steps in this case is not as significant. All results are in agreement with Sections~\ref{sec:comparison_and_hybrid_approach} and~\ref{sec:correction_steps}. In all cases, the inference is better at lower $T$'s, as in Fig.~\ref{fig:inferred_vs_real_all_no_corrections}.

\begin{figure*}
\centerline{
\subfigure[~$T=0.05$.]{\includegraphics[width=1.9in, height=1.3in]{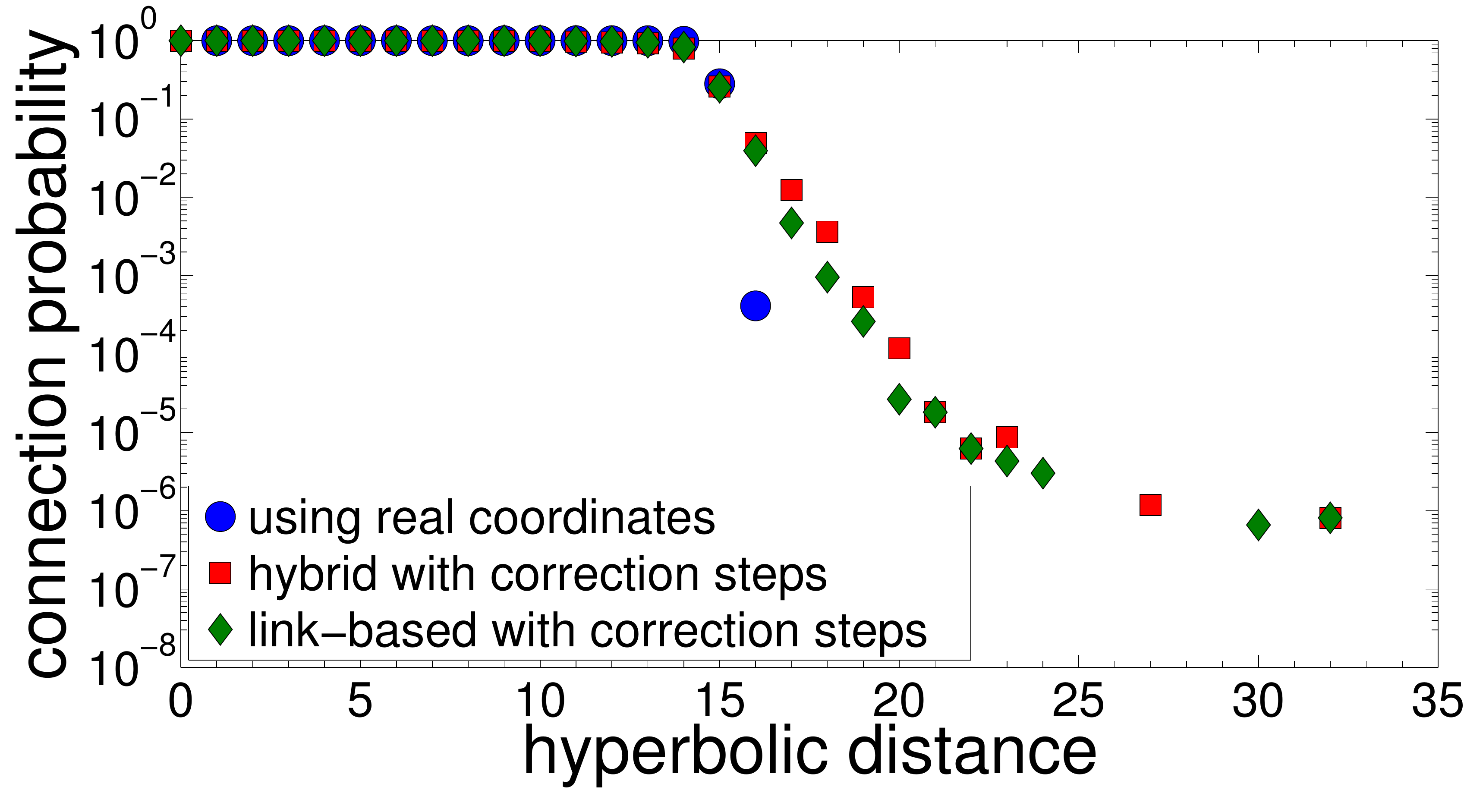}}
\subfigure[~$T=0.4$.]{\includegraphics[width=1.9in, height=1.3in]{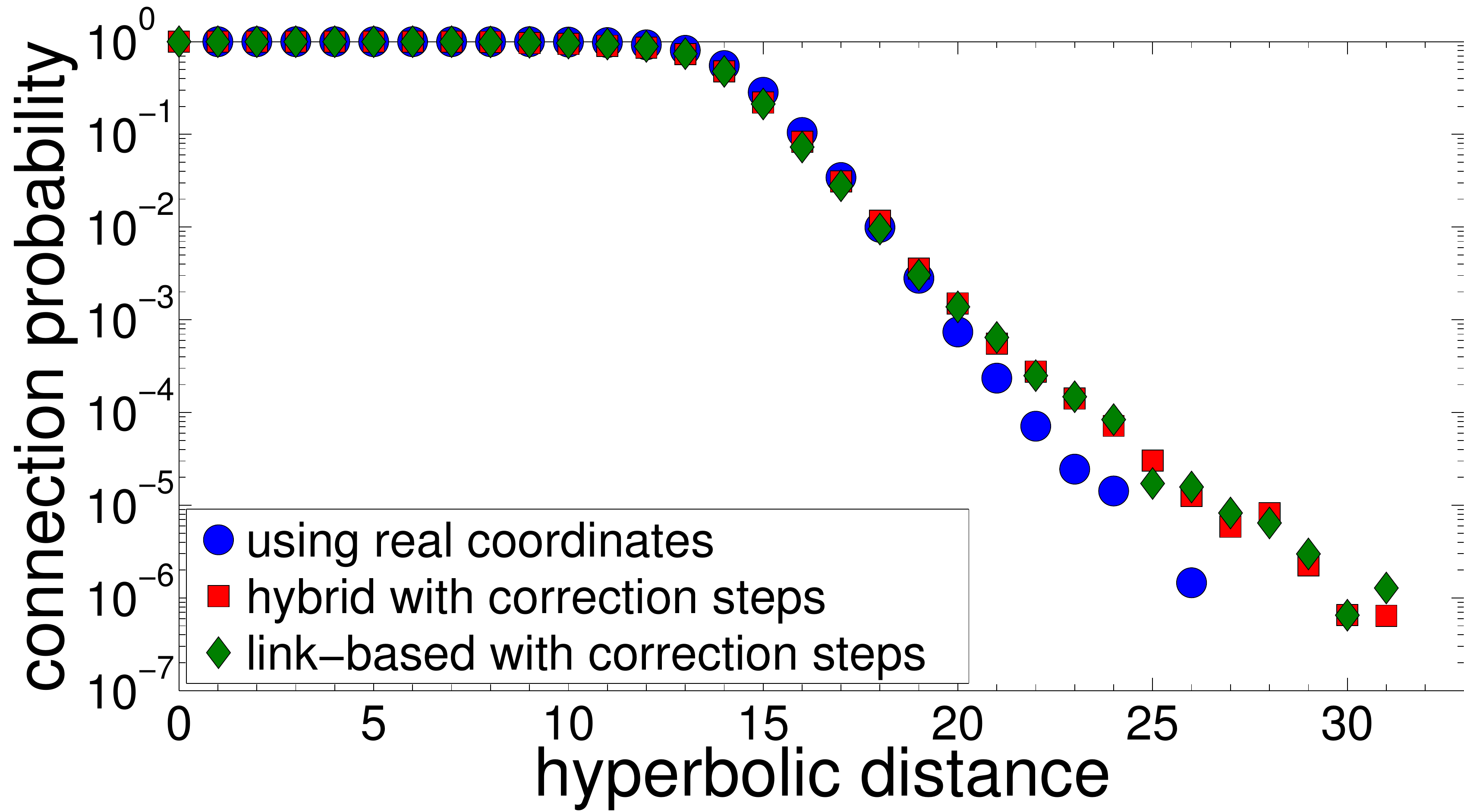}}
\subfigure[~$T=0.7$.]{\includegraphics[width=1.9in, height=1.3in]{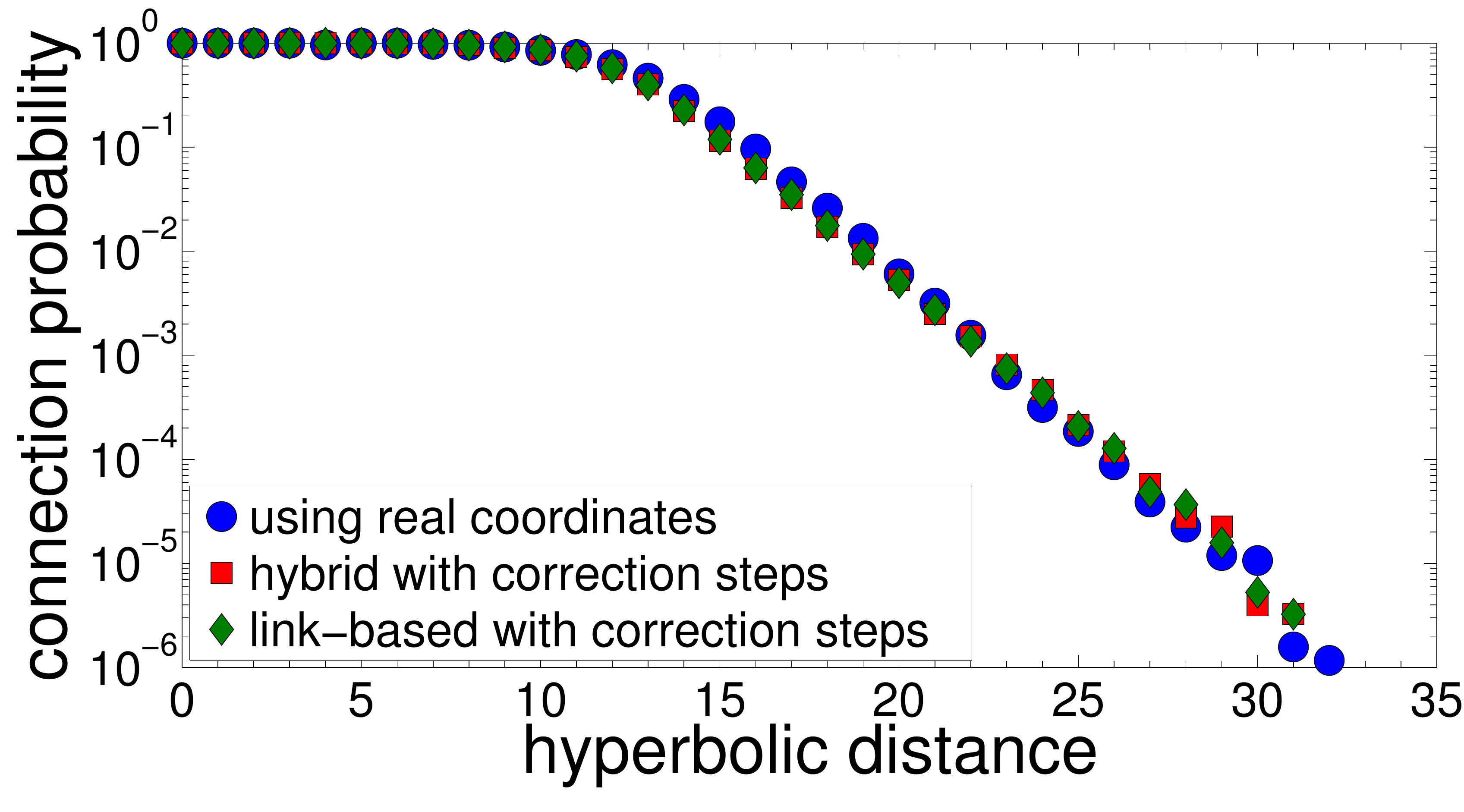}}
}
\caption{Connection probabilities with inferred (radial and angular) node coordinates obtained by the hybrid and link-based methods, and with real node coordinates. The results correspond to the mappings of Fig.~\ref{fig:inferred_vs_real_all_with_corrections}.
\label{fig:connection_probability_with_corrections}}
\end{figure*}

The corresponding connection probabilities are shown in Fig.~\ref{fig:connection_probability_with_corrections}. Compared to Fig.~\ref{fig:connection_probability_no_corrections}, we observe that correction steps can help to better capture the connection probability tail, in both the hybrid and link-based methods. Finally, the logarithmic losses and the performance of GR are reported in Tables~\ref{tab:logloss_with_corrections} and~\ref{tab:navigation_with_corrections}. In Table~\ref{tab:logloss_with_corrections}, we observe that all logarithmic losses are smaller compared to those in Table~\ref{tab:logloss_no_corrections}, and even closer to the logarithmic losses obtained with the real coordinates. This means that correction steps improve the quality of the obtained mappings in all cases. The improvement is quite significant for the link-based method, as expected, which at lower temperatures yields even lower logarithmic losses than the hybrid method. From Table~\ref{tab:navigation_with_corrections}, we see that the efficiency of GR is better compared to the results in Table~\ref{tab:navigation_no_corrections}, especially for the link-based method. We also note from Tables~\ref{tab:navigation_with_corrections} and~\ref{tab:navigation_no_corrections} that in some high-temperature cases, GR with inferred node coordinates performs even better than GR with real node coordinates. A possible explanation for this effect is given in Section VIII of~\cite{hypermap_ton}.
\begin{table*}
\begin{center}
\caption{Logarithmic Losses in the mappings of Fig.~\ref{fig:inferred_vs_real_all_with_corrections}
\label{tab:logloss_with_corrections}}
\begin{tabular}{|c|c|c|c|c|c|c|}
\hline Network & $LL^{\textnormal{real}}$ &  $LL^{\textnormal{inf}}$, hybrid & $LL^{\textnormal{inf}}$, link-based & $LL^{\textnormal{rand}}$ & $r_{LL}$, hybrid & $r_{LL}$, link-based \\
\hline $T=0.05$ & $1.1 \times 10^4$ & $5.5 \times 10^4$ & $3.9\times 10^4$ & $123 \times 10^4$ &$e^{1175000}$&$e^{1191000}$\\
\hline $T=0.4$ & $2.4 \times 10^4$ & $2.9 \times 10^4$ & $2.8 \times 10^4$ & $17 \times 10^4$&$e^{141000}$&$e^{142000}$\\
\hline $T=0.7$ & $4.1 \times 10^4$ & $4.1 \times 10^4$ & $4.1 \times 10^4$ & $11 \times 10^4$&$e^{69000}$&$e^{69000}$\\
\hline
\end{tabular}
\end{center}
\end{table*}
\begin{table*}
\begin{center}
\caption{Success ratio $p_s$ and average hop-length $\bar{h}$ of greedy paths in the mappings of Fig.~\ref{fig:inferred_vs_real_all_with_corrections}
\label{tab:navigation_with_corrections}}
\begin{tabular}{|c|c|c|c|}
\hline Network & using real coordinates & using inferred coordinates, hybrid & using inferred coordinates, link-based\\
\hline $T=0.05$ & $p_s=0.99, \bar{h}=3.0$ & $p_s=0.97, \bar{h}=3.1$ & $p_s=0.98, \bar{h}=3.1$\\
\hline $T=0.4$ & $p_s=0.94, \bar{h}=3.2$ & $p_s=0.96, \bar{h}=3.3$ & $p_s=0.97, \bar{h}=3.3$ \\
\hline $T=0.7$ & $p_s=0.77, \bar{h}=3.5$ & $p_s=0.93, \bar{h}=3.7$ & $p_s=0.93, \bar{h}=3.7$ \\
\hline
\end{tabular}
\end{center}
\end{table*}

\begin{figure*}
\centerline{
\subfigure{\includegraphics[width=2.2in, height=1.5in]{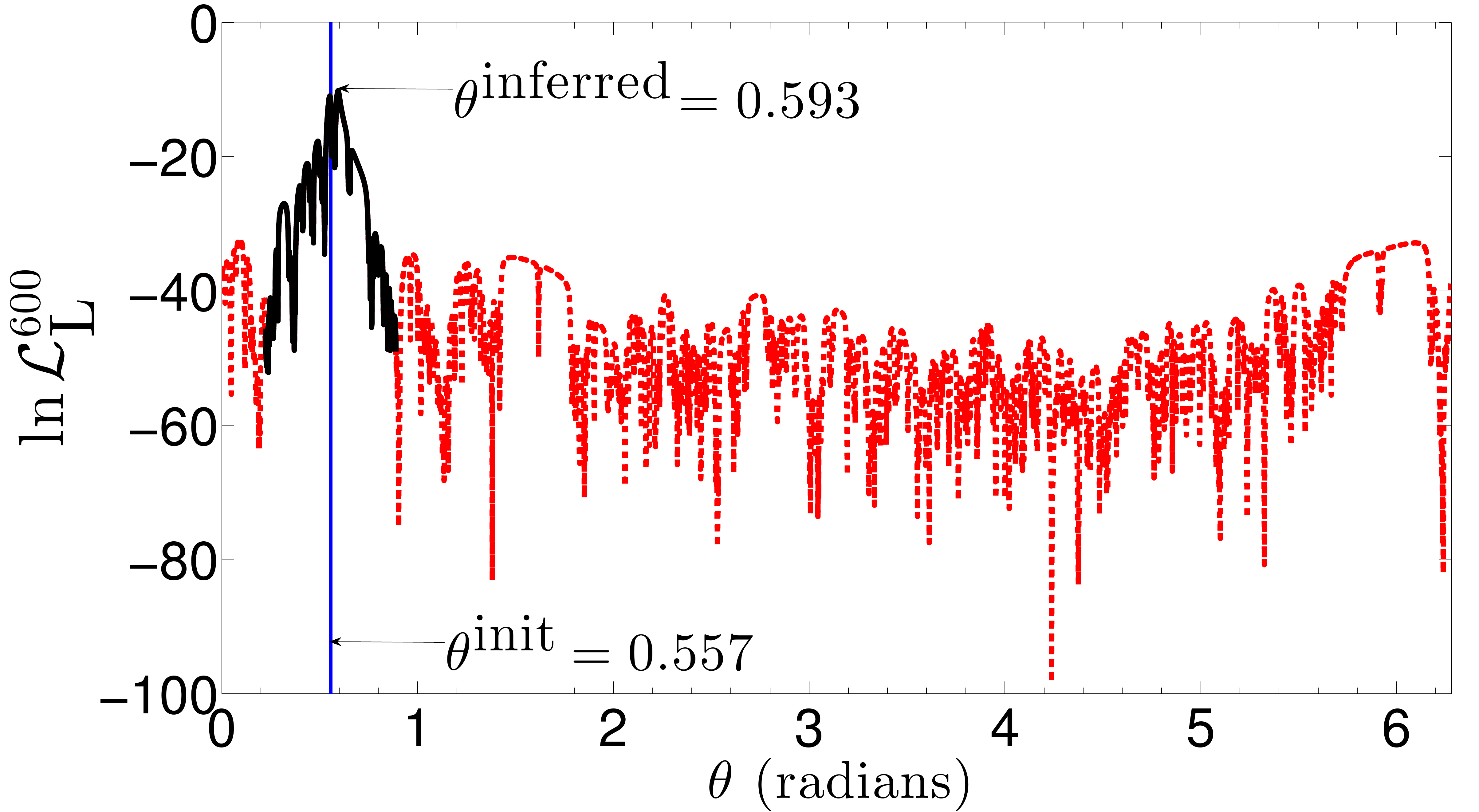}}
\subfigure{\includegraphics[width=2.2in, height=1.5in]{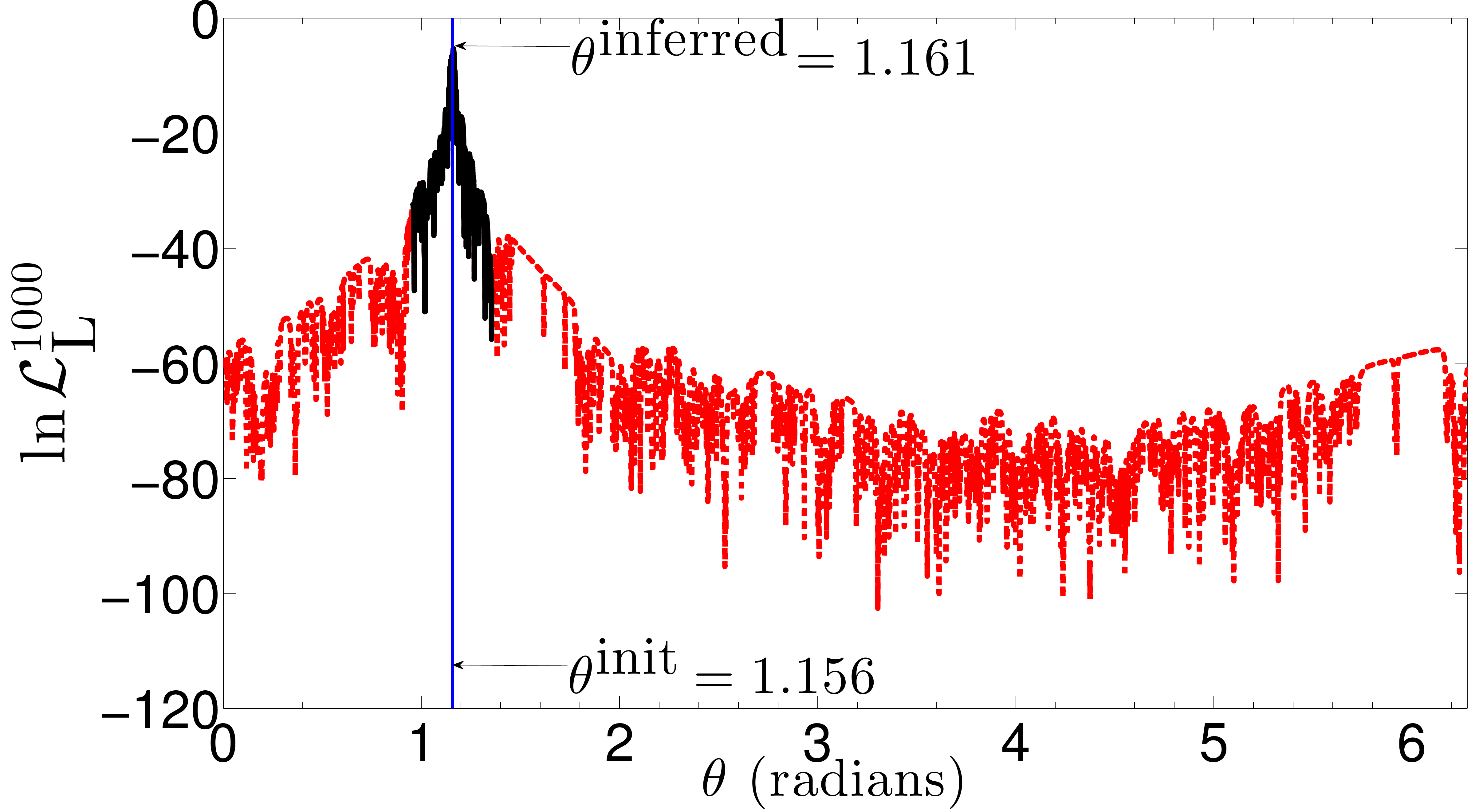}}
\subfigure{\includegraphics[width=2.2in, height=1.5in]{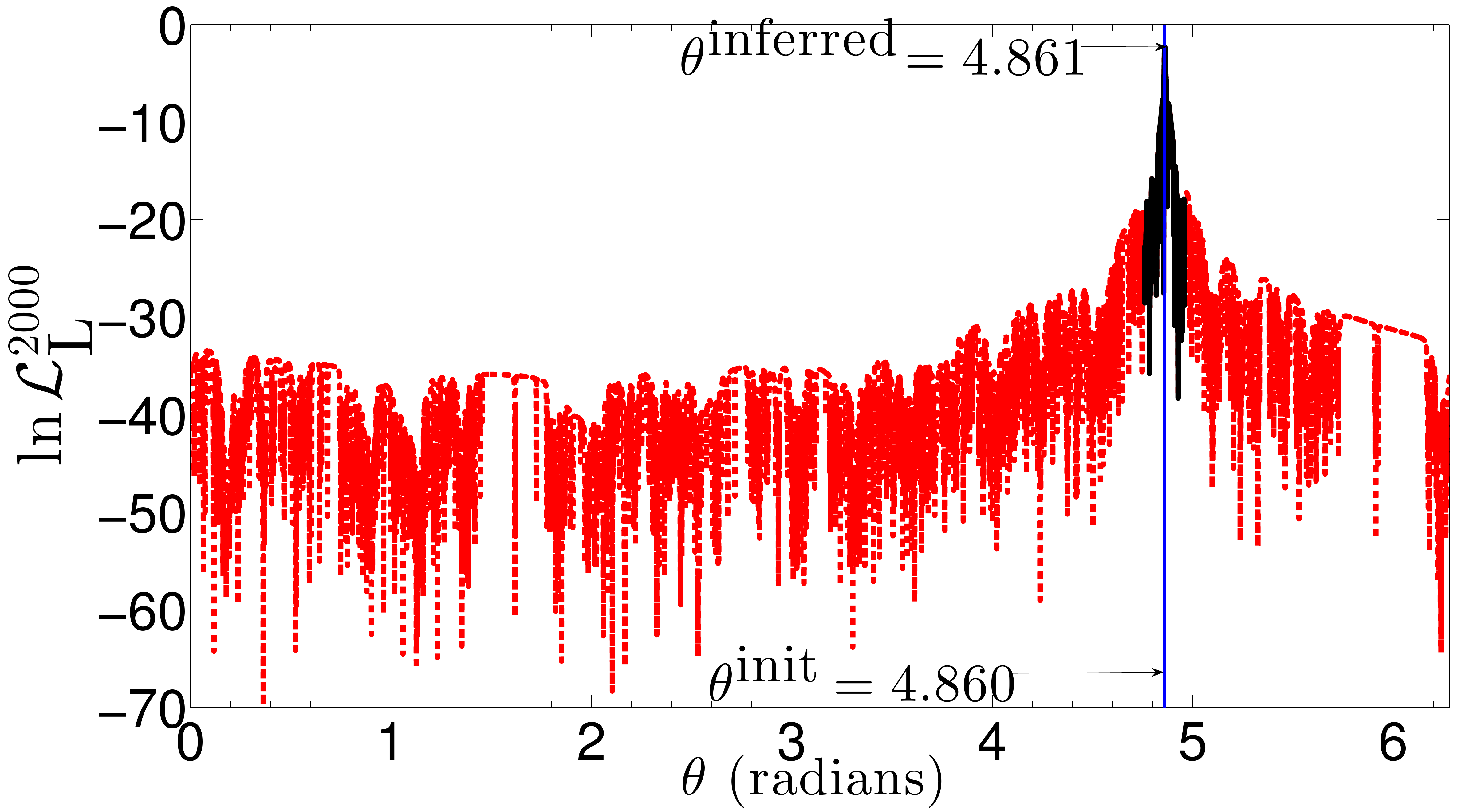}}}
\caption{Likelihood landscapes for different nodes in a synthetic network with $t=5000$ nodes and parameters $m=1.5, L=2.5, \gamma=2.1, T=0.4$. The plots show the log-likelihoods  $\ln{\mathcal{L}^{i}_{\textnormal{L}}}$, $i=600, 1000, 2000$, with the original version of the method that samples the likelihood over the whole $[0, 2\pi]$ domain (dashed red line), and with its fast version that samples the likelihood only over the $\theta$-region shown by the solid black line. The vertical line in each plot shows the initial estimate for the angle, $\theta^{\textnormal{init}}$, while $\theta^{\textnormal{inferred}}$ is the final inferred angle.
\label{fig:likelihood_landscapes_fast}}
\end{figure*}

\begin{figure*}
\centerline{
\subfigure[~$T=0.05$, hybrid with correction steps.]{\includegraphics[width=2.1in, height=1.5in]{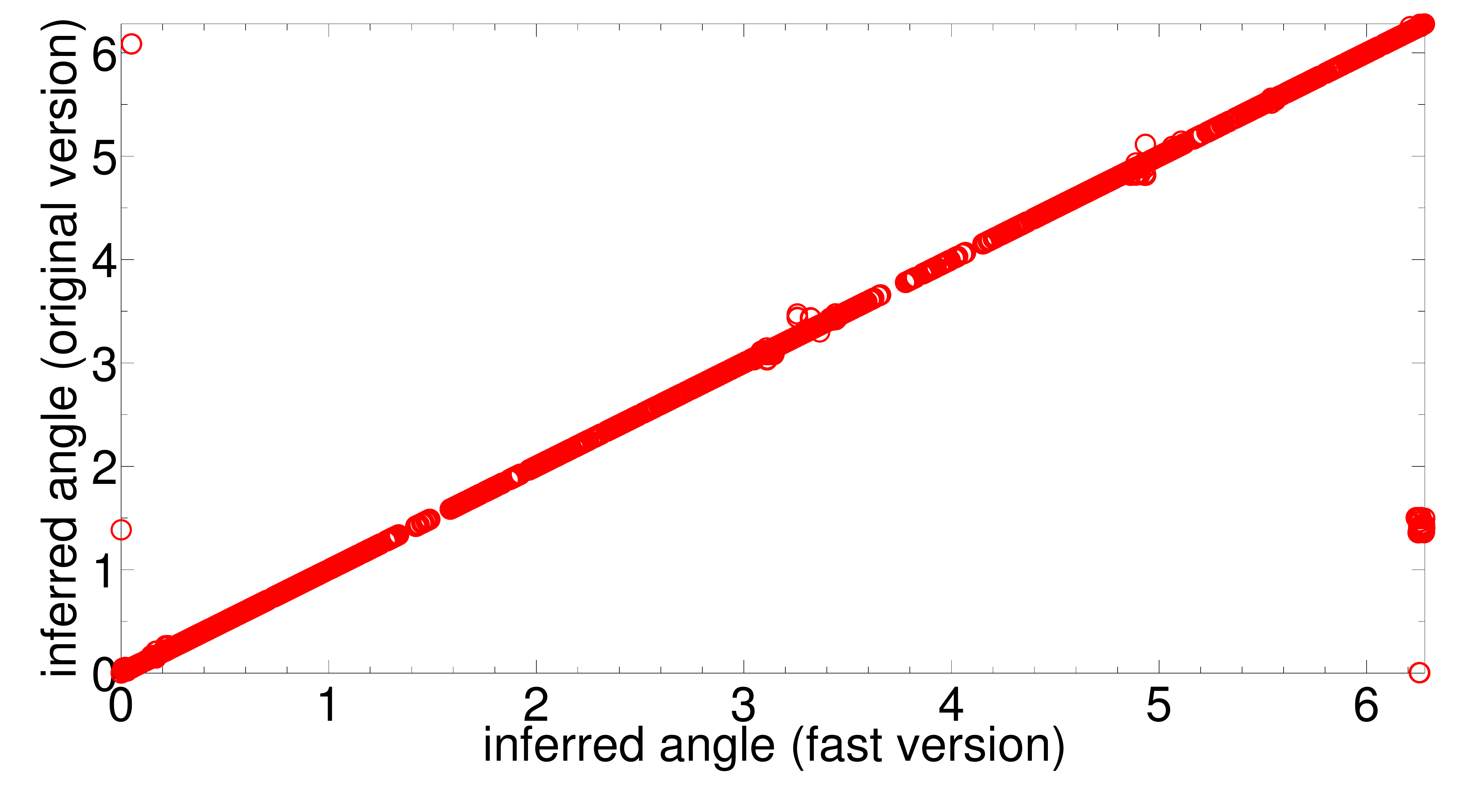}}
\subfigure[~$T=0.4$, hybrid with correction steps.]{\includegraphics[width=2.1in, height=1.5in]{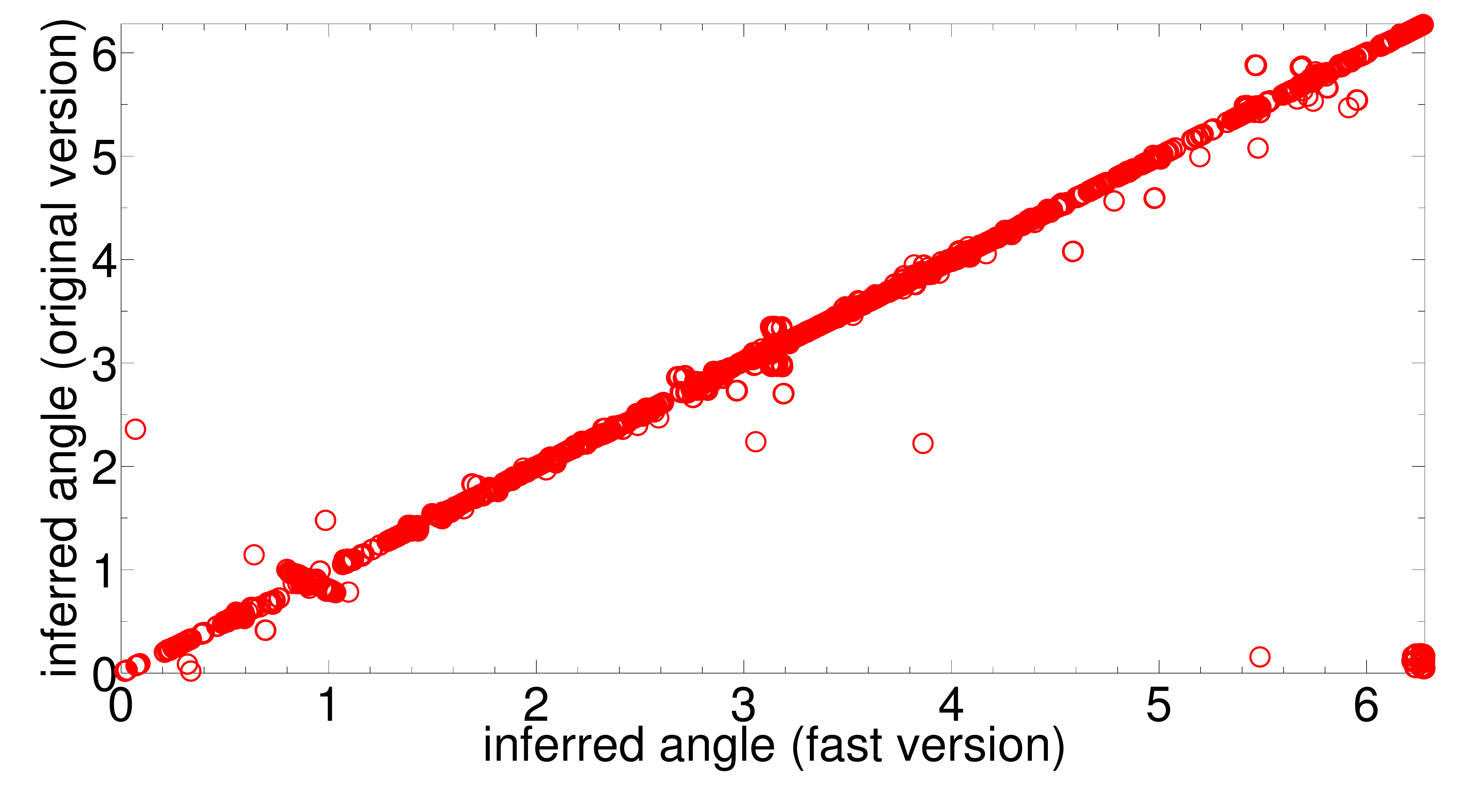}}
\subfigure[~$T=0.7$, hybrid with correction steps.]{\includegraphics[width=2.1in, height=1.5in]{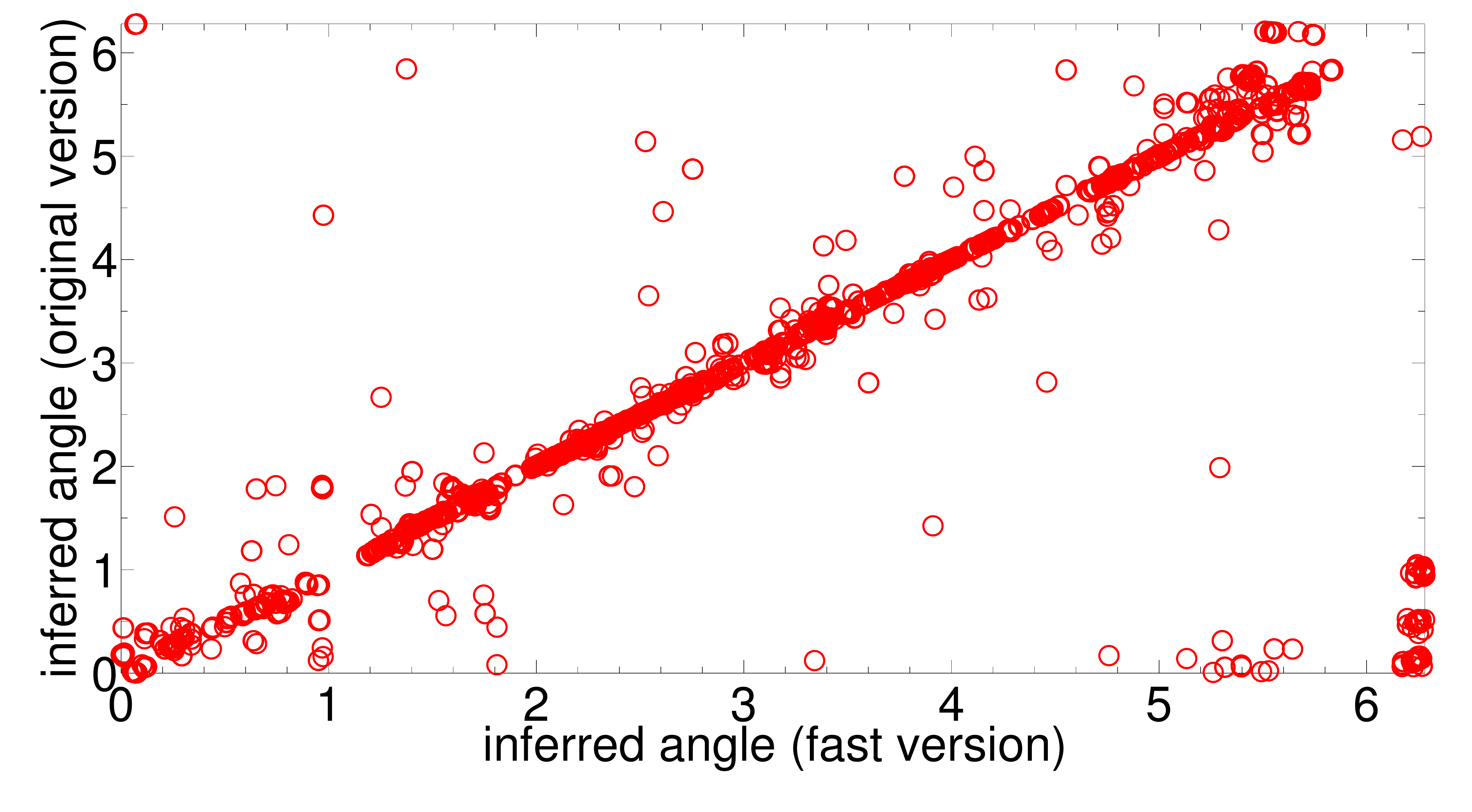}}
}
\caption{Inferred angles (in radians) with the original and fast versions of the hybrid method for synthetic networks with $t=5000$ nodes, $m=1.5, L=2.5, \gamma=2.1$, and $T$ as shown in the captions.
\label{fig:inferred_vs_real_fast}}
\end{figure*}

\textbf{Fast methods.} Finally, we present results with the speedup heuristic described in Section~\ref{sec:embedding_speedup}, where we set constant $C=200$ (Section~\ref{sec:embedding_speedup}). We consider the hybrid and link-based methods with correction steps as before, and we run the speedup heuristic for all nodes with degrees $k < k_{\textnormal{speedup}}=10$. We call these versions of the methods \emph{fast versions}. Figure~\ref{fig:likelihood_landscapes_fast} shows likelihood landscapes sampled by the link-based method, and the corresponding regions of the likelihoods sampled by its fast version. We observe that the fast version infers the same angle as the original version, which always corresponds to the maximum of the likelihood. We also observe that the initial estimate of the angle is very close to the final inferred angle, as expected. Figure~\ref{fig:inferred_vs_real_fast} juxtaposes the inferred angles with the original and fast version of the hybrid method for all the network nodes. Similar results hold for the link-based method. From the figure, we observe a very good match for almost all the node angles, especially at lower temperatures. Tables~\ref{tab:logloss_fast} and~\ref{tab:navigation_with_corrections_fast} show the logarithmic losses and the performance of GR, where the results are very similar to those in Tables~\ref{tab:logloss_with_corrections} and~\ref{tab:navigation_with_corrections}.
\begin{table}
\begin{center}
\caption{Logarithmic Losses obtained by the fast version of the methods (with correction steps)
\label{tab:logloss_fast}}
\begin{tabular}{|c|c|c|}
\hline Network &  $LL^{\textnormal{inf}}$, fast hybrid & $LL^{\textnormal{inf}}$, fast link-based\\
\hline $T=0.05$ & $6.2 \times 10^4$ & $4.0\times 10^4$\\
\hline $T=0.4$ & $3.0 \times 10^4$ & $2.9 \times 10^4$\\
\hline $T=0.7$ & $4.2 \times 10^4$ & $4.1 \times 10^4$\\
\hline
\end{tabular}
\end{center}
\end{table}
\begin{table}
\begin{center}
\caption{Success ratio $p_s$ and average hop-length $\bar{h}$ of greedy paths obtained by the fast version of the methods (with correction steps)
\label{tab:navigation_with_corrections_fast}}
\begin{tabular}{|c|c|c|}
\hline Network  & fast hybrid & fast link-based\\
\hline $T=0.05$ & $p_s=0.97, \bar{h}=3.1$ & $p_s=0.98, \bar{h}=3.1$\\
\hline $T=0.4$ & $p_s=0.96, \bar{h}=3.3$ & $p_s=0.97, \bar{h}=3.3$ \\
\hline $T=0.7$ &  $p_s=0.91, \bar{h}=3.7$ & $p_s=0.92, \bar{h}=3.7$ \\
\hline
\end{tabular}
\end{center}
\end{table}

\textbf{Summary of the results.} To summarize, in this section we have validated that: (i) the common-neighbors method is more accurate than the link-based method for nodes appearing at early MLE times; (ii) at larger MLE times, the two methods yield approximately the same results; (iii) the hybrid method performs significantly better from the link-based method if correction steps are not used; (iv) if correction steps are used, then hybrid and link-based methods perform similarly; (v) correction steps can help improve the quality of the obtain mappings in all cases, but their effect on the hybrid method is not as significant as in the link-based method; and (vi) the fast and original versions of the methods perform almost the same. Our results indicate that the best options are the fast versions of either the hybrid or link-based methods with correction steps. However, we note that the correction steps are an ad-hoc and computational intensive heuristic, requiring $O(i^3)$ computations if run at time~$i$. We have observed that these steps are beneficial when run at relatively small times $i$, not exceeding a few hundred nodes~\cite{hypermap_ton}. But being a heuristic, there are no universal guidelines of when exactly they should be invoked on a given real network to be embedded with the best results. Since correction steps do not have a significant effect on the hybrid approach, \emph{the fast hybrid method without correction steps} might be the best option in general in terms of accuracy and computational complexity tradeoffs.

\section{Application to the Internet}
\label{sec:internet}

We now consider the Autonomous Systems (AS) Internet topology extracted from the data collected by the Archipelago active
measurement infrastructure (ARK) developed by CAIDA~\cite{ClHy09}, which is available at~\cite{as_topo_data}. The connections in the topology are not physical but logical, representing AS relationships~\cite{as_topo_data}. Specifically, an AS is a part of the Internet infrastructure administrated by a single company or organization. Pairs of ASes peer to exchange traffic. These peering relationships in the AS graph are represented as links between AS nodes. CAIDA's IPv4 Routed /24 AS Links Dataset~\cite{as_topo_data} provides regular snapshots of AS links derived from ongoing traceroute-based IP-level topology measurements. A detailed description of the measurement process is given in~\cite{as_topo_data}. The AS topology has a stable power law degree distribution with exponent $\gamma=2.1$, average node (AS) degree $\bar{k} \approx 5$ and average clustering $\bar{c} \approx 0.6$. We consider $6$ snapshots of the topology spaced by $3$-month intervals from September 2009 to December 2010. These snapshots consist respectively of $t=24091, 25910, 26307, 26756, 28353, 29333$ ASes.

\begin{figure*}
\centerline{
\subfigure[]{\includegraphics[width=2.2in, height=1.5in]{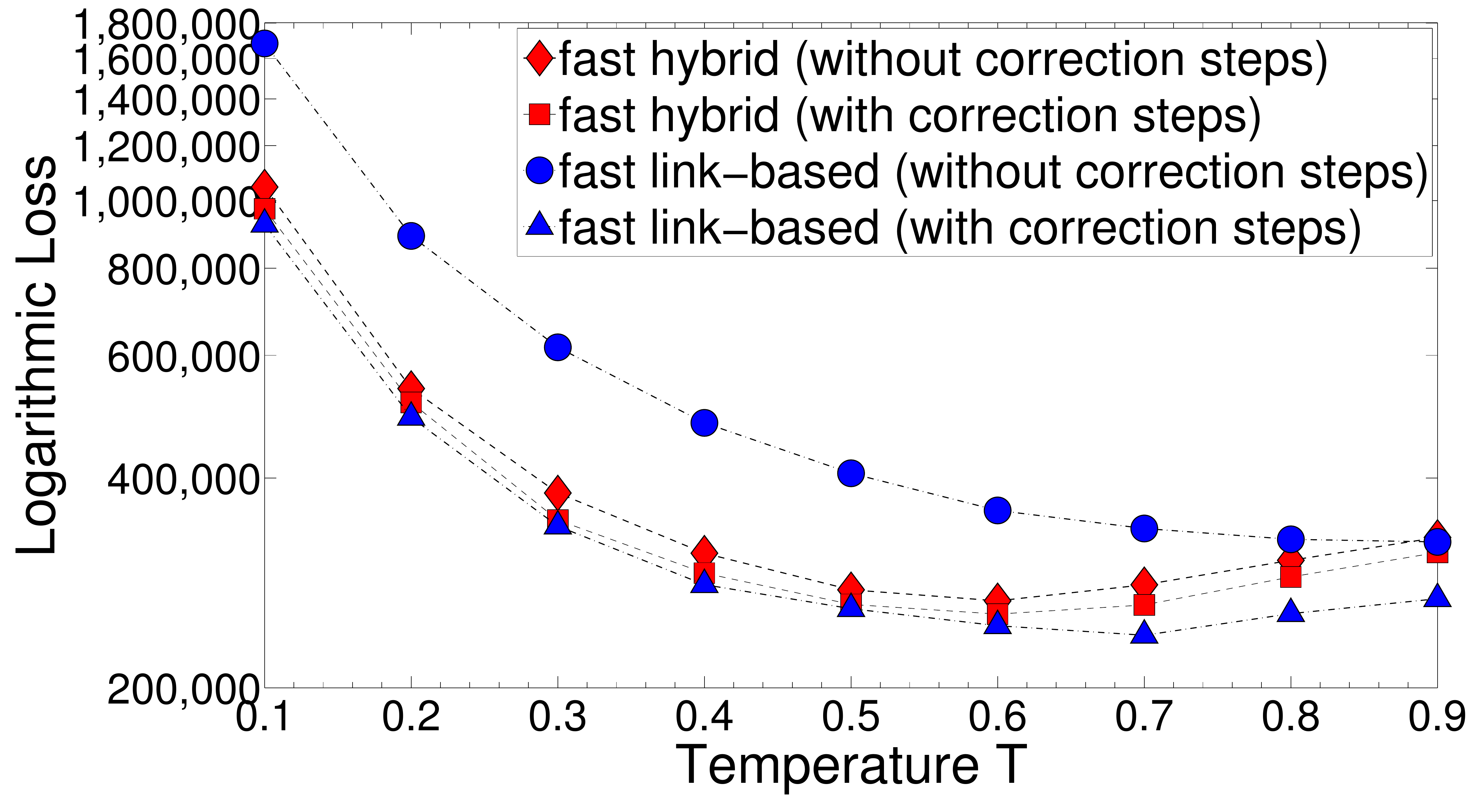}}
\subfigure[]{\includegraphics[width=2.2in, height=1.5in]{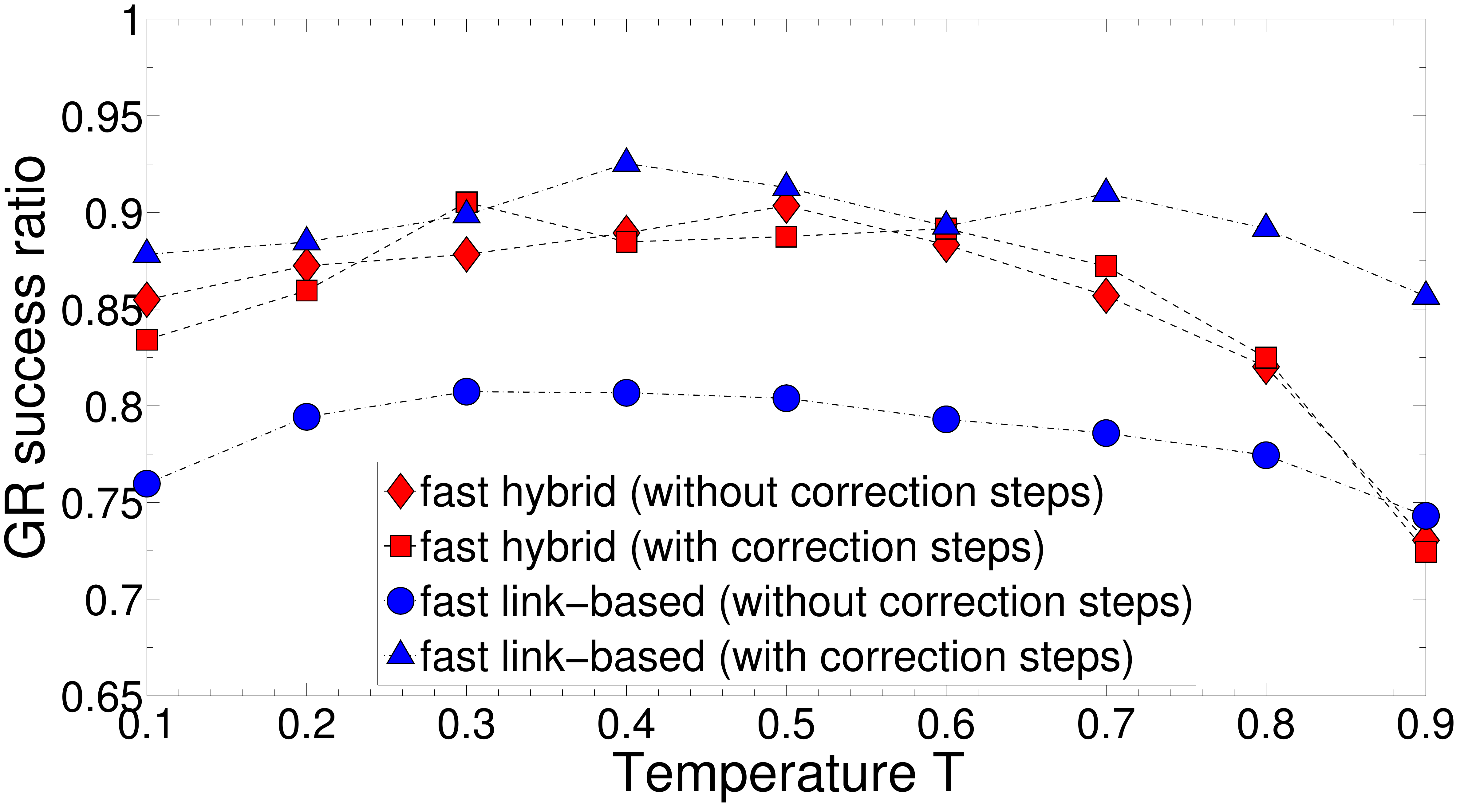}}
\subfigure[]{\includegraphics[width=2.2in, height=1.55in]{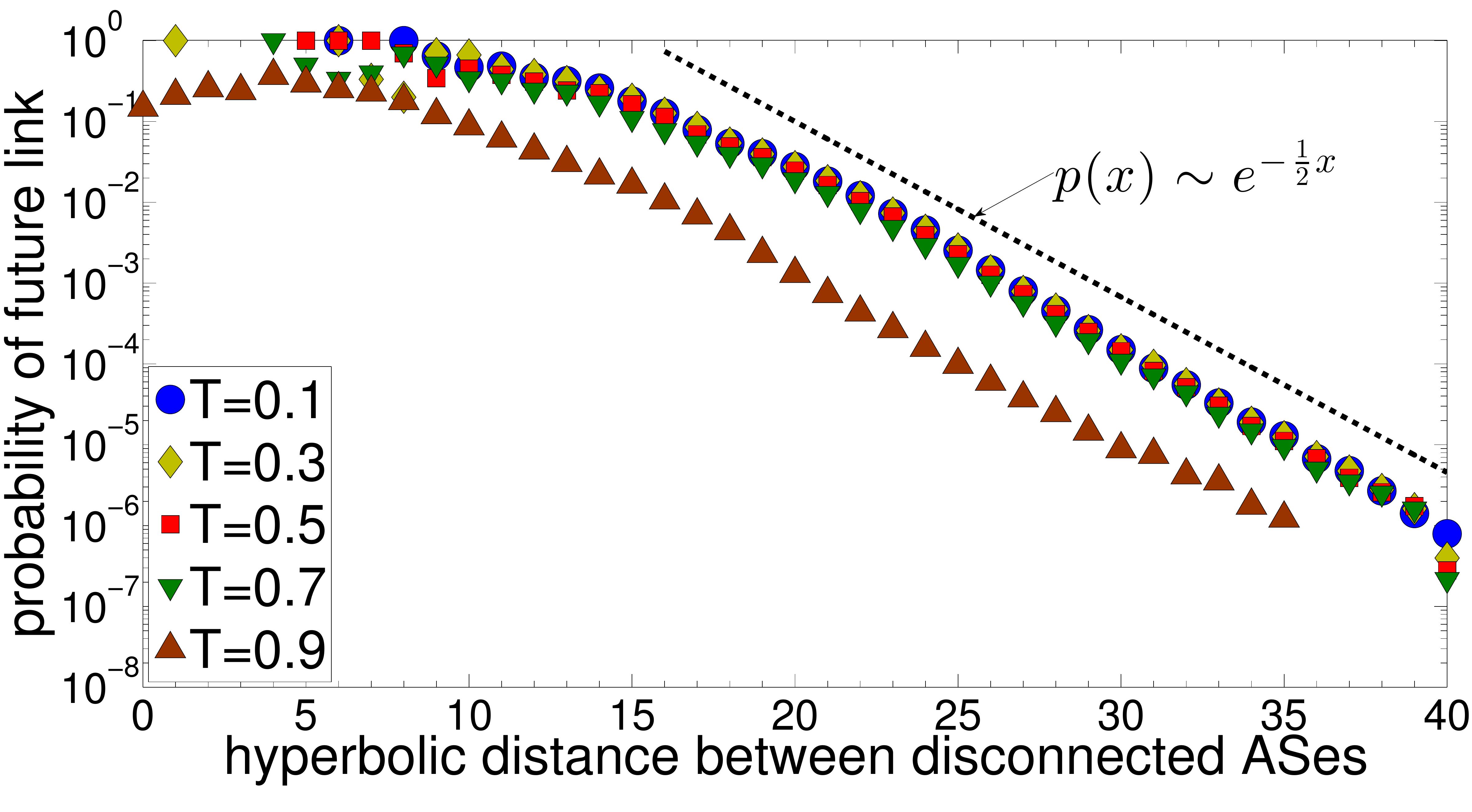}}
}
\caption{Logarithmic Loss ($LL^{\textnormal{inf}}$), GR success ratio ($p_s$), and future-link probability in a mapped snapshot of the AS Internet (Sept.~2009 snapshot). In plots (a) and (b) the results are obtained by the fast hybrid and link-based methods, with and without correction steps. The results in plot (c) are obtained by the fast hybrid method with correction steps. In all cases $k_{\textnormal{speedup}} = 3$, and the results are shown for different values of the temperature parameter $T$.
\label{fig:metrics}}
\end{figure*}

\textbf{Logarithmic Loss and greedy routing efficiency.} In Fig.~\ref{fig:metrics} we mapped the Sept.~2009 snapshot using the fast hybrid and link-based methods with and without correction steps. The correction steps were applied as described in the previous section. In all cases we used the estimated $m=1.5$, $L=\frac{\bar{k}-2m}{2}=1$, $\gamma=2.1, \zeta=1$, and different values of $T$ in $[0.1, 0.9]$. The speedup heuristic was applied for all nodes with degrees $k < k_{\textnormal{speedup}}=3$. Figs.~\ref{fig:metrics}(a),(b) show the obtained logarithmic losses and the efficiency of greedy routing (GR) in all cases. We observe that correction steps do not have a significant effect on the hybrid method whose lowest logarithmic loss is obtained at $T=0.6$. This value is close to the value $T=0.45$--$0.5$ required to construct synthetic networks with the same clustering $\bar{c}$ as in the Internet~\cite{hypermap_ton}. The link-based method without correction steps yields significantly higher logarithmic losses than the hybrid method, for almost all temperature values $T$. These losses decrease when correction steps are used and become similar to the ones in the hybrid method. These results agree with our observations in the previous section on synthetic networks, which indicated that the link-based method without correction steps is not as accurate at inferring the angular coordinates of nodes, while correction steps are not as important for the hybrid method, cf. Figs.~\ref{fig:inferred_vs_real_all_no_corrections}, \ref{fig:inferred_vs_real_all_with_corrections} and Tables~\ref{tab:logloss_no_corrections}, \ref{tab:logloss_with_corrections}.

GR is also very efficient. In the hybrid method, with or without correction steps, the success ratios are close to $90\%$ for a wide range of $T$ in $[0.3, 0.6]$. In the link-based method without correction steps the success ratios are smaller, and become similar to the hybrid method's only if correction steps are used. These results agree again with our previous observations on synthetic networks, cf. Tables~\ref{tab:navigation_no_corrections}, \ref{tab:navigation_with_corrections}.

\textbf{Prediction of future links.} Fig.~\ref{fig:metrics}(c) shows the empirical probability that a future link appears between two disconnected ASes as a function of their hyperbolic distance in Sept.~2009.  To compute this probability, we consider all disconnected AS pairs in Sept.~2009 and all \emph{future links} that appear between these pairs in the period Sept.~2009--Dec.~2010 ($48119$ new links). We then bin the range of hyperbolic distances between these pairs from zero to the maximum distance into small bins. For each bin we find all the disconnected pairs located at the hyperbolic distances falling within the bin. The percentage of pairs in this set of pairs that get connected with a future link, is the value of the empirical future-link probability at the bin. From Fig.~\ref{fig:metrics}(c), we observe that this probability decreases with the hyperbolic distance between disconnected ASes, as expected. Furthermore, this decrease is exponential at large distances. We note that the shape of this probability is similar to the connection probability in our model, cf. Fig.~\ref{fig:connection_probability_with_corrections}, but it has a slope that does not depend on $T$; in fact, different values of $T \leq 0.7$ yield very similar results.

To provide a deeper insight on the ability of the fast hybrid and link-based methods to predict future links, we also compute the \emph{Area Under the Receiver Operating Characteristic Curve (AUC)}~\cite{LuZhou11}. The AUC here is defined as the probability that a randomly selected link from the set of our future links is given a better score (i.e., a higher existense likelihood) than a randomly selected nonexistent link, where the ``nonexistent links'' are the disconnected AS pairs in Sept.~2009 that never get connected in Sept.~2009--Dec.~2010. The score $s_{ij}$ between two disconnected ASes $(i, j)$ is the hyperbolic distance $x_{ij}$ between them. The smaller this score, i.e., the smaller the hyperbolic distance between two disconnected ASes, the more likely it is that these two ASes will get connected, cf. Fig.~\ref{fig:metrics}(c). The degree to which the AUC exceeds $0.5$ indicates how much better the method performs than pure chance, while $\textnormal{AUC} = 1$ is the best possible AUC.

\begin{figure*}
\centerline{
\subfigure[]{\includegraphics[width=2.2in, height=1.5in]{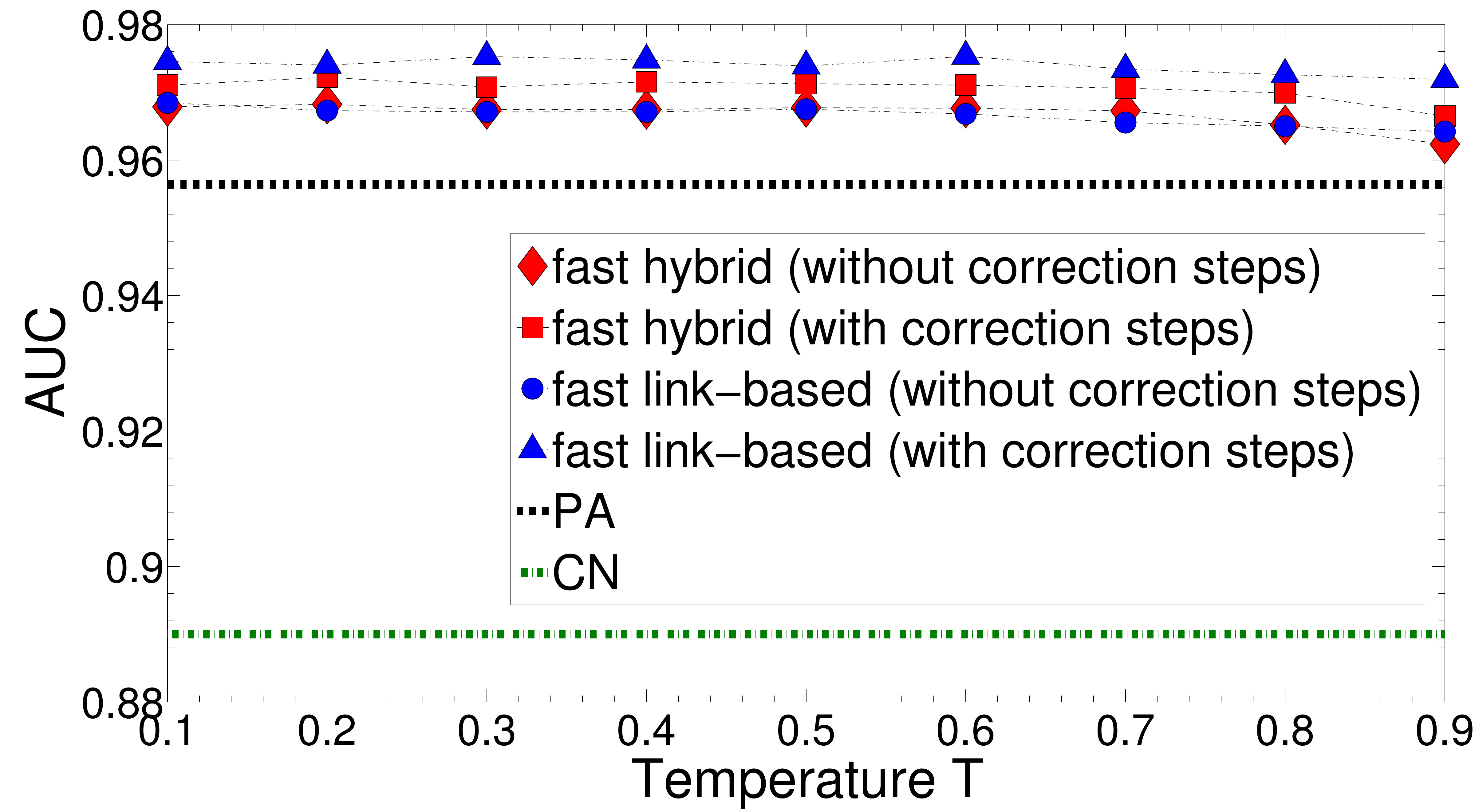}}
\subfigure[]{\includegraphics[width=2.2in, height=1.52in]{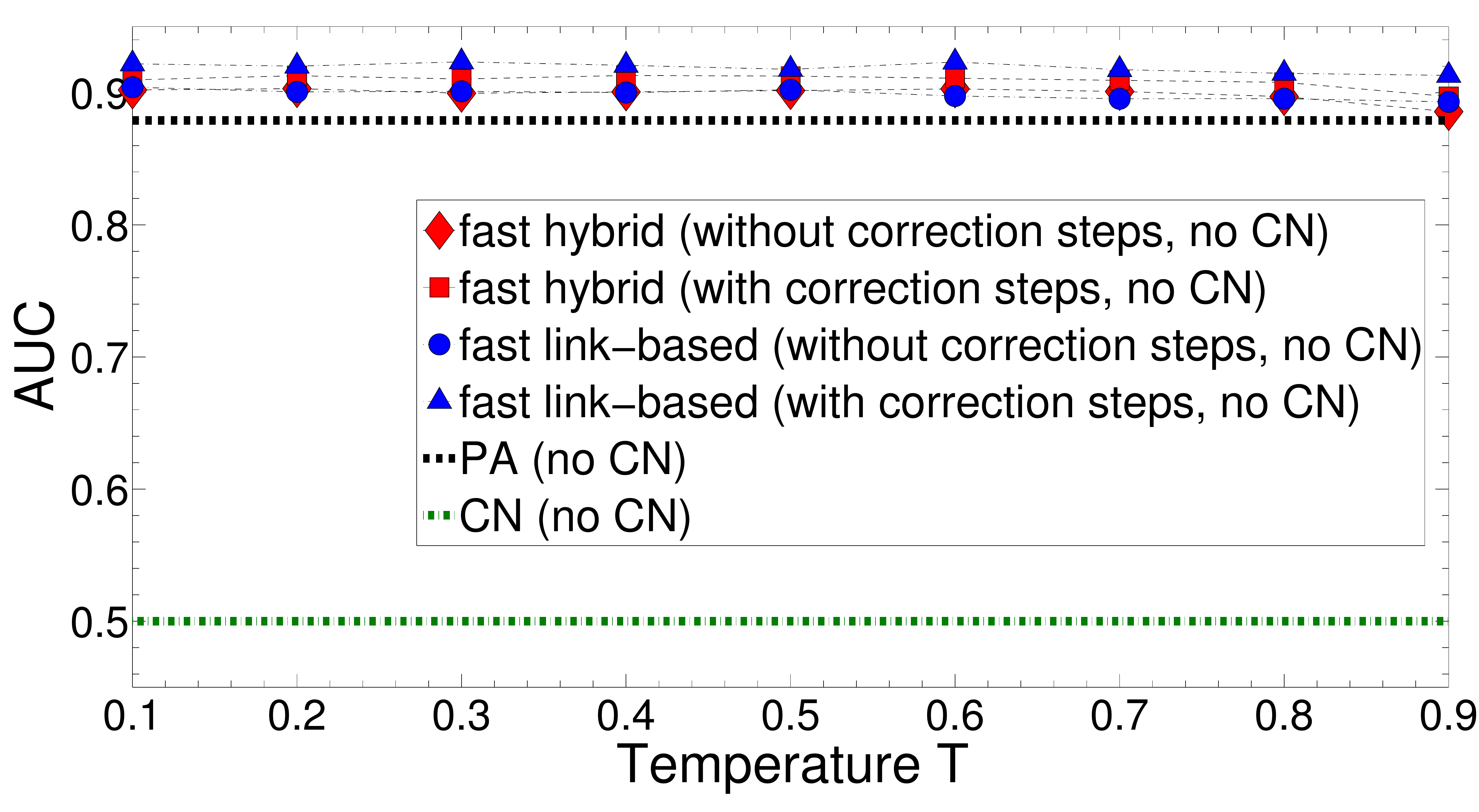}}
\subfigure[]{\includegraphics[width=2.2in, height=1.5in]{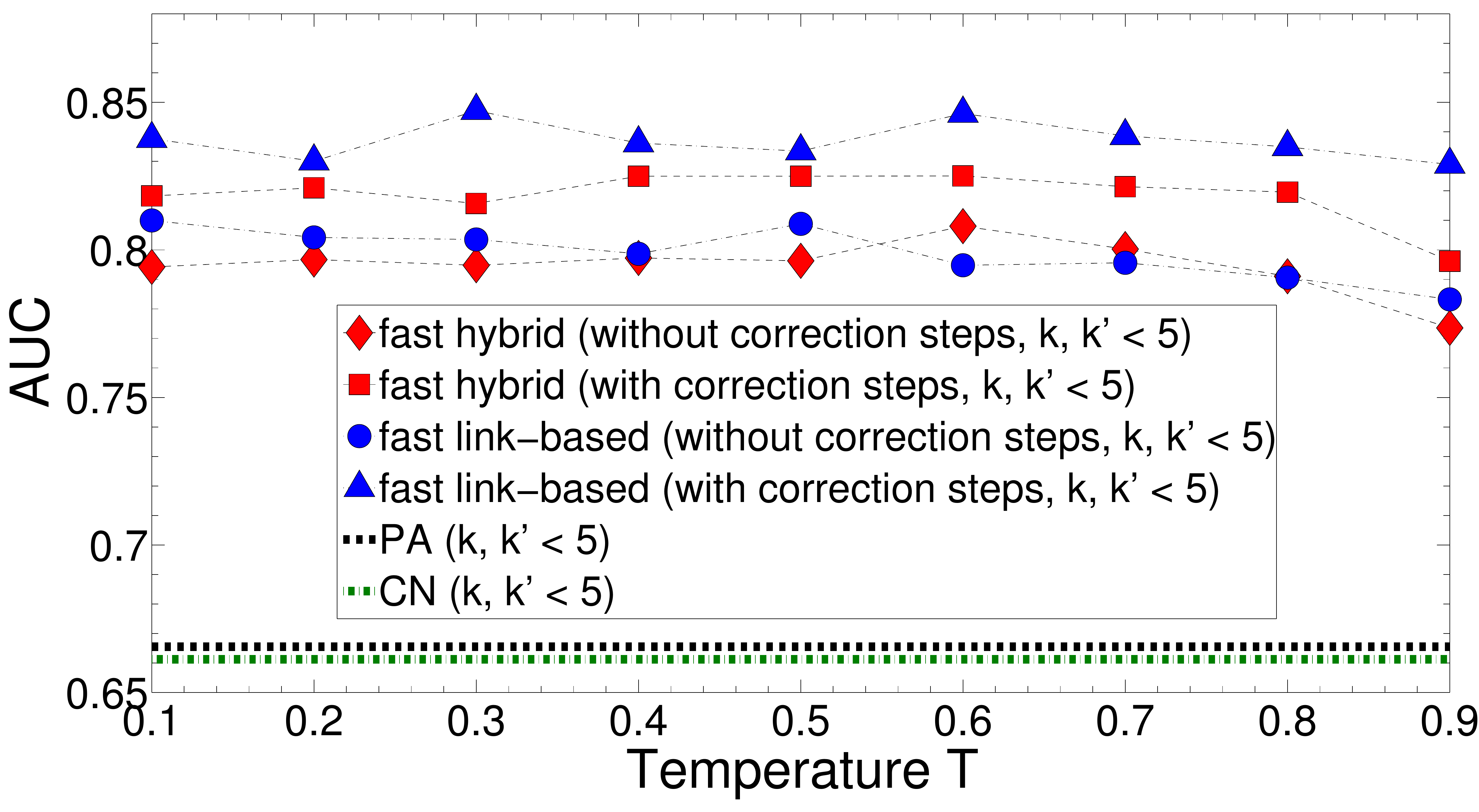}}
}
\caption{Performance of future-link prediction in the AS Internet with the fast hybrid and link-based methods ($k_{\textnormal{speedup}} = 3$), and comparison to the Preferential Attachment (PA) and Common-Neighbors (CN) heuristics. In each case, the AS snapshot of Sept.~2009 is considered. In plot (a), the AUC is computed over all disconnected AS pairs and the new links that appear between them in Sept.~2009--Dec.~2010 ($48119$ new links); in plot (b), the AUC is computed only over the disconnected AS pairs that have no common neighbors ($95\%$ of all disconnected pairs) and the new links between them ($9279$ new links); and in plot (c), the AUC is computed only over the disconnected AS pairs with degrees $k, k' < \bar{k}=5$ ($72\%$ of all disconnected pairs) and the new links between them ($2050$ new links).
\label{fig:aucs}}
\end{figure*}

The results are shown in Fig.~\ref{fig:aucs}(a) for different values of $T$, and are juxtaposed to the results obtained with the Preferential Attachment (PA) and Common-Neighbors (CN) heuristics~\cite{LuZhou11}. In PA, the score between two disconnected ASes $(i, j)$ is $s_{ij}=k_i \times k_j$, where $k_i, k_j$ are the degrees of the ASes, while in CN $s_{ij}=n_{ij}$, where $n_{ij}$ is the number of common neighbors between the ASes. The higher these scores the higher the chance of a future link between the disconnected ASes. From Fig.~\ref{fig:aucs}(a), we observe that the fast hybrid and link-based methods yield very high AUC values, around $0.97$ for almost all $T$, outperforming the PA and CN heuristics. Note that hybrid and link-based methods perform similarly with respect to this performance metric. This is not surprising, since as we have seen in the previous section, the resulting connection probabilities in the two methods are quite similar, cf.~Figs.~\ref{fig:connection_probability_no_corrections},\ref{fig:connection_probability_with_corrections}. In particular, even though the link-based method without correction steps is not as accurate at inferring the real angular coordinates of nodes, cf. Figs.~\ref{fig:inferred_vs_real_all_no_corrections}(d-f), its resulting connection probabilities are close to the ones obtained by the hybrid method, cf. Fig.~\ref{fig:connection_probability_no_corrections}. That is, these results also agree with our previous observations on synthetic networks. In Fig.~\ref{fig:aucs}(b), we compute the AUC by considering only disconnected AS pairs with no common neighbors and the future links among these pairs. In this case, CN performs as good as pure chance since it assigns a zero score to all the pairs, while the fast hybrid and link-based methods still perform remarkably well, with $\textnormal{AUC}$ values between $0.89$ and $0.92$. Finally, in Fig.~\ref{fig:aucs}(c), we compute the AUC by considering only disconnected AS pairs with low degrees, less than the average degree $\bar{k}=5$, and the future links among these pairs. The figure shows that the methods still perform very well, with $\textnormal{AUC}$ values between $0.79$ and $0.85$ for $T\leq0.8$, significantly outperforming the PA and CN heuristics.

To summarize, our results indicate that our methods have a very strong predictive power. Specifically, they perform remarkably well not only in predicting the ``easy-to-predict" future links, i.e., the links that appear among nodes with high degrees or many common neighbors, but also in predicting the ``hard-to-predict" future links, i.e., the links that appear among nodes with low degrees or no common neighbors.  In that sense one can say that the measure of proximity (hyperbolic distances) between nodes in our approach reflects reality more accurately than the PA and CN approaches do, and that our methods can infer these distances in the real Internet quite accurately. The predictive power of our methods is not very sensitive to the exact value of $T$, with the best results obtained for $T \leq 0.8$, cf. Fig.~\ref{fig:aucs}.

\begin{figure*}
\centerline{
\subfigure[~September 2009]{\includegraphics[width=2.3in, height=1.3in]{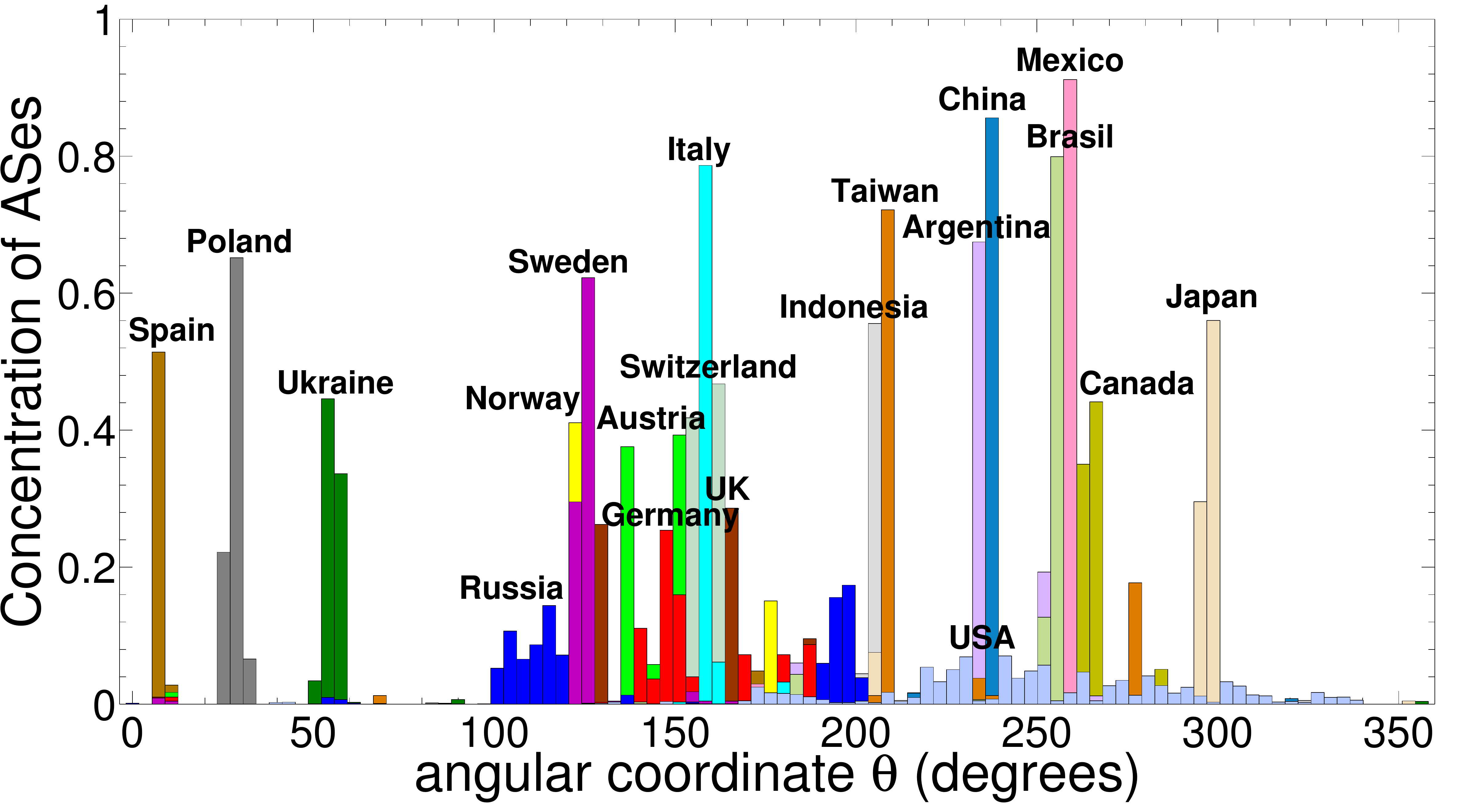}}
\subfigure[~December 2009]{\includegraphics[width=2.3in, height=1.3in]{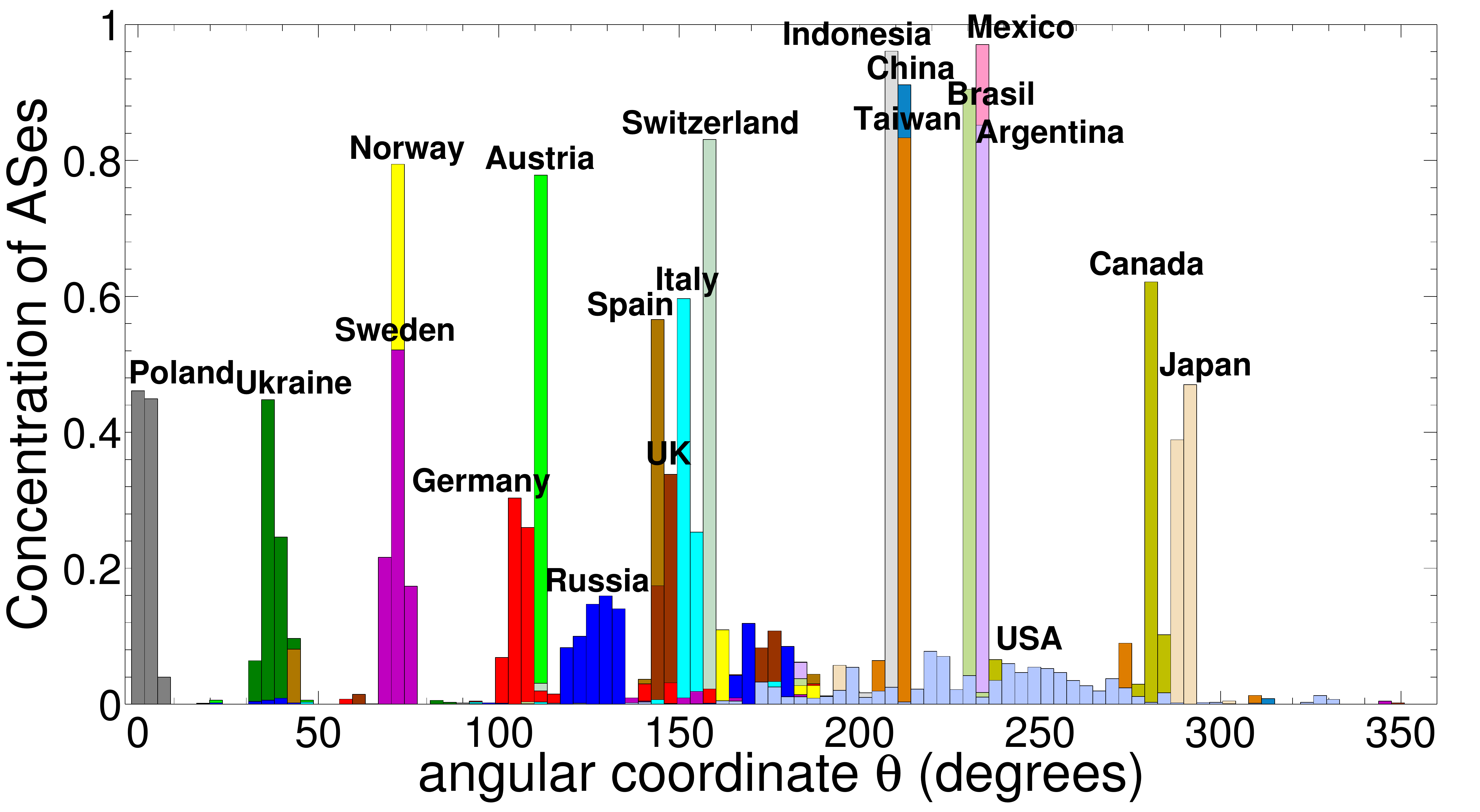}}
\subfigure[~March 2010]{\includegraphics[width=2.3in, height=1.3in]{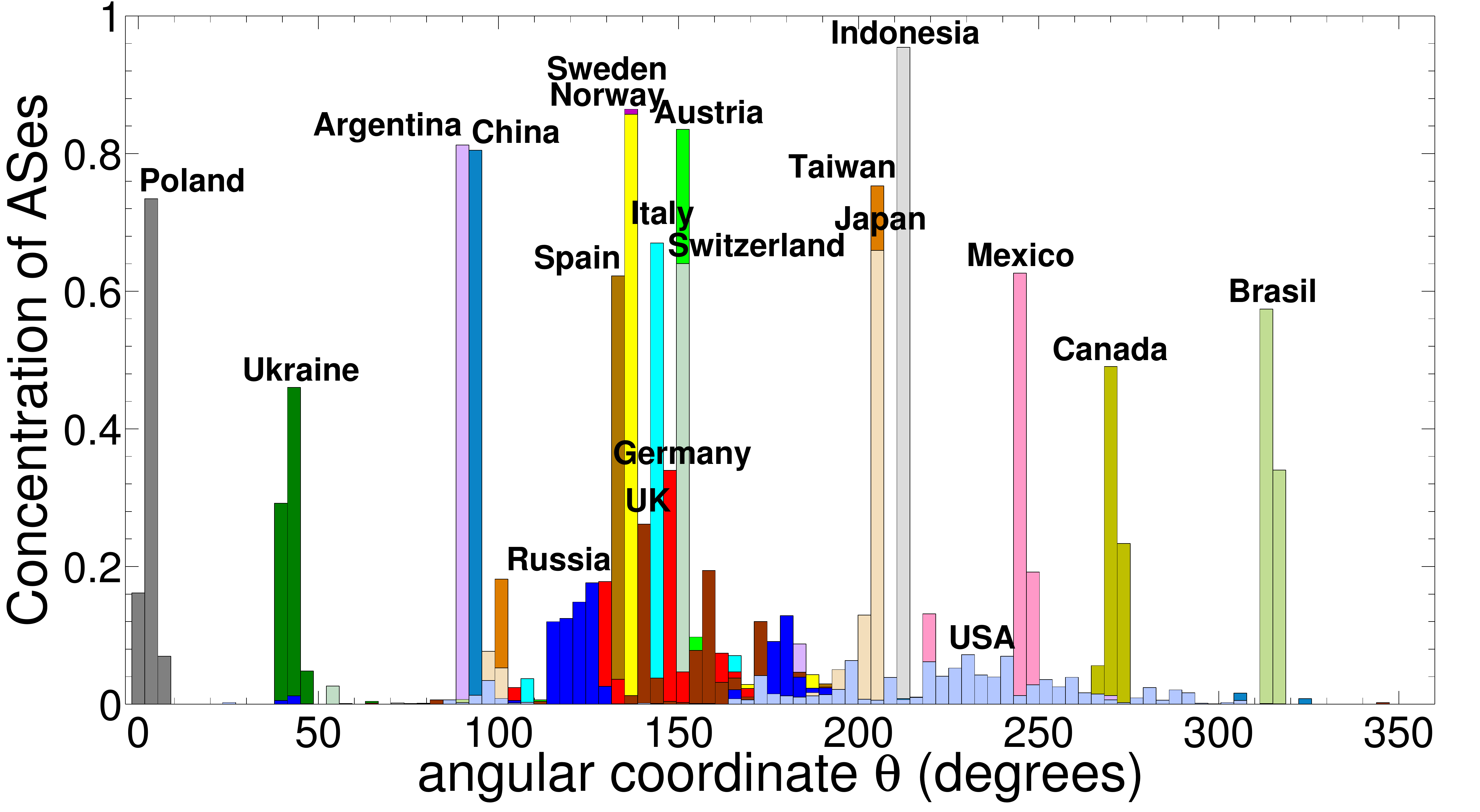}}
}
\centerline{
\subfigure[~June 2010]{\includegraphics[width=2.3in, height=1.3in]{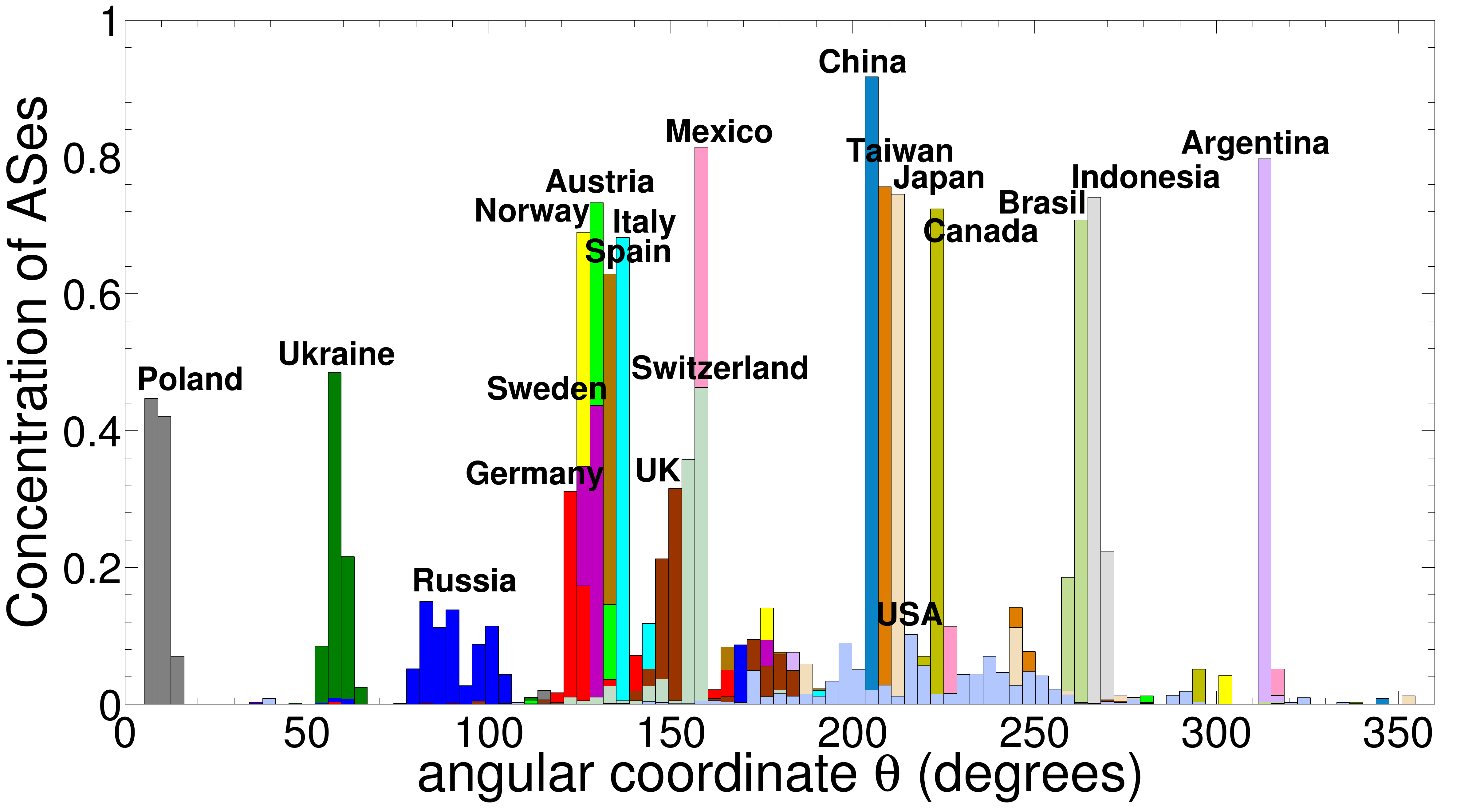}}
\subfigure[~September 2010]{\includegraphics[width=2.3in, height=1.3in]{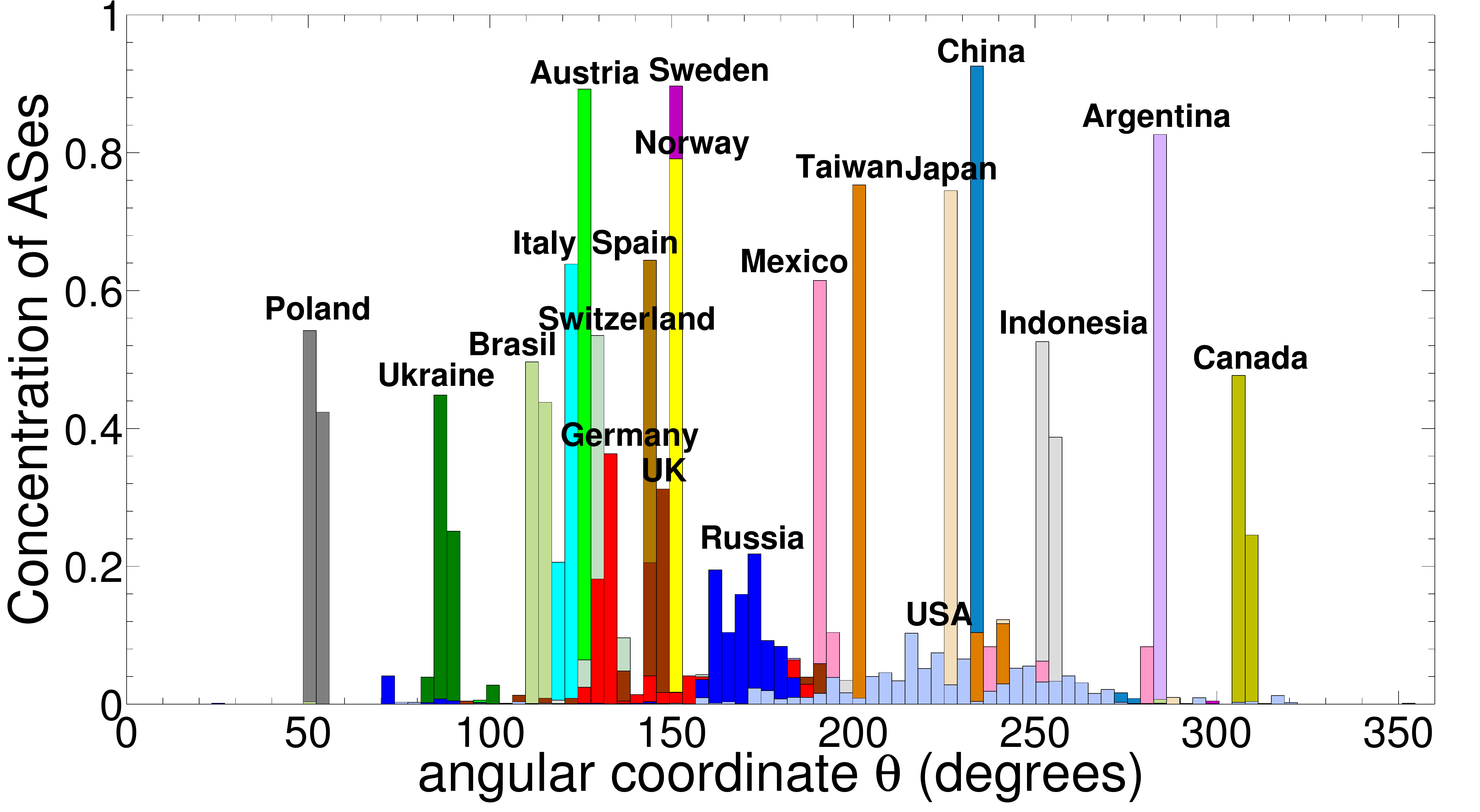}}
\subfigure[~December 2010]{\includegraphics[width=2.3in, height=1.3in]{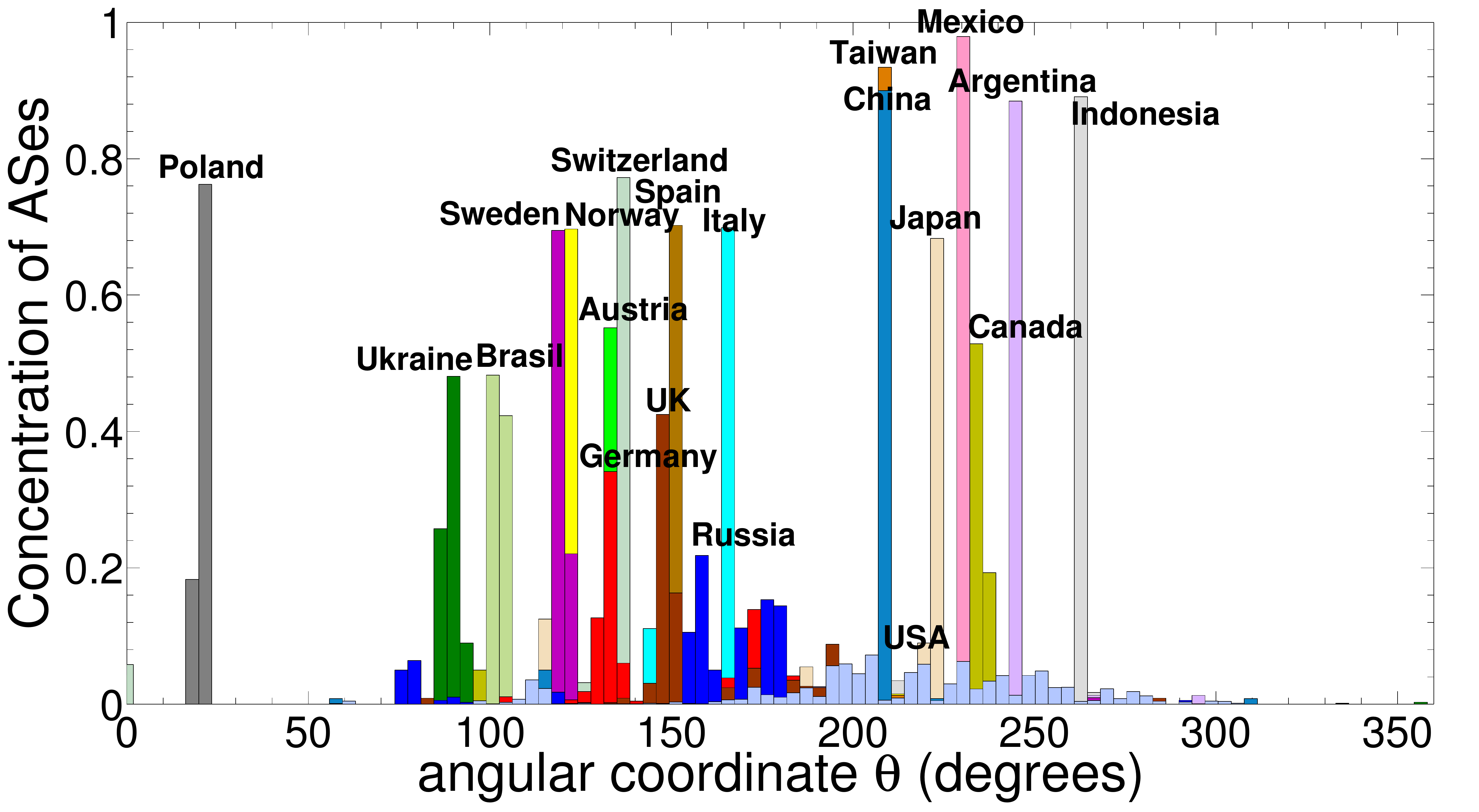}}
}
\centerline{
\subfigure[]{\includegraphics[width=2.25in, height=1.4in]{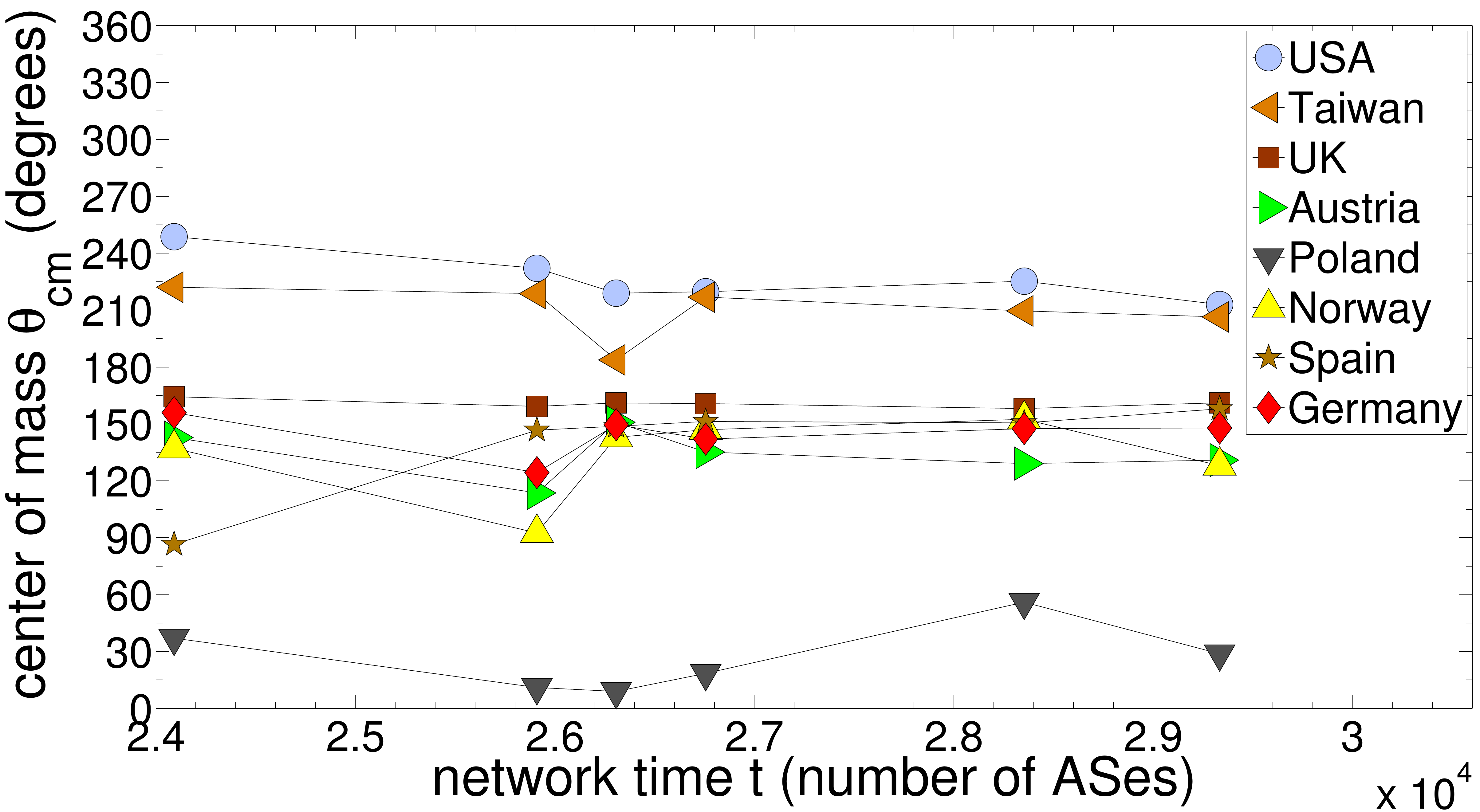}}
\subfigure[]{\includegraphics[width=2.25in, height=1.4in]{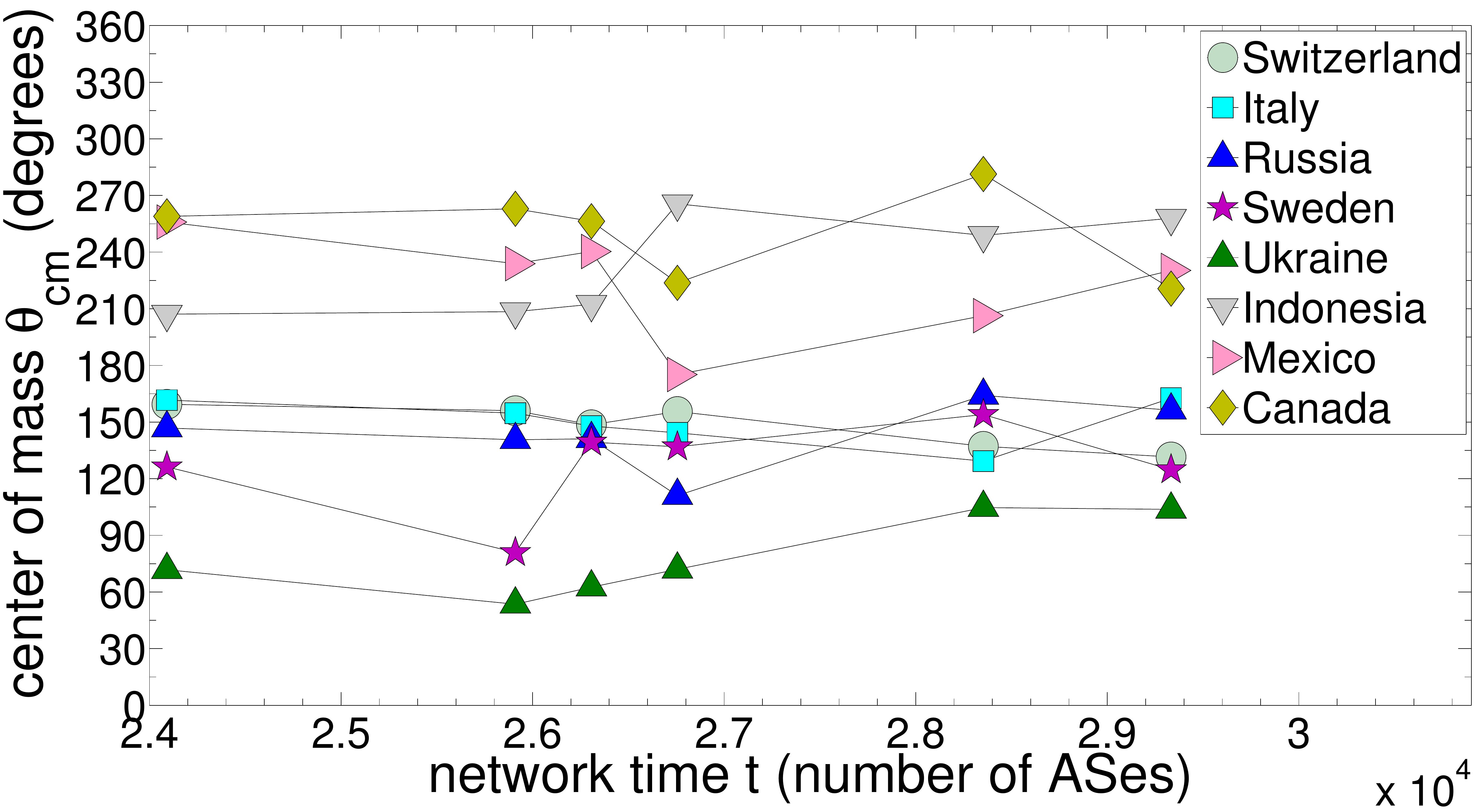}}
\subfigure[]{\includegraphics[width=2.25in, height=1.4in]{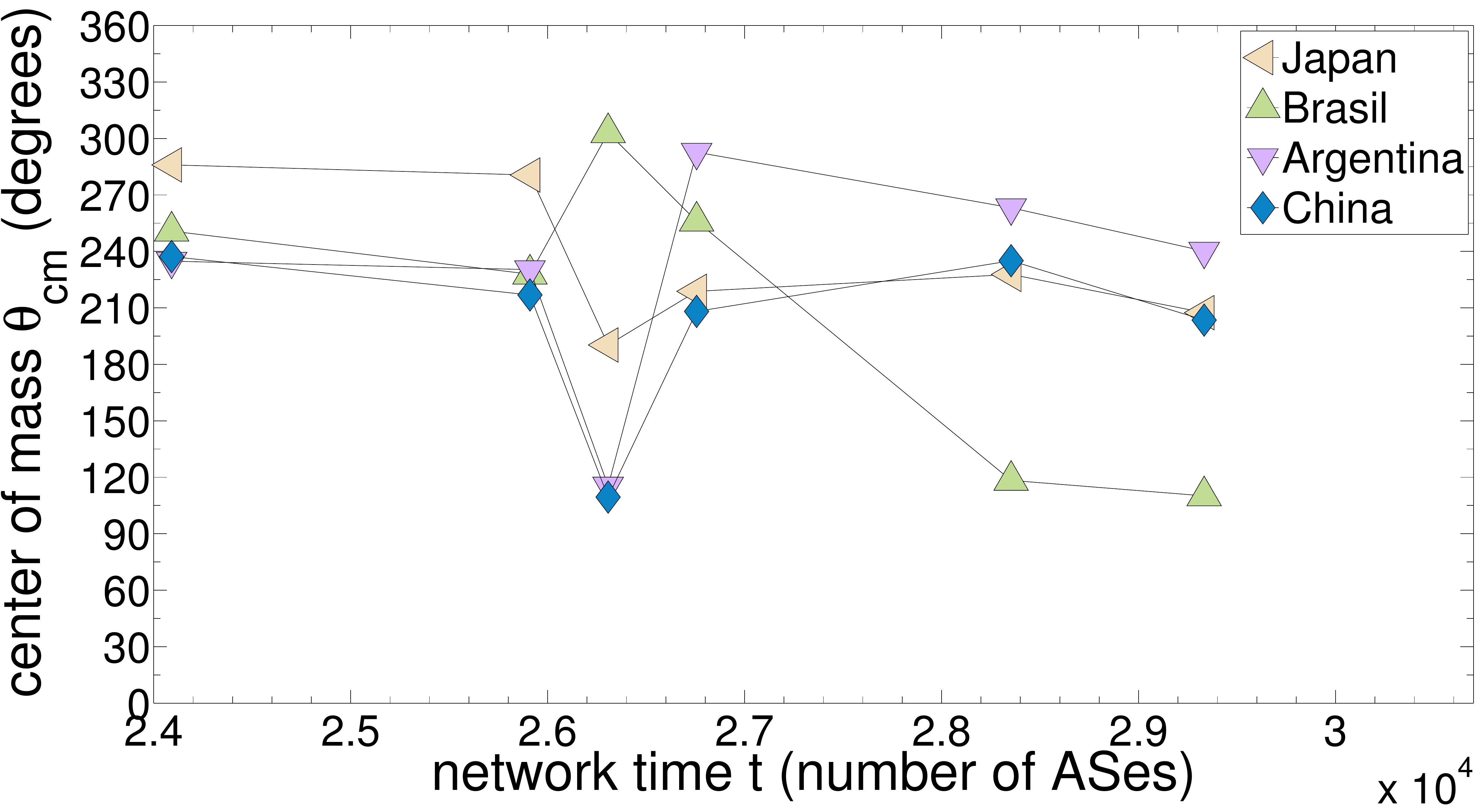}}
}
\caption{Distributions of angular coordinates of ASes belonging to the same country during September~2009--December~2010 (a-f), and the evolution of the angular center of masses of the corresponding communities over time (g-i). For each snapshot in (a-f), the angular center of mass of each country is $\theta_{cm}=(1/n)\sum_{b}\theta(b)n(b)$, where $n$ is the number of ASes belonging to the country, $n(b)$ is the number of such ASes falling within bin $b$, $\theta(b)$ is the value of $\theta$ in the bin, and the summation is over all the bins. For each country, $\theta_{cm}$ is shown (g-i) as a function of the network time $t$, i.e., as a function of the number of ASes in the snapshots (a-f), $t=24091, 25910, 26307, 26756, 28353, 29333$, respectively.
\label{fig:as_angle_evolution}}
\end{figure*}

\textbf{Evolution of soft AS communities.} In Fig.~\ref{fig:as_angle_evolution} we map our $6$ AS snapshots, using the fast hybrid method with correction steps as before, with $T=0.6$ that yielded the lowest logarithmic loss and $k_{\textnormal{speedup}} = 3$. In all cases, the angle $\theta_1$ of node $i=1$ (see step 3 of Fig.~\ref{fig:the_method}) is fixed to $\theta_1=\pi$. Figs.~\ref{fig:as_angle_evolution}(a-f) show that the method produces meaningful mappings, in the sense that the method infers soft communities of ASes belonging to the same country, where by soft communities we mean groups of nodes located close to each other in the space. For each mapped snapshot, Fig.~\ref{fig:as_angle_evolution} shows the angular distribution of ASes belonging to the same country for $20$ different countries. For comparison among the distributions, for each snapshot after Sept.~2009 we consider only the ASes that were also present in Sept.~2009. The $x$-axis in Figs.~\ref{fig:as_angle_evolution}(a-f) (angular coordinate) uses bins of size $3.6^{o}$. The AS-to-country mapping is taken from the CAIDA AS ranking project~\cite{DiKrFo06}. We observe that the fast hybrid method places ASes belonging to the same country close to each other in the angular space. The reason for this is that ASes belonging to the same country tend to connect more densely to each other than to the rest of the world. Connected ASes are attracted to each other, while disconnected ASes repel, and the fast hybrid method feels these attraction/repulsion forces, placing groups of densely connected ASes in narrow regions, close to each other. As expected, due to significant geographic spread in ASes belonging to the US, these ASes are more widespread. We note that other reasons besides geographic proximity may affect the connectivity between ASes, such as economical, political, and performance related reasons. The mapping method does not favor any specific reason but relies only on the connectivity between ASes in order to place the ASes at the right angular (and consequently hyperbolic) distances.

Figs.~\ref{fig:as_angle_evolution}(g-i) also show how the angular center of masses of the considered AS communities evolve in the similarity space during the period Sept.~2009--Dec.~2010. We observe that some communities, e.g., USA and several European countries,  have a more stable position in this space than others, e.g., Argentina and Brazil. The observed dynamics in the similarity space is likely due to a combination of two classes of factors: 1)~stochastic fluctuations and noise coming from the data (our mapping does not introduce any additional randomness since the algorithm is deterministic), and 2)~real dynamics of nodes and communities in the similarity space, caused by new connections and disconnections within and across the communities. Similar results hold for the link-based method with correction steps.

\section{Other related work}
\label{sec:other_work}

A different mapping of the AS Internet to the hyperbolic plane was performed in~\cite{ShaTa08}. The authors found that the hop lengths of the shortest
AS paths in the Internet can be embedded into the hyperbolic plane with low distortion, and that the resulting embedding can be used for efficient overlay
network construction and accurate path distance estimation. Our work is different from~\cite{ShaTa08}, in that hyperbolic distances between ASes in our case
are not directly defined by their ``observable" AS path lengths. Instead, they are defined by ``hidden" popularity and similarity node coordinates that manifest
themselves indirectly via the nodes' connections and disconnections. As indicated by the performance of greedy routing in Section~\ref{sec:validation}, short paths follow well the underlying hyperbolic geodesics in our mappings. However, nodes at short path distances are not always hyperbolically closer than nodes separated by longer paths. For the same reason, our approach differs from multidimensional scaling (MDS) techniques, which try to compute coordinates for points in low dimensional geometric spaces, e.g.,~\cite{curved_learning}, such that the distances between the points in these spaces match as closely as possible some given distances between the points.

Besides~\cite{hypermap_ton,BoPa10}, perhaps the most relevant earlier work is~\cite{sarkar_embedding}. In this work, the authors considered a model of social networks where nodes reside in a latent Euclidean space~\cite{raftery_model}. Nodes that are sufficiently close in this space have higher chances of being connected. Based on this model, the authors presented a combined MDS and maximum likelihood estimation (MLE) procedure for inferring the node coordinates in the latent space. The procedure can take into consideration previously estimated node positions, e.g.,  estimated node positions in a previous closely-spaced network snapshot, and penalize large displacements from these positions, in an attempt to yield more accurate embeddings. The authors applied this procedure to create embeddings for link prediction, and to illustrate how relationships between authors in co-authorship data change over time. The main difference of our work from~\cite{sarkar_embedding} is that in our case the latent  space is not Euclidean but hyperbolic, the latter providing a more accurate reflection of the geometry of real networks~\cite{KrPa10,BoPa10,PaBoKr11}. In contrast to earlier work on latent network geometry inference, here we have departed from the traditional link-based inference methods, and based our inference entirely on a higher-order similarity statistics---the statistics of the number of common neighbors between nodes.

\section{Conclusion}
\label{sec:conclusion}

In summary, we have introduced and explored a new method for inferring node similarity coordinates based on the number of common neighbors between nodes, and have released the software package implementing this network mapping method to public~\cite{hypermap-cn}.
We have shown that this approach is more accurate than the link-based approach~\cite{hypermap_ton}, unless heuristic periodic adjustments (or correction steps) are used. The common-neighbors approach is more computationally intensive, but we have devised a hybrid method that combines the common-neighbors and link-based approaches, and showed how to reduce its running time to $O(t^2)$. The correction steps can be used in this hybrid approach as well, but their effect is not significant. Therefore they can be entirely avoided to reduce running time. We have validated this method on synthetic model networks, and applied it to the evolving AS Internet. Taken altogether, our results advance our understanding of how to efficiently and accurately map real networks to their underlying hyperbolic spaces.

An interesting open problem is whether more computationally efficient but also more sophisticated numerical optimization methods~\cite{optimization_book} can be applied to the latent network geometry inference problem. Such methods may expedite the maximization of the likelihoods $\mathcal{L}_{\textnormal{L}}^i$ and $\mathcal L_{\textnormal{CN}}^{i}$  in Eqs.~(\ref{eq:local_likelihood_links}), (\ref{eq:local_likelihood_CN}), without sacrificing the embedding quality. We note that our ``brute-force" approach of sampling the likelihoods at small $\Delta \theta$ intervals in order to find their global maximum appears currently to be the best option among all other methods that we have investigated. These methods~\cite{optimization_book} tend to work reliably only if the function to maximize is relatively smooth, has only one easily detectable global maximum, or only few local maxima. In contrast, the likelihood profiles we have to deal with, Figs.~\ref{fig:likelihood_landscapes},\ref{fig:likelihood_landscapes_fast}, are very rugged and rough, abundant with sharp local maxima, rendering unusable all the other methods that we have experimented with.
\begin{figure*}
\centerline{
\subfigure[~$T=0.1$.]{\includegraphics[width=1.75in, height=1.3in]{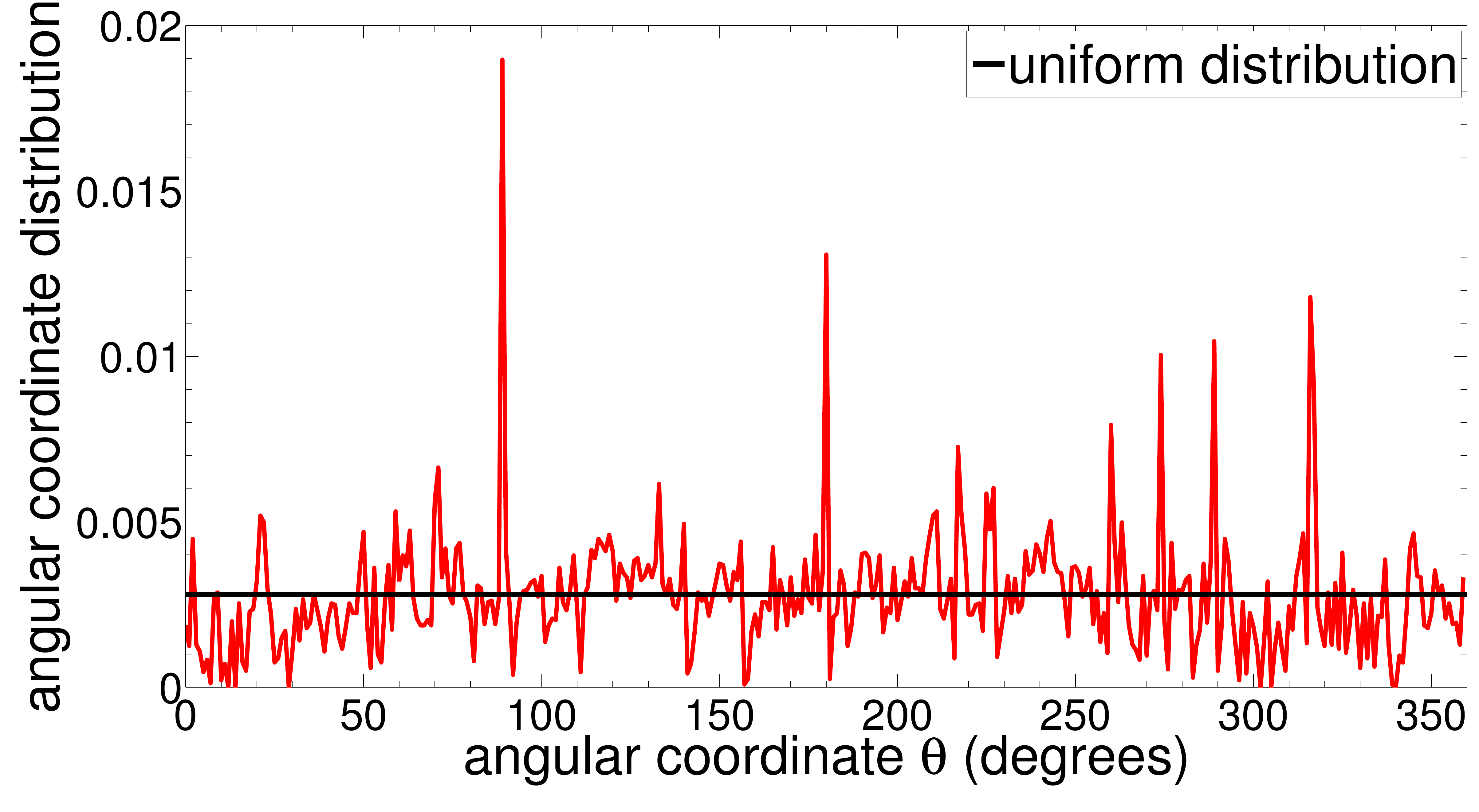}}
\subfigure[~$T=0.3$.]{\includegraphics[width=1.75in, height=1.3in]{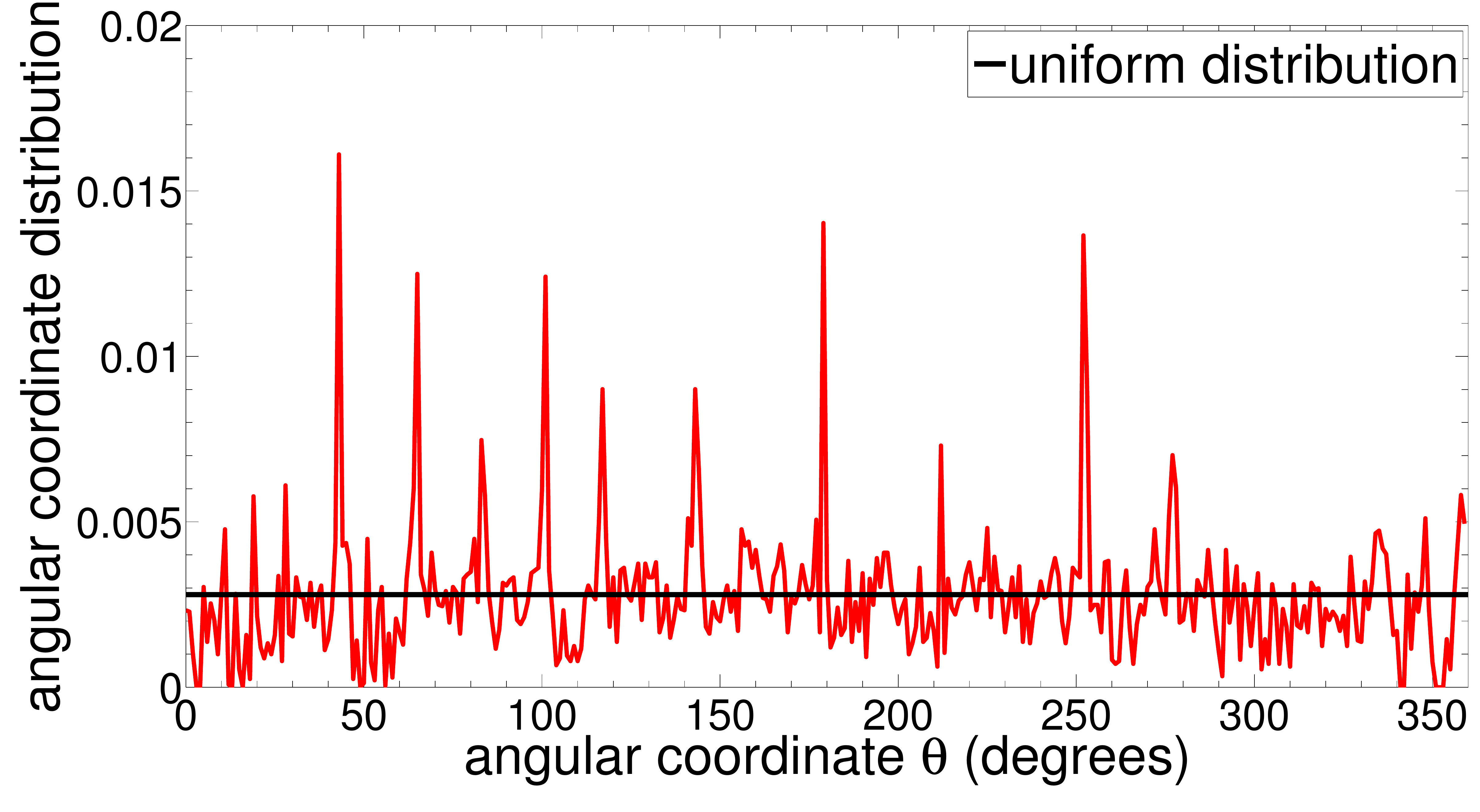}}
\subfigure[~$T=0.6$.]{\includegraphics[width=1.75in, height=1.3in]{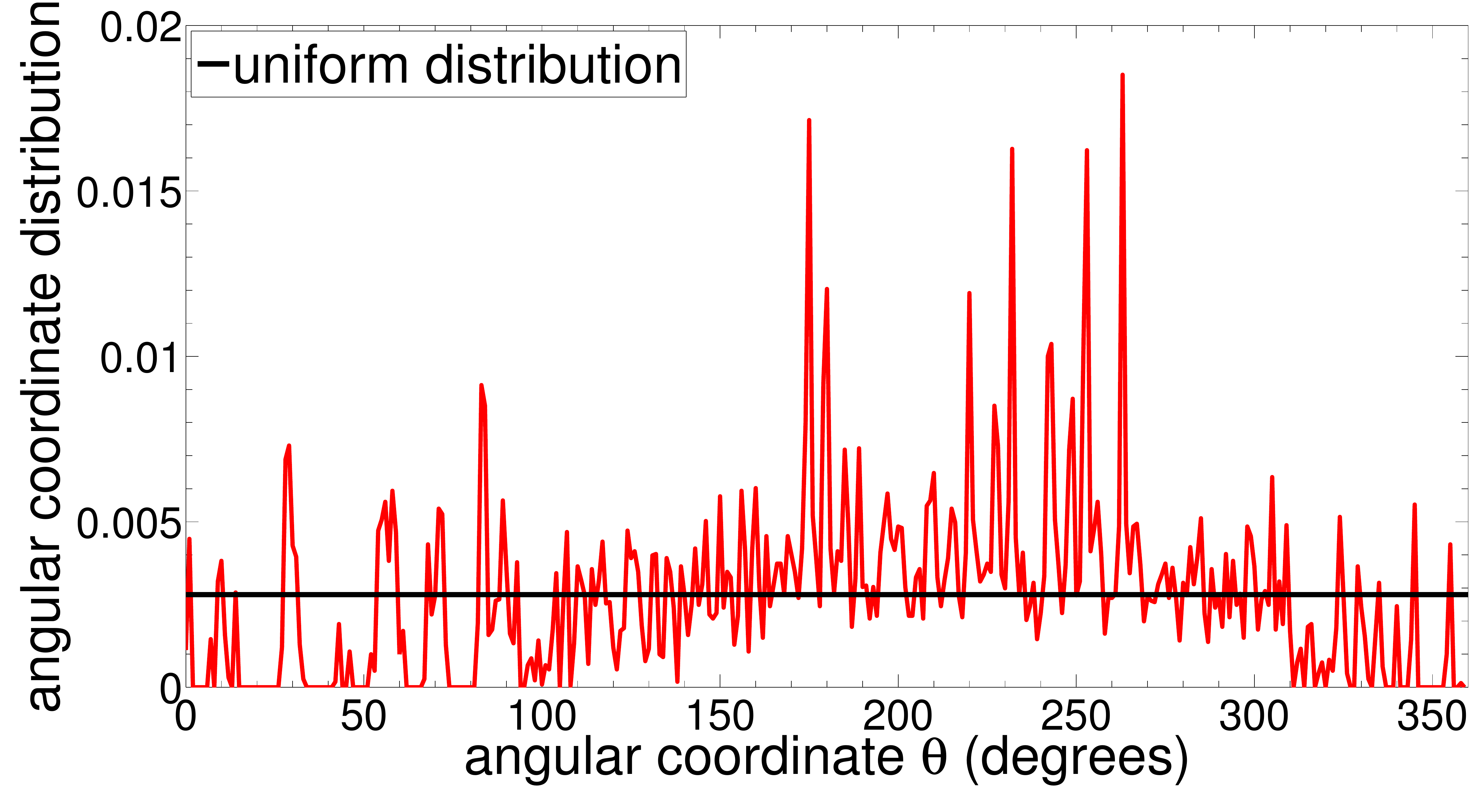}}
\subfigure[~$T=0.8$.]{\includegraphics[width=1.75in, height=1.3in]{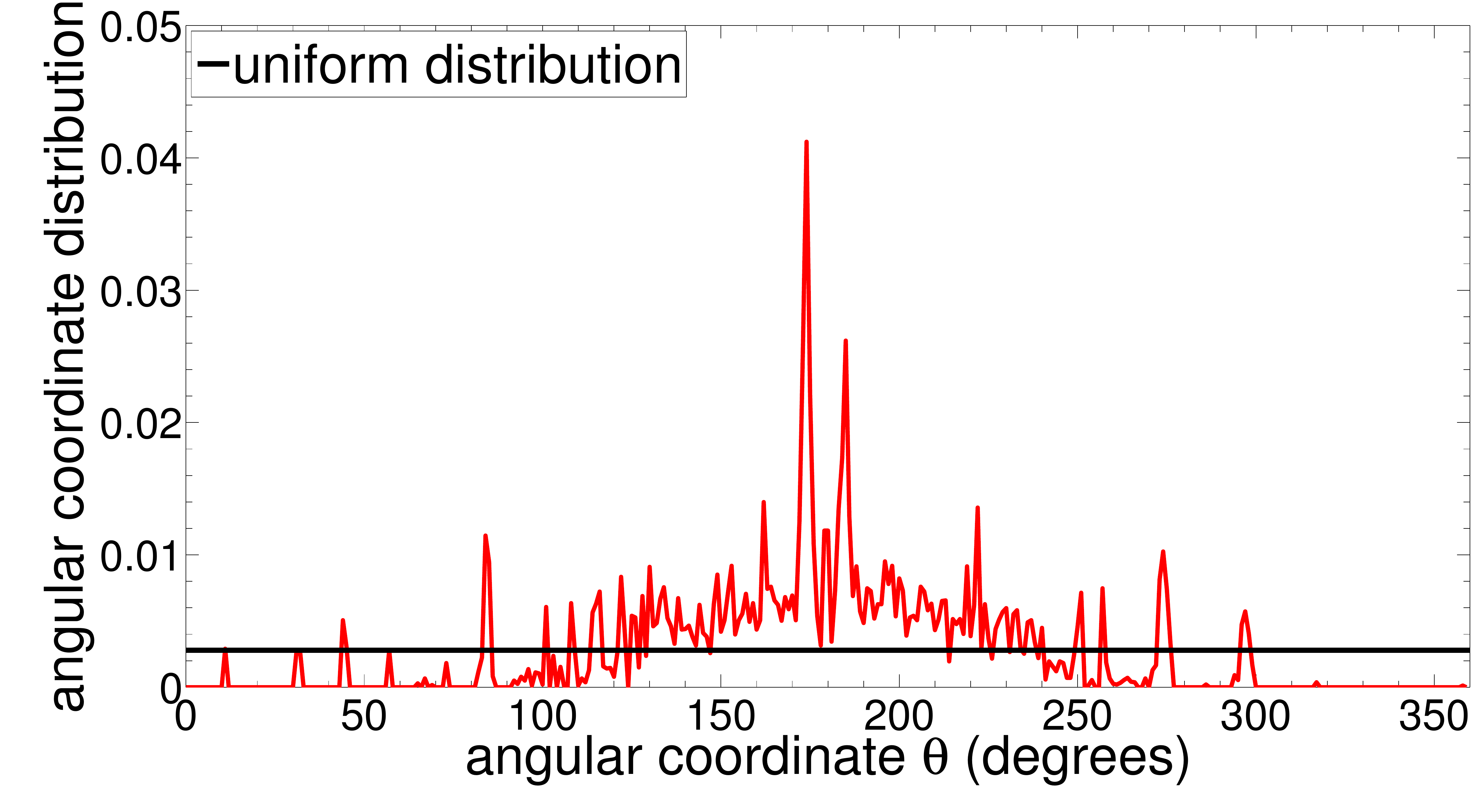}}
}
\caption{Distribution of the inferred AS angles (Sept.~2009 snapshot) with the fast hybrid method ($k_{\textnormal{speedup}} = 3$) and different values of the temperature parameter $T$.
\label{fig:AS_angular_distribution}}
\end{figure*}

All the inference methods presented here and in~\cite{hypermap_ton,BoPa10} use the uniform distribution as the prior~\cite{priors} for the angular distribution of nodes, Eq.~(\ref{eq:local_likelihood_1}). This means that the methods do not make any prior assumption about the node angular positions. Instead, they assume that all positions are equiprobable, and let the given data, i.e., the given network adjacency matrix, to determine the positions. The distributions of the inferred angular coordinates can then be non-uniform in mappings of real networks produced by these methods, since many real networks tend to have some nontrivial community structure. For example, Figure~\ref{fig:AS_angular_distribution} shows the distribution of the inferred AS angles in September 2009. The lowest logarithmic loss (Fig.~\ref{fig:metrics}(a)) is achieved at $T=0.6$, and the corresponding distribution of angular coordinates is clearly non-uniform. In this context, an interesting open problem is to consider extensions of network geometry models that are capable of explaining the emergence of soft community structure in networks and non-uniform distribution of nodes in the similarity space, e.g., \cite{ZuBo15}, and to develop mapping methods for such models that would use non-uniform priors.

Finally, given an efficient and accurate method to map real complex networks into their underlying hyperbolic spaces, one of the most interesting open problems is to decipher the laws that govern the dynamics of nodes in these spaces, Fig.~\ref{fig:as_angle_evolution}. As real networks are characterized by a hierarchical organization and nontrivial community structure~\cite{newman03c-review, Dorogovtsev10-book}, we expect this dynamics to be also highly nontrivial, but definitely not random. This observation suggests that it might be possible to accurately predict the future positions of nodes in the underlying hyperbolic spaces. The precise knowledge of this spatial dynamics of nodes can then be used to predict \emph{fine-grained} network dynamics, forecasting future connections and disconnections among nodes over different timescales.

\begin{acknowledgments}
We thank M.~Kitsak, M.~Bogu{\~n}{\'a}, and C.~Psomas  for useful discussions and suggestions. This work was supported by a Marie Curie International Reintegration Grant within the 7th European Community Framework Programme, by an AWS in Education grant award, by DARPA grant No.\ HR0011-12-1-0012; NSF grants No.\ CNS-1344289,
CNS-1442999, CNS-0964236, CNS-1441828, CNS-1039646, and CNS-1345286; and by Cisco Systems.
\end{acknowledgments}


\begin{thebibliography}{26}
\expandafter\ifx\csname natexlab\endcsname\relax\def\natexlab#1{#1}\fi
\expandafter\ifx\csname bibnamefont\endcsname\relax
  \def\bibnamefont#1{#1}\fi
\expandafter\ifx\csname bibfnamefont\endcsname\relax
  \def\bibfnamefont#1{#1}\fi
\expandafter\ifx\csname citenamefont\endcsname\relax
  \def\citenamefont#1{#1}\fi
\expandafter\ifx\csname url\endcsname\relax
  \def\url#1{\texttt{#1}}\fi
\expandafter\ifx\csname urlprefix\endcsname\relax\def\urlprefix{URL }\fi
\providecommand{\bibinfo}[2]{#2}
\providecommand{\eprint}[2][]{\url{#2}}

\bibitem[{\citenamefont{Barab\'{a}si and Albert}(1999)}]{BarAlb99}
\bibinfo{author}{\bibfnamefont{A.-L.} \bibnamefont{Barab\'{a}si}}
  \bibnamefont{and} \bibinfo{author}{\bibfnamefont{R.}~\bibnamefont{Albert}},
  \bibinfo{journal}{Science} \textbf{\bibinfo{volume}{286}},
  \bibinfo{pages}{509} (\bibinfo{year}{1999}).

\bibitem[{\citenamefont{Dorogovtsev et~al.}(2000)\citenamefont{Dorogovtsev,
  Mendes, and Samukhin}}]{DoMe00pop}
\bibinfo{author}{\bibfnamefont{S.~N.} \bibnamefont{Dorogovtsev}},
  \bibinfo{author}{\bibfnamefont{J.}~\bibnamefont{Mendes}}, \bibnamefont{and}
  \bibinfo{author}{\bibfnamefont{A.}~\bibnamefont{Samukhin}},
  \bibinfo{journal}{arXiv:cond-mat/0009090}  (\bibinfo{year}{2000}).

\bibitem[{\citenamefont{McPherson et~al.}(2001)\citenamefont{McPherson,
  Smith-Lovin, and Cook}}]{McPh01}
\bibinfo{author}{\bibfnamefont{M.}~\bibnamefont{McPherson}},
  \bibinfo{author}{\bibfnamefont{L.}~\bibnamefont{Smith-Lovin}},
  \bibnamefont{and} \bibinfo{author}{\bibfnamefont{J.~M.} \bibnamefont{Cook}},
  \bibinfo{journal}{Annu Rev Sociol} \textbf{\bibinfo{volume}{27}},
  \bibinfo{pages}{415} (\bibinfo{year}{2001}).

\bibitem[{\citenamefont{Papadopoulos et~al.}(2012)\citenamefont{Papadopoulos,
  Kitsak, Serrano, Bogu\~{n}\'{a}, and Krioukov}}]{PaBoKr11}
\bibinfo{author}{\bibfnamefont{F.}~\bibnamefont{Papadopoulos}},
  \bibinfo{author}{\bibfnamefont{M.}~\bibnamefont{Kitsak}},
  \bibinfo{author}{\bibfnamefont{M.~A.} \bibnamefont{Serrano}},
  \bibinfo{author}{\bibfnamefont{M.}~\bibnamefont{Bogu\~{n}\'{a}}},
  \bibnamefont{and} \bibinfo{author}{\bibfnamefont{D.}~\bibnamefont{Krioukov}},
  \bibinfo{journal}{Nature} \textbf{\bibinfo{volume}{489}}
  (\bibinfo{year}{2012}).

\bibitem[{\citenamefont{Papadopoulos et~al.}(2015)\citenamefont{Papadopoulos,
  Psomas, and Krioukov}}]{hypermap_ton}
\bibinfo{author}{\bibfnamefont{F.}~\bibnamefont{Papadopoulos}},
  \bibinfo{author}{\bibfnamefont{C.}~\bibnamefont{Psomas}}, \bibnamefont{and}
  \bibinfo{author}{\bibfnamefont{D.}~\bibnamefont{Krioukov}},
  \bibinfo{journal}{IEEE/ACM Transactions on Networking}
  \textbf{\bibinfo{volume}{23}}, \bibinfo{pages}{198} (\bibinfo{year}{2015}).

\bibitem[{\citenamefont{Bogu\~{n}\'{a}
  et~al.}(2010)\citenamefont{Bogu\~{n}\'{a}, Papadopoulos, and
  Krioukov}}]{BoPa10}
\bibinfo{author}{\bibfnamefont{M.}~\bibnamefont{Bogu\~{n}\'{a}}},
  \bibinfo{author}{\bibfnamefont{F.}~\bibnamefont{Papadopoulos}},
  \bibnamefont{and} \bibinfo{author}{\bibfnamefont{D.}~\bibnamefont{Krioukov}},
  \bibinfo{journal}{Nature Communications} \textbf{\bibinfo{volume}{1}},
  \bibinfo{pages}{62} (\bibinfo{year}{2010}).

\bibitem[{\citenamefont{Krioukov et~al.}(2010)\citenamefont{Krioukov,
  Papadopoulos, Kitsak, Vahdat, and Bogu\~{n}\'{a}}}]{KrPa10}
\bibinfo{author}{\bibfnamefont{D.}~\bibnamefont{Krioukov}},
  \bibinfo{author}{\bibfnamefont{F.}~\bibnamefont{Papadopoulos}},
  \bibinfo{author}{\bibfnamefont{M.}~\bibnamefont{Kitsak}},
  \bibinfo{author}{\bibfnamefont{A.}~\bibnamefont{Vahdat}}, \bibnamefont{and}
  \bibinfo{author}{\bibfnamefont{M.}~\bibnamefont{Bogu\~{n}\'{a}}},
  \bibinfo{journal}{Physical Review E} \textbf{\bibinfo{volume}{82}},
  \bibinfo{pages}{36106} (\bibinfo{year}{2010}).

\bibitem[{hyp()}]{hypermap-cn}
\emph{\bibinfo{title}{{HyperMap-CN Software Package}}},
  \bibinfo{note}{\url{https://bitbucket.org/dk-lab/2015_code_hypermap}}.

\bibitem[{\citenamefont{Sarkar et~al.}(2011)\citenamefont{Sarkar, Chakrabarti,
  and Moore}}]{sarkar11}
\bibinfo{author}{\bibfnamefont{P.}~\bibnamefont{Sarkar}},
  \bibinfo{author}{\bibfnamefont{D.}~\bibnamefont{Chakrabarti}},
  \bibnamefont{and} \bibinfo{author}{\bibfnamefont{A.~W.} \bibnamefont{Moore}},
  in \emph{\bibinfo{booktitle}{Proceedings of the Twenty-Second international
  joint conference on Artificial Intelligence}} (\bibinfo{publisher}{AAAI
  Press}, \bibinfo{year}{2011}), pp. \bibinfo{pages}{2722--2727}.

\bibitem[{\citenamefont{Kitsak and Krioukov}(2011)}]{maksim_bipartite}
\bibinfo{author}{\bibfnamefont{M.}~\bibnamefont{Kitsak}} \bibnamefont{and}
  \bibinfo{author}{\bibfnamefont{D.}~\bibnamefont{Krioukov}},
  \bibinfo{journal}{Phys. Rev. E} \textbf{\bibinfo{volume}{84}},
  \bibinfo{pages}{026114} (\bibinfo{year}{2011}).

\bibitem[{\citenamefont{Dorogovtsev}(2010)}]{Dorogovtsev10-book}
\bibinfo{author}{\bibfnamefont{S.~N.} \bibnamefont{Dorogovtsev}},
  \emph{\bibinfo{title}{{Lectures on Complex Networks}}}
  (\bibinfo{publisher}{Oxford University Press}, \bibinfo{address}{Oxford},
  \bibinfo{year}{2010}).

\bibitem[{\citenamefont{Bonahon}(2009)}]{Bonahon09-book}
\bibinfo{author}{\bibfnamefont{F.}~\bibnamefont{Bonahon}},
  \emph{\bibinfo{title}{{Low-Dimensional Geometry}}} (\bibinfo{publisher}{AMS},
  \bibinfo{address}{Providence}, \bibinfo{year}{2009}).

\bibitem[{\citenamefont{Ash and
  Dol\'{e}ans-Dade}(1999)}]{probability-and-measure}
\bibinfo{author}{\bibfnamefont{R.~B.} \bibnamefont{Ash}} \bibnamefont{and}
  \bibinfo{author}{\bibfnamefont{C.~A.} \bibnamefont{Dol\'{e}ans-Dade}},
  \emph{\bibinfo{title}{{Probability \& Measure Theory, Second Edition}}}
  (\bibinfo{publisher}{Academic Press}, \bibinfo{year}{1999}).

\bibitem[{\citenamefont{Bogu\~{n}\'{a}
  et~al.}(2009)\citenamefont{Bogu\~{n}\'{a}, Krioukov, and claffy}}]{BoKrKc08}
\bibinfo{author}{\bibfnamefont{M.}~\bibnamefont{Bogu\~{n}\'{a}}},
  \bibinfo{author}{\bibfnamefont{D.}~\bibnamefont{Krioukov}}, \bibnamefont{and}
  \bibinfo{author}{\bibfnamefont{K.}~\bibnamefont{claffy}},
  \bibinfo{journal}{Nature Physics} \textbf{\bibinfo{volume}{5}},
  \bibinfo{pages}{74} (\bibinfo{year}{2009}).

\bibitem[{\citenamefont{Claffy et~al.}(2009)\citenamefont{Claffy, Hyun, Keys,
  Fomenkov, and Krioukov}}]{ClHy09}
\bibinfo{author}{\bibfnamefont{K.}~\bibnamefont{Claffy}},
  \bibinfo{author}{\bibfnamefont{Y.}~\bibnamefont{Hyun}},
  \bibinfo{author}{\bibfnamefont{K.}~\bibnamefont{Keys}},
  \bibinfo{author}{\bibfnamefont{M.}~\bibnamefont{Fomenkov}}, \bibnamefont{and}
  \bibinfo{author}{\bibfnamefont{D.}~\bibnamefont{Krioukov}}, in
  \emph{\bibinfo{booktitle}{CATCH}} (\bibinfo{publisher}{IEEE Computer
  Society}, \bibinfo{year}{2009}),
  \urlprefix\url{http://www.caida.org/projects/ark/}.

\bibitem[{as_()}]{as_topo_data}
\emph{\bibinfo{title}{{IPv4 Routed /24 AS Links Dataset}}},
  \bibinfo{note}{\url{http://www.caida.org/data/active/ipv4_routed_topology_aslinks_dataset.xml}}.

\bibitem[{\citenamefont{Lu and Zhou}(2011)}]{LuZhou11}
\bibinfo{author}{\bibfnamefont{L.}~\bibnamefont{Lu}} \bibnamefont{and}
  \bibinfo{author}{\bibfnamefont{T.}~\bibnamefont{Zhou}},
  \bibinfo{journal}{Physica A: Statistical Mechanics and its Applications}
  \textbf{\bibinfo{volume}{390}}, \bibinfo{pages}{1150} (\bibinfo{year}{2011}).

\bibitem[{\citenamefont{Dimitropoulos et~al.}(2007)\citenamefont{Dimitropoulos,
  Krioukov, Fomenkov, Huffaker, Hyun, claffy, and Riley}}]{DiKrFo06}
\bibinfo{author}{\bibfnamefont{X.}~\bibnamefont{Dimitropoulos}},
  \bibinfo{author}{\bibfnamefont{D.}~\bibnamefont{Krioukov}},
  \bibinfo{author}{\bibfnamefont{M.}~\bibnamefont{Fomenkov}},
  \bibinfo{author}{\bibfnamefont{B.}~\bibnamefont{Huffaker}},
  \bibinfo{author}{\bibfnamefont{Y.}~\bibnamefont{Hyun}},
  \bibinfo{author}{\bibfnamefont{K.}~\bibnamefont{claffy}}, \bibnamefont{and}
  \bibinfo{author}{\bibfnamefont{G.}~\bibnamefont{Riley}},
  \bibinfo{journal}{Comput Commun Rev} \textbf{\bibinfo{volume}{37}},
  \bibinfo{pages}{29} (\bibinfo{year}{2007}).

\bibitem[{\citenamefont{Shavitt and Tankel}(2008)}]{ShaTa08}
\bibinfo{author}{\bibfnamefont{Y.}~\bibnamefont{Shavitt}} \bibnamefont{and}
  \bibinfo{author}{\bibfnamefont{T.}~\bibnamefont{Tankel}},
  \bibinfo{journal}{IEEE/ACM Transactions on Networking}
  \textbf{\bibinfo{volume}{16}} (\bibinfo{year}{2008}).

\bibitem[{\citenamefont{Begelfor and Werman}(2005)}]{curved_learning}
\bibinfo{author}{\bibfnamefont{E.}~\bibnamefont{Begelfor}} \bibnamefont{and}
  \bibinfo{author}{\bibfnamefont{M.}~\bibnamefont{Werman}},
  \bibinfo{journal}{Tech. Rep. HUJI-CSE-LTR-2006-191, School of Engineering and
  Computer Science, Hebrew University of Jerusalem}  (\bibinfo{year}{2005}),
  \bibinfo{note}{\url{http://www.cs.huji.ac.il/~werman/Papers/cmds.pdf}}.

\bibitem[{\citenamefont{Sarkar and Moore}(2005)}]{sarkar_embedding}
\bibinfo{author}{\bibfnamefont{P.}~\bibnamefont{Sarkar}} \bibnamefont{and}
  \bibinfo{author}{\bibfnamefont{A.~W.} \bibnamefont{Moore}},
  \bibinfo{journal}{SIGKDD Explor. Newsl.} \textbf{\bibinfo{volume}{7}},
  \bibinfo{pages}{31} (\bibinfo{year}{2005}).

\bibitem[{\citenamefont{Hoff et~al.}(2002)\citenamefont{Hoff, Raftery, and
  Handcock}}]{raftery_model}
\bibinfo{author}{\bibfnamefont{P.~D.} \bibnamefont{Hoff}},
  \bibinfo{author}{\bibfnamefont{A.~E.} \bibnamefont{Raftery}},
  \bibnamefont{and} \bibinfo{author}{\bibfnamefont{M.~S.}
  \bibnamefont{Handcock}}, \bibinfo{journal}{J. Amer. Stat. Assoc.}
  \textbf{\bibinfo{volume}{97}} (\bibinfo{year}{2002}).

\bibitem[{\citenamefont{Nocedal and Wright}(2000)}]{optimization_book}
\bibinfo{author}{\bibfnamefont{J.}~\bibnamefont{Nocedal}} \bibnamefont{and}
  \bibinfo{author}{\bibfnamefont{S.}~\bibnamefont{Wright}},
  \emph{\bibinfo{title}{{Numerical Optimization}}}
  (\bibinfo{publisher}{Springer}, \bibinfo{year}{2000}).

\bibitem[{\citenamefont{Jaynes}(1968)}]{priors}
\bibinfo{author}{\bibfnamefont{E.}~\bibnamefont{Jaynes}},
  \bibinfo{journal}{IEEE Transactions on Systems Science and Cybernetics}
  \textbf{\bibinfo{volume}{4}}, \bibinfo{pages}{227} (\bibinfo{year}{1968}).

\bibitem[{\citenamefont{Zuev et~al.}(2015)\citenamefont{Zuev, Bogu\~{n}\'{a},
  Bianconi, and Krioukov}}]{ZuBo15}
\bibinfo{author}{\bibfnamefont{K.}~\bibnamefont{Zuev}},
  \bibinfo{author}{\bibfnamefont{M.}~\bibnamefont{Bogu\~{n}\'{a}}},
  \bibinfo{author}{\bibfnamefont{G.}~\bibnamefont{Bianconi}}, \bibnamefont{and}
  \bibinfo{author}{\bibfnamefont{D.}~\bibnamefont{Krioukov}},
  \bibinfo{journal}{arXiv:1501.06835}  (\bibinfo{year}{2015}).

\bibitem[{\citenamefont{Newman}(2003)}]{newman03c-review}
\bibinfo{author}{\bibfnamefont{M.~E.~J.} \bibnamefont{Newman}},
  \bibinfo{journal}{SIAM Rev} \textbf{\bibinfo{volume}{45}},
  \bibinfo{pages}{167} (\bibinfo{year}{2003}).

\end{thebibliography}

\end{document}